\documentclass[useAMS,usenatbib]{mn2e}
\usepackage{txfonts}
\usepackage{graphicx}
\usepackage{natbib}
\usepackage{amssymb}

\def\del#1{{}}

\newcommand{\dd}{\mathrm{d}}
\newcommand{\bra}{\langle}
\newcommand{\ket}{\rangle}
\newcommand{\ltsima}{$\; \buildrel < \over \sim \;$}
\newcommand{\lsim}{\lower.5ex\hbox{\ltsima}}
\newcommand{\gtsima}{$\; \buildrel > \over \sim \;$}
\newcommand{\gsim}{\lower.5ex\hbox{\gtsima}}

\newcommand{\e}{\rmn{e}}
\newcommand{\p}{\rmn{p}}

\newcommand{\CR}{{\rm CR}}
\renewcommand{\th}{{\rm th}}

\newcommand{\M}{{\mathcal M}}

\newcommand{\eff}{\mathrm{eff}}
\newcommand{\vir}{\mathrm{vir}}
\newcommand{\eps}{\varepsilon}
\newcommand{\vecbf}{\mathbfit}
\newcommand{\vel}{\upsilon}

\title[Cosmic rays in clusters of galaxies -- I. Effects on thermal cluster observables]
{Simulating cosmic rays in clusters of galaxies -- I. Effects on the
Sunyaev-Zel'dovich effect and the X-ray emission}

\author[C.~Pfrommer et al.]
  {C.~Pfrommer,$^1$\thanks{e-mail: pfrommer@cita.utoronto.ca (CP);
  ensslin@mpa-garching.mpg.de (TAE); volker@mpa-garching.mpg.de (VS);
  jubelgas@mpa-garching.mpg.de (MJ); kdolag@mpa-garching.mpg.de (KD)} 
  T.~A.~En{\ss}lin,$^2$\footnotemark[1]
  V.~Springel,$^2$\footnotemark[1] M.~Jubelgas,$^2$\footnotemark[1]
  K.~Dolag$^2$\footnotemark[1] \\ 
  $^1$Canadian Institute for Theoretical Astrophysics, University of Toronto,
  60 St. George Street, Toronto, Ontario, M5S 3H8, Canada \\
  $^2$Max-Planck-Institut f\"ur Astrophysik, Karl-Schwarzschild-Stra{\ss}e 1,
  Postfach 1317, 85741 Garching, Germany }

\begin{document}
\pagerange{\pageref{firstpage}--\pageref{lastpage}} \pubyear{2006}
\maketitle
\label{firstpage}

\begin{abstract}
  We performed high-resolution simulations of a sample of 14 galaxy clusters
  that span a mass range from $5\times 10^{13} h^{-1}\, \rmn{M}_\odot$ to
  $2\times 10^{15} h^{-1}\, \rmn{M}_\odot$ to study the effects of cosmic rays
  (CRs) on thermal cluster observables such as X-ray emission and the
  Sunyaev-Zel'dovich effect.  We analyse the CR effects on the intra-cluster
  medium while simultaneously taking into account the cluster's dynamical state
  as well as the mass of the cluster.  The modelling of the cosmic ray physics
  includes adiabatic CR transport processes, injection by supernovae and
  cosmological structure formation shocks, as well as CR thermalization by
  Coulomb interaction and catastrophic losses by hadronic interactions. While
  the relative pressure contained in CRs within the virial radius is of the
  order of 2 per cent in our non-radiative simulations, their contribution
  rises to 32 per cent in our simulations with dissipative gas physics
  including radiative cooling, star formation, and supernova feedback.  The
  relative CR pressure rises towards the outer regions due to a combination of
  the following effects: CR acceleration is more efficient at the peripheral
  strong accretion shocks compared to weak central flow shocks, adiabatic
  compression of a composite of CRs and thermal gas disfavours the CR pressure
  relative to the thermal pressure due to the softer equation of state of CRs,
  and CR loss processes are more important at the dense centres. Interestingly,
  in the radiative simulations the relative CR pressure reaches high values of
  the order of equipartition with the thermal gas in each cluster galaxy due to
  the fast thermal cooling of gas which diminishes the thermal pressure support
  relative to that in CRs.  This also leads to a lower effective adiabatic
  index of the composite gas that increases the compressibility of the
  intra-cluster medium. This effect slightly increases the central density,
  thermal pressure and the gas fraction. While the X-ray luminosity in low mass
  cool core clusters is boosted by up to 40 per cent, the integrated
  Sunyaev-Zel'dovich effect appears to be remarkably robust and the total flux
  decrement only slightly reduced by typically 2 per cent. The resolved
  Sunyaev-Zel'dovich maps, however, show a larger variation with an increased
  central flux decrement.
\end{abstract}


\section{Introduction}

Cosmic ray protons (hereafter referred to as CRs) play a decisive role within
the interstellar medium of our own Galaxy.  CRs and magnetic fields each
contribute roughly as much energy and pressure to the galactic ISM as the
thermal gas does, CRs trace past energetic events such as supernovae, and they
reveal the underlying structure of the baryonic matter distribution through
their interactions.  In contrast, it is unknown how much pressure support is
provided by CRs to the thermal plasma of clusters of galaxies. A substantial CR
pressure contribution might have a major impact on the properties of the
intra-cluster medium (ICM) and potentially modify thermal cluster observables
such as the X-ray emission and the Sunyaev-Zel'dovich (SZ) effect
\citep[][]{1972CoASP...4..173S}. This effect arises since photons from the
cosmic microwave background experience inverse Compton up-scattering by thermal
electrons of the ICM. CR induced modifications might imprint themselves in
changes of cluster scaling relations or modify their intrinsic scatter and thus
change the effective mass threshold of X-ray or SZ surveys. Neglecting such a
CR component in reference simulations might introduce biases in the
determination of cosmological parameters. Finally, simulating realistic CR
distributions within galaxy clusters enables us to predict the non-thermal
cluster emission resulting from CR interactions with thermal gas protons. The
resulting model signatures of the expected radio synchrotron and $\gamma$-ray
emission can then be tested with the upcoming new generation of $\gamma$-ray
instruments (e.g.~imaging atmospheric \v{C}erenkov telescopes and the
GLAST\footnote{Gamma-ray Large Area Space Telescope,
http://glast.gsfc.nasa.gov/} satellite) and radio telescopes
(e.g.~LOFAR\footnote{LOw Frequency ARray, http://www.lofar.org/} and extended
Very Large Array\footnote{http://www.aoc.nrao.edu/evla/}).

A thorough dynamical treatment of CR transport and loss processes in
numerical simulations of cosmological large-scale structure formation
has not yet been attempted due to the very complex CR physics
involved, despite the potential dynamical importance of CRs.  There
have been pioneering efforts to study the CR acceleration into the
intergalactic medium (IGM) on large scales \citep{2001ApJ...559...59M,
2002ApJ...579..337K, 2003JKAS...36..105R, 2004JKAS...37..477R,
2005ApJ...620...44K}. However, these codes were quite limited in their
adaptive resolution capability and neglected dissipative gas physics
including radiative cooling, star formation, and supernova feedback.
There have also been numerical modelling of discretised CR energy
spectra on top of grid-based cosmological
simulations~\citep{2001CoPhC.141...17M}, but so far these
implementations neglected the hydrodynamic pressure of the CR
component and were unable to resolve the observationally accessible,
dense central regions of clusters.

First important results underscoring the importance of CR populations in
clusters were obtained in non-radiative cosmological simulations and predicted
relative CR pressure contributions of $P_\CR / P_\th \sim 20$ per cent up to 50
per cent \citep{2001ApJ...559...59M}, which also motivated suggestions that CRs
may slow-down cluster cooling flows \citep{2005ApJ...620..191C}. Since the CR
acceleration depends crucially on the shock strength of cosmological structure
formation shocks, there has been a series of efforts undertaken to develop
schemes that are capable of measuring the Mach numbers in cosmological
simulations \citep{2000ApJ...542..608M, 2003ApJ...593..599R,
2006MNRAS.367..113P}. To allow studies of the dynamical effects of CRs in
radiatively cooling galactic and cluster environments, we have developed a CR
formalism that is based on smoothed particle hydrodynamical representation of
the equations of motion.  We introduced a number of approximations to reduce
the complexity of the problem while still being able to capture as many
physical properties and peculiarities of CR populations as possible.  In our
model, the emphasis is given to the dynamical impact of CRs on hydrodynamics,
and not on an accurate spectral representation of the CRs. The guiding
principles are energy, pressure, and particle number conservations, as well as
adiabatic invariants.  Non-adiabatic processes are mapped onto modifications of
these principles \citep{2006...Ensslin, 2006...Jubelgas, 2006MNRAS.367..113P}.

\begin{figure}
\resizebox{\hsize}{!}{\includegraphics{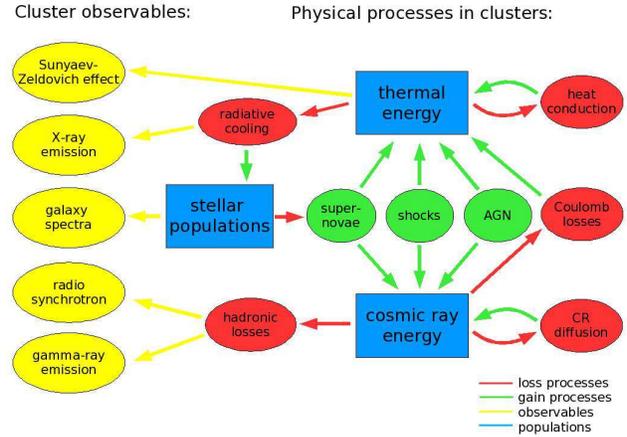}}
\caption{Overview over the relevant physical processes in galaxy clusters. The
  right-hand side shows the interplay of different physical processes while the
  left-hand side shows observables that inform about the properties of
  cluster and their dynamical state. Gain processes are denoted in green and
  loss respectively redistribution processes are denoted in red.}
\label{fig:CR_flowchart}
\end{figure}

A simplified overview over the relevant physical processes in galaxy clusters
is given in Fig.~\ref{fig:CR_flowchart}. The {\em upper central part} shows
standard processes which are usually considered in simulations. Radiative
cooling of the gas leads eventually to star formation in the densest regions
that exceed a certain density threshold. This happens in the central cluster
regions and within substructures leading to individual galaxies.  Once the
nuclear energy has been used up, massive stars explode in supernovae that drive
strong shock waves into the ambient interstellar medium (ISM) which resupply
thermal and turbulent energy. On larger scales, structure formation shock waves
dissipate gravitational energy associated with hierarchical clustering into
thermal energy of the gas, thus supplying the intra-cluster medium (ICM) with
entropy and thermal pressure support.  There are three main observables
associated with these processes: the hot ICM emits thermal bremsstrahlung
radiation with an X-ray luminosity that depends on the square of the electron
number density.  The amplitude of the Sunyaev-Zel'dovich effect
\citep[][]{1972CoASP...4..173S} depends on the pressure of the thermal electron
population integrated along the line-of-sight through the cluster.  Finally,
galaxy spectra probe directly the stellar populations of intra-cluster galaxies
and indirectly the cluster's potential through their velocity dispersion
\citep[for reviews see][]{1988xrec.book.....S, 2005RMP...77...207V}.

The {\em lower part} of Fig.~\ref{fig:CR_flowchart} sketches the cosmic ray
physics within clusters. CRs behave differently compared to the thermal
gas. Their equation of state is softer, they are able to travel actively over
macroscopic distances, and their energy loss time-scales are typically larger
than the thermal ones.  Besides thermalization, collisionless shocks are also
able to accelerate ions of the high-energy tail of the Maxwellian through
diffusive shock acceleration \citep[for reviews see][]{1983RPPh...46..973D,
1987PhR...154....1B, 2001RPPh...64..429M}. These energetic ions are reflected
at magnetic irregularities through magnetic resonances between the gyro-motion
and waves in the magnetised plasma and are able to gain energy in moving back
and forth through the shock front. This acceleration process typically yields a
cosmic ray proton (CR) population with a power-law distribution of the particle
momenta. CRs are accelerated on galactic scales through supernova shocks while
they are injected by structure formation shock waves on even larger scales up
to tens of Mpc. 

The ubiquitous cosmic magnetic fields couple the otherwise dynamically
independent ingredients like the ICM plasma, and the CR gas into a single,
however complex fluid. Magnetic fields prevent charged relativistic particles
to travel macroscopic distances with their intrinsic velocity close to the
speed of light. Instead, the particles gyrate around, and travel slowly along
magnetic field lines. Occasionally, they get scattered on magnetic
irregularities.  On macroscopic scales, the transport can often be described as
a diffusion process if the gyro-radius can be regarded to be small. Thus, CR
diffusion redistributes the CR energy density macroscopically. Thermal heat
conduction is an analogous process that reallocates the thermal energy of the
ICM. The CR energy reservoir suffers two main loss processes: (1) CR energy is
transferred into the thermal energy reservoir through individual electron
scatterings in the Coulomb field of the CR particle as well as by small
momentum transfers through excitations of quantised plasma oscillations. We
refer to the sum of both effects as Coulomb losses.  (2) Provided the CR
momentum exceeds the threshold $p \simeq 0.8\,\mbox{GeV} / c$ for the hadronic
reaction with ambient protons, they produce pions which decay into secondary
electrons (and neutrinos) and $\gamma$-rays:
\begin{eqnarray}
  \pi^\pm &\rightarrow& \mu^\pm + \nu_{\mu}/\bar{\nu}_{\mu} \rightarrow
  e^\pm + \nu_{e}/\bar{\nu}_{e} + \nu_{\mu} + \bar{\nu}_{\mu}\nonumber\\
  \pi^0 &\rightarrow& 2 \gamma \,.\nonumber
\end{eqnarray}
Only CR protons above this kinematic threshold are therefore visible through
their decay products via radiative processes, making them directly
observationally detectable. These hadronically produced relativistic electrons
and positrons can emit a halo of radio synchrotron emission in the presence of
ubiquitous intra-cluster magnetic fields. However, there are other processes
that accelerate relativistic electrons such as resonant and non-resonant
interactions of mildly relativistic electrons with magneto-hydrodynamic waves
or turbulent spectra. Since the distribution of magnetic field strengths with
cluster radius is also not well known, radio synchrotron emission alone has
limited predictive power. Future $\gamma$-ray satellites should be able to
detect the associated hadronically induced $\gamma$-ray emission resulting from
neutral pion decay and allow stronger conclusions on the parent CR population
in clusters.

So far, we have neglected feedback from active galactic nuclei (AGN) in our
simulations despite its importance \citep[for first models of the thermal
feedback within cosmological simulations, see][]{2006MNRAS.366..397S}.
Gravitational energy associated with the accretion onto super-massive black
holes is converted into large-scale jets and eventually dissipated into thermal
and CR energy. A reliable model would be required to self-regulate and respond
to the instantaneous cooling rate.

This paper studies the interplay of thermal gas and CRs and their effect on the
observables of the thermal gas such as X-ray emission and the
Sunyaev-Zel'dovich effect. Two companion papers are studying directly the CR
related multi-frequency observables and the systematic bias and scatter of
cluster scaling relations that are associated with CRs and how these
modifications effect the determination of cosmological parameters. The outline
of the paper is as follows.  Section~\ref{sec:sims} describes the general setup
of the simulations, the different physical processes which we simulated, and
our cluster sample.  In Sect.~\ref{sec:results}, we present the results on the
relative CR pressure integrated over the clusters, and show the profiles of
thermodynamical and CR-related quantities separately for non-radiative and
radiative simulations. Both types of simulation models will be concluded by an
emerging picture of CR-induced changes to the ICM.
Section~\ref{sec:physical_processes} analyses physical processes related to CRs
in detail and complements the previous section. Finally,
Sect.~\ref{sec:thermal_obs} studies the consequences of these processes for
thermal observables such as X-ray emission and the Sunyaev-Zel'dovich effect.

\section{The simulations}
\label{sec:sims}

\subsection{General setup}

Simulations were performed using the ``concordance'' cosmological cold dark
matter model with a cosmological constant ($\Lambda$CDM). The cosmological
parameters of our model are: $\Omega_\rmn{m} = \Omega_\rmn{DM} +
\Omega_\rmn{b} = 0.3$, $\Omega_\rmn{b} = 0.039$, $\Omega_\Lambda = 0.7$, $h =
0.7$, $n = 1$, and $\sigma_8 = 0.9$. Here, $\Omega_\rmn{m}$ denotes the total
matter density in units of the critical density for geometrical closure today,
$\rho_\rmn{crit} = 3 H_0^2 / (8 \upi G)$. $\Omega_\rmn{b}$ and
$\Omega_\Lambda$ denote the densities of baryons and the cosmological constant
at the present day. The Hubble constant at the present day is parametrized as
$H_0 = 100\,h \mbox{ km s}^{-1} \mbox{Mpc}^{-1}$, while $n$ denotes the
spectral index of the primordial power-spectrum, and $\sigma_8$ is the {\em
  rms} linear mass fluctuation within a sphere of radius $8\,h^{-1}$Mpc
extrapolated to $z=0$. This model yields a reasonable fit to current
cosmological constraints. 

Our simulations were carried out with an updated and extended version of the
distributed-memory parallel TreeSPH code GADGET-2 \citep{2005MNRAS.364.1105S,
2001NewA....6...79S} that includes self-consistent cosmic ray physics
\citep{2006...Ensslin, 2006...Jubelgas, 2006MNRAS.367..113P}. Gravitational
forces were computed using a combination of particle-mesh and tree algorithms.
Hydrodynamic forces are computed with a variant of the smoothed particle
hydrodynamics (SPH) algorithm that conserves energy and entropy where
appropriate, i.e. outside of shocked regions \citep{2002MNRAS.333..649S}.

\begin{table*}
\caption{\scshape: Different physical processes used in our simulation models:}
\begin{tabular}{l | c c c  c c c }
\hline
\hline
Simulated physics$^{(1)}$ & \multicolumn{3}{c}{non-radiative simulation models$^{(1)}$:} 
                        & \multicolumn{3}{c}{radiative simulation models$^{(1)}$:} \\
& reference mod. & shock-CR mod. & simple shock-CR mod.
& reference mod. & shock-CR mod. & complete CR mod. \\
\hline
thermal shock heating & \checkmark & \checkmark & \checkmark & \checkmark & \checkmark & \checkmark \\
radiative cooling     & & & & \checkmark & \checkmark & \checkmark \\
star formation        & & & & \checkmark & \checkmark & \checkmark \\
Coulomb CR losses     & & \checkmark & \checkmark & & \checkmark & \checkmark \\
hadronic CR losses     & & \checkmark & \checkmark & & \checkmark & \checkmark \\
shock-CRs with $\zeta = \mbox{const}^{(2)}$ & & & \checkmark & & & \\
shock-CRs with $\zeta(\M)^{(2)}$ & & \checkmark & & & \checkmark & \checkmark \\
supernova-CRs              & & & & & & \checkmark \\
\hline
\end{tabular}   
\begin{quote} 
 {\scshape Notes:}\\ 
(1) This table serves as an overview over our simulated models. The first column
 shows the simulated physics, the next three columns specify our non-radiative
 simulation models, while the last three columns show our radiative simulation
 models. \\
(2) We employ two different CR injection schemes at structure formation shocks:
 The realistic one uses a Mach number dependent energy injection efficiency
 $\zeta(\M)$ while the simplified scheme uses a constant energy injection
 efficiency, $\zeta = \mbox{const}$. The latter scheme is a
 simplification that exaggerates the CR effects for better visibility.  For
 further details and references please refer to Sec.~\ref{sec:models}.\\
\end{quote}
\label{tab:models}
\end{table*} 

We have performed high-resolution hydrodynamic simulations of the
formation of 14 galaxy clusters. The clusters span a mass range from
$5 \times 10^{13}\, h^{-1}\, \rmn{M}_\odot$ to $2 \times 10^{15}\, h^{-1}\,
\rmn{M}_\odot$ and have originally been selected from a low-resolution
dark-matter- (DM)-only simulation \citep{2001MNRAS.328..669Y} with
box-size $479\, h^{-1} \,\mbox{Mpc}$ of a flat $\Lambda$CDM model with
the parameters mentioned above. Using the `zoomed initial conditions'
technique \citep{1993ApJ...412..455K, 1997MNRAS.286..865T}, we then
re-simulated the clusters with higher mass and force resolution by
adding short-wavelength modes within the Lagrangian regions in the
initial conditions that will evolve later-on into the structures of
interest \citep[using initial conditions from][]{2005MNRAS.364..753D}.
In these high-resolution regions, dark matter particles of the parent
simulation are split into a dark matter and gaseous part, with the
mass ratio reflecting the value of the cosmic baryon fraction.  Force
and mass resolution are then gradually degraded in regions larger than
at least 3-5 virial radii from the clusters to limit the computational
cost while providing a correct representation of the large scale tidal
gravitational field. Using an iterative method, the size of the cluster
region that is re-simulated at high-resolution was always chosen to be
large enough to prevent any contamination with heavy dark matter
particles.

In high-resolution regions, the dark matter particles had masses of $m_\rmn{DM}
= 1.13 \times 10^9\,h^{-1}\,\rmn{M}_\odot$ and SPH particles $m_\rmn{gas} = 1.7\times
10^8\,h^{-1}\,\rmn{M}_\odot$ so each individual cluster is resolved by $8 \times
10^4$ to $4\times 10^6$ particles, depending on its final mass. For details of
cluster properties, see Sect.~\ref{sec:clusters}.  The SPH densities were
computed from 48 neighbours, allowing the SPH smoothing length to drop at most
to half of the value of the gravitational softening length of the gas
particles. This choice of the SPH smoothing length leads to our minimum gas
resolution of approximately $8\times 10^{9}\,h^{-1}\,\rmn{M}_\odot$.  For the initial
redshift we chose $1+z_\rmn{init}=60$.  The gravitational force softening was
of a spline form \citep[e.g.,][]{1989ApJS...70..419H} with a Plummer equivalent
softening length that is assumed to have a constant comoving scale down to $z =
5$, and a constant value of $5\,h^{-1}$kpc in physical units at later epochs.

\subsection{The models}
\label{sec:models}

\begin{figure}
\resizebox{\hsize}{!}{\includegraphics{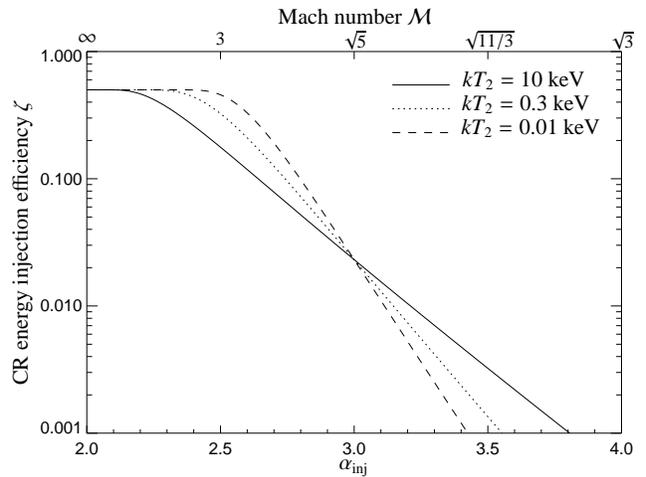}}
\caption{CR energy injection efficiency for the diffusive shock acceleration
  process. Shown is the CR energy injection efficiency $\zeta =
  \eps_\rmn{CR,inj}/\eps_\rmn{diss}$ for the three post-shock temperatures
  $kT_2/\mbox{keV} = 0.01, 0.3$, and 10. For this illustrative figure, we
  consider only CRs above the kinematic momentum threshold $q_\rmn{threshold} =
  p_\CR / (m_\p c) = 0.83$ of the hadronic CR p-p interaction which are able to
  produce pions that successively decay into secondary electrons, neutrinos,
  and $\gamma$-rays. We choose equipartition between injected CR energy and
  dissipated thermal non-relativistic energy as plausible saturation value of the CR
  energy injection efficiency, i.e. $\zeta_\rmn{max} = 0.5$
  \citep{2003ApJ...593..599R, 2006...Ensslin}.}
\label{fig:zeta}
\end{figure}

For each galaxy cluster we ran six different simulations separated into two
sets of three simulations each (cf.~Table~\ref{tab:models}). One set uses only
non-radiative gas physics while the other set consists of three simulations
that include radiative cooling and star formation. The first simulation of the
non-radiative set is our reference simulation with non-radiative physics only,
i.e.~the gas is transported adiabatically unless it experiences structure
formation shock waves that supply the gas with entropy and thermal pressure
support. The second simulation of the set allows for a simplified model of CR
acceleration at structure formation shocks with a constant energy injection
efficiency of $\zeta = \eps_\rmn{CR,inj}/\eps_\rmn{diss} = 0.5$, independent of
the shock strength, while reducing at the same time the amount of thermal
dissipation to ensure energy conservation at the shock. Here,
$\eps_\rmn{CR,inj}$ denotes the injected CR energy density and
$\eps_\rmn{diss}$ the total dissipated energy at the shock.  This model
represents an extreme simplification that exaggerates the CR effects for better
visibility. All CR models include Coulomb and hadronic loss processes. The
third simulation of the non-radiative set uses a more realistic model for the
CR acceleration at structure formation shocks which is based on the thermal
leakage model \citep[e.g.,][]{1984ApJ...286..691E, 1995ApJ...447..944K,
2006...Ensslin}. It employs the fact that the most energetic ions in the
exponential tail of the Maxwellian of the post-shock regime are able to
participate in the process of diffusive shock acceleration and injects a
power-law in momentum that depends on the instantaneous shock strength in the
simulation \citep[for details, cf.][]{2006MNRAS.367..113P}. With this thermal
leakage model, we are able to derive a CR energy injection efficiency,
$\zeta(\M)$, for the diffusive shock acceleration process. It depends only on
the Mach number of the shock, $\M= \vel_\rmn{s} / c_1$ (shock velocity
$\vel_\rmn{s}$ divided by pre-shock sound speed $c_1$), the post-shock
temperature $T_2$, and a saturation value of the CR energy injection efficiency
at high values of the Mach number, $\zeta_\rmn{max} = 0.5$ as shown in
Fig.~\ref{fig:zeta} \citep{2003ApJ...593..599R, 2006...Ensslin}.

The first simulation of our radiative set is a reference simulation
with radiative cooling and star formation, but no CR physics. The
second radiative simulation accounts for CR acceleration at structure
formation shocks and allows for all CR loss processes. Finally, the
last simulation additionally assumes that a constant fraction
$\zeta_\rmn{SN} = 0.3$ of the kinetic energy of a supernova ends up in
the CR population, which is motivated by TeV $\gamma$-ray observations
of a supernova remnant that find an energy fraction of $\zeta_\rmn{SN}
\simeq 0.1 - 0.3$ when extrapolating the CR distribution function
\citep{2006Natur.439..695A}. We choose a maximum value for this
supernova energy efficiency owing to the large uncertainties and our
aim to bracket the realistic case with the two radiative CR
simulations.

Radiative cooling was computed assuming an optically thin gas of primordial
composition (mass-fraction of $X = 0.76$ for hydrogen and $1-X = 0.24$ for
helium) in collisional ionisation equilibrium, following
\citet{1996ApJS..105...19K}. We have also included heating by a photo-ionising,
time-dependent, uniform ultraviolet (UV) background expected from a population
of quasars \citep{1996ApJ...461...20H}, which reionizes the Universe at $z
\simeq 6$.  The effect of a photo-ionising background is that of preventing the
collapse of overdense regions of gas and inhibiting subsequent star formation
within halos of sub-$L_*$ galaxies \citep[e.g.,][]{2002MNRAS.333..156B}, thus
having a secondary effect on the resolution of our simulation.  If not
counteracted by any feedback process, radiative cooling in high-resolution
simulations is well known to give rise to a central cooling flow that
overproduces the amount of stars not only in the group/cluster overdense
environment but also in the average environment \citep[e.g.,][ and references
therein]{2001MNRAS.326.1228B, 2004MNRAS.348.1078B, 2006MNRAS.367.1641B}.  As
discussed by \citet{2003MNRAS.339..289S}, their multiphase ISM model with
supernovae feedback alone does however not fully resolve this problem, despite
its ability to regulate the consumption of cold gas into stars within the
ISM. The reason for this is because the cooling rates within halos remain
essentially unaffected in the model, i.e.~the supply of gas to the dense
star-forming ISM is largely unchanged, while by construction the phases of the
ISM remain coupled to each other, preventing baryons to leave the ISM (except
for dynamical effects like gas stripping in mergers). Being mainly interested
in global properties of the ICM, we do not include the metallicity effect in
the cooling function due both to the approximate treatment of metal generation
and diffusion and in order not to exaggerate the cooling problem of the cluster
cores since we also do not follow the AGN feedback that seems to be essential
to reliably model the dense cluster centres
\citep[e.g.,][]{2006MNRAS.366..397S}.

Star formation is treated using the hybrid multiphase model for the
interstellar medium introduced by \citet{2003MNRAS.339..289S}. In short, the
ISM is pictured as a two-phase fluid consisting of cold clouds that are
embedded at pressure equilibrium in an ambient hot medium.  The clouds form
from the cooling of high density gas, and represent the reservoir of baryons
available for star formation. When stars form, according to a Salpeter initial
mass function (IMF) \citep{1955ApJ...121..161S}, the energy released by
supernovae heats the ambient hot phase of the ISM, and in addition, clouds in
supernova remnants are evaporated. These effects establish a tightly
self-regulated regime for star formation in the ISM.  In practice, the scheme
is numerically implemented as a sub-resolution model, i.e.~cold clouds are not
resolved individually. Instead, only their total mass fraction in each element
of the ISM is computed, otherwise they are treated in a stochastic fashion with
their collective effect on the ISM dynamics being described by an effective
equation of state.  The numerical implementation of this multiphase model
describes each gas particle as composed by a hot component, having its own mass
and density, and a cold neutral component. The relative amount of these two
phases is determined by the local value of gas density and temperature.

Cosmic ray physics was computed by using a new formulation that follows the
most important injection and loss processes self-consistently while accounting
for the CR pressure in the equations of motion \citep{2006...Ensslin,
2006...Jubelgas, 2006MNRAS.367..113P}. We refer to these papers for a detailed
description of the formalism, providing here only a short summary of the model.
In our methodology, the non-thermal cosmic ray population of each gaseous fluid
element is approximated by a simple power law spectrum in particle momentum,
characterised by an amplitude, a low-momentum cut-off, and a fixed slope
$\alpha = 2.3$. Adiabatic CR transport processes such as compression and
rarefaction, and a number of physical source and sink terms which modify the
cosmic ray pressure of each particle are modelled. The most important sources
considered are injection by supernovae (in our radiative simulations) and
diffusive shock acceleration at cosmological structure formation shocks, while
the primary sinks are thermalization by Coulomb interactions, and catastrophic
losses by hadronic interactions.

\subsection{Simulated clusters}
\label{sec:clusters}

\begin{table}
\caption{\scshape: Cluster sample}
\begin{tabular}{l l l l r r}
\hline
\hline
Cluster & sim.'s & dyn. state$^{(1)}$ & $M_{200}^{(2)}$ & $R_{200}^{(2)}$ & $kT_{200}^{(3)}$ \\
& & & [$h^{-1}\,\rmn{M}_\odot$] & [$h^{-1}\,$Mpc] & [keV] \\
\hline
1  & g8a  & CC    & $1.8\times 10^{15}$ & 2.0  & 13.1 \\
2  & g1a  & CC    & $1.3\times 10^{15}$ & 1.8  & 10.6 \\
3  & g72a & PostM & $1.1\times 10^{15}$ & 1.7  & 9.4  \\
4  & g51  & CC    & $1.1\times 10^{15}$ & 1.7  & 9.4  \\
                                                    
5  & g1b  & M     & $3.7\times 10^{14}$ & 1.2  & 4.7  \\
6  & g72b & M     & $1.5\times 10^{14}$ & 0.87 & 2.4  \\
7  & g1c  & M     & $1.4\times 10^{14}$ & 0.84 & 2.3  \\
8  & g8b  & M     & $1.0\times 10^{14}$ & 0.76 & 1.9  \\
9  & g1d  & M     & $9.2\times 10^{13}$ & 0.73 & 1.7  \\
                                                    
10 & g676 & CC    & $8.8\times 10^{13}$ & 0.72 & 1.7  \\
11 & g914 & CC    & $8.5\times 10^{13}$ & 0.71 & 1.6  \\
12 & g1e  & M     & $6.4\times 10^{13}$ & 0.65 & 1.3  \\
13 & g8c  & M     & $5.9\times 10^{13}$ & 0.63 & 1.3  \\
14 & g8d  & PreM  & $5.4\times 10^{13}$ & 0.61 & 1.2  \\
\hline
\end{tabular}   
\begin{quote} 
  {\scshape Notes:}\\
  (1) The dynamical state has been classified through a combined criterion
  invoking a merger tree study and the visual inspection of the X-ray
  brightness maps. The labels for the clusters are M--merger, PostM--post
  merger (slightly elongated X-ray contours, weak cool core region
  developing), PreM--pre-merger (sub-cluster already within the virial
  radius), CC--cool core cluster with extended cooling region (smooth X-ray
  profile).\\
  (2) The virial mass and radius are related by $M_\Delta(z) = \frac{4}{3}
  \pi\, \Delta\, \rho_\rmn{crit}(z) R_\Delta^3 $, where $\Delta=200$ denotes a
  multiple of the critical overdensity $\rho_\rmn{crit}(z) = 3 H (z)^2/ (8\pi
  G)$. \\  
  (3) The virial temperature is defined by $kT_\Delta = G M_\Delta \, \mu\,
  m_\p / (2 R_\Delta)$, where $\mu$ denotes the mean molecular weight.
\end{quote}
\label{tab:sample}
\end{table} 

\begin{table*}
\caption{\scshape: Volume averaged values for $\bra X_\CR \ket = \bra P_\CR
  \ket / \bra P_\rmn{th} \ket$ and $E_\CR / E_\rmn{th}$ within the virial radius:}
\begin{tabular}{l l c c c c}
\hline
\hline
relative CR quantity & sample$^{(1)}$ & radiative, & radiative, & non-radiative,    
& non-radiative,\\
& & shock-CRs$^{(2)}$  & shock- \& SNe-CRs$^{(2)}$ & shock-CRs$^{(2)}$
& shock-CRs with $\zeta = \mbox{const}^{(2)}$\\
\hline
$\bra X_\CR \ket = \bra P_\CR \ket / \bra P_\rmn{th} \ket$&
  small CC & 0.045 & 0.216 & 0.022 & 0.321 \\
& large CC & 0.034 & 0.068 & 0.015 & 0.369 \\
& merger   & 0.161 & 0.414 & 0.022 & 0.249 \\
& all      & 0.120 & 0.321 & 0.021 & 0.272 \\
\hline
$E_\CR / E_\rmn{th}$ &
  small CC & 0.083 & 0.358 & 0.037 & 0.541 \\
& large CC & 0.057 & 0.114 & 0.025 & 0.609 \\
& merger   & 0.266 & 0.684 & 0.036 & 0.410 \\
& all      & 0.199 & 0.531 & 0.034 & 0.449 \\
\hline
\end{tabular}   
\begin{quote} 
 {\scshape Notes:}\\ 
(1) The different cluster samples consist of small cool core (CC) clusters
 (cluster 10, 11), large cool core clusters (cluster 1, 2, 4), merging clusters
 (other clusters), and all 14 clusters.\\
(2) The third column derives from our radiative simulations that account only
 for structure formation CRs while the values of the forth column
 correspond to the radiative simulations that take additionally supernova CRs into
 account. The values in the fifth column derive from non-radiative simulations
 that accelerate CRs at structure formation shocks using a Mach number dependent
 energy injection efficiency $\zeta(\M)$, while the last column derives from the 
 scheme with a constant energy injection efficiency, $\zeta = \mbox{const}$.\\ 
\end{quote}
\label{tab:XCR}
\end{table*} 

We chose our sample of 14 galaxy clusters such that they span a wide
range in mass ($5\times 10^{13} h^{-1}\, \rmn{M}_\odot$ to $2\times 10^{15}
h^{-1}\, \rmn{M}_\odot$) and show a variety of dynamical states ranging from
relaxed cool core (CC) clusters to violent merging clusters
(cf. Table~\ref{tab:sample}). Since the CR injection crucially depends
on the strength of shock waves we expect the impact of CR effects to
depend on both, the cluster size and its dynamical state. To
investigate this dependence, we re-simulated three isolated clusters
(cluster 4, 10, and 11) and three super-cluster regions which are each
dominated by a large cluster (cluster 1, 2, and 3) and surrounded by
satellite clusters (cluster 5 to 9 and 12 to 14).

We analysed the clusters with a halo-finder based on spherical
overdensity followed by a merger tree analysis in order to get the
mass accretion history of the main progenitor. We also produced
projections of the X-ray emissivity at redshift $z=0$ in order to get
a visual impression of the cluster morphology. The dynamical state of
a cluster is defined by a combined criterion: (i) if the cluster did
not experience a major merger with a progenitor mass ratio 1:3 or
larger after $z=0.8$ (corresponding to a look-back time of $\sim 5\,
h^{-1}\,$Gyr) and (ii) if the visual impression of the cluster's X-ray
morphology is relaxed, it was defined to be a cool core cluster. The
merging clusters are subdivided into currently merging clusters,
pre-mergers that show a merging sub-cluster already within the virial
radius, and post-mergers with slightly elongated X-ray contours in
combination with a weak cool core region developing. While this
classification may not always be unambiguous in individual cases, it
provides robust samples of clusters in these different dynamical
states. Interestingly, we note that each satellite cluster within our three
super-cluster regions shows signs of merging activity, presumably
triggered by the high ambient density that is necessary for the
development of a super-cluster region.

The spherical overdensity definition of the virial mass of the cluster is
given by the material lying within a sphere centred on a local density
maximum, whose radial extend $R_\Delta$ is defined by the enclosed threshold
density condition $M (< R_\Delta) / (4 \pi R_\Delta^3 / 3) =
\rho_\rmn{thres}$. We chose the threshold density $\rho_\rmn{thres}(z) =
\Delta\, \rho_\rmn{cr} (z)$ to be a multiple $\Delta=200$ of the critical
density of the universe $\rho_\rmn{cr} (z) = 3 H (z)^2/ (8\pi G)$. For a
better comparison with observations and reasons laid out in detail by
\citet{2004cgpc.symp....1E}, we assume a constant $\Delta=200$ although some
treatments employ a time-varying $\Delta$ in cosmologies with $\Omega_\rmn{m}
\ne 1$ \citep{1996MNRAS.282..263E}. In the reminder of the paper, we use the
terminology $R_\rmn{vir}$ instead of $R_{200}$.

\section{Results}
\label{sec:results}

\subsection{Relative CR pressure}
\label{sec:rel_CR_pressure}

Which galaxy clusters are mostly affected by CRs? And does cooling and star
formation change the generic picture of CRs in clusters? To answer these two
questions it is instructive to consider the CR pressure $P_\CR$ relative
to the thermal pressure $P_\rmn{th}$ integrated over the cluster. This is a
good measure of the importance of dynamical effects of CRs on the ICM.
Table~\ref{tab:XCR} shows the volume averaged values for $\bra X_\CR \ket =
\bra P_\CR \ket / \bra P_\rmn{th} \ket$ within the virial radius
$R_\rmn{vir}$ and the ratio of the CR-to-thermal energy within the galaxy
cluster. 

The main findings are the following. (1) Among the non-radiative simulations,
$\bra X_\CR \ket$ is much smaller ($\bra X_\CR \ket \sim 2$ per cent) in our improved
scheme where CRs are accelerated at structure formation shocks using Mach
number dependent energy injection efficiency $\zeta(\M)$ compared to the
simplified scheme where CRs have been injected with a high constant energy injection
efficiency of $\zeta=0.5$. This shows the importance of a Mach number dependent CR injection
scheme to study CR effects within the ICM.  (2) Neglecting for the moment the
simplified scheme, we notice that $\bra X_\CR \ket$ is much larger in our
radiative simulations compared to our non-radiative simulations for reasons
that will be analysed in the next sections. (3) Among our radiative
simulations, the complete model that accounts for CRs from structure formation
shocks on large scales and supernova shocks within the ISM of individual
galaxies shows a larger relative CR pressure compared to the model that only
accounts for structure formation CRs. Note that the complete model opens simply another
channel for CR injection. (4) Comparing different dynamical states of clusters,
the relative CR pressure contribution is largest in merging clusters, followed
by small cool core clusters, and large cool core clusters.  Strong merger shock
waves efficiently inject CRs at the external cluster regions and mix the highly
CR-enriched intergalactic medium outside clusters with the ICM. An effect that
boosts the relative CR pressure during a merger. Weak virialisation shocks
traversing the cluster after the merger thermalize random gas motions and yield
a decrease of $\bra X_\CR \ket$.  (5) The volume averages for $X_\CR$ are
dominated by the outer regions due to the rising profile of $X_\CR$ with
increasing radius. In other words, the absolute value of $\bra X_\CR \ket$
depends on the definition of the virial radius. A larger virial radius $R_{100}
> R_{200}$ would thus lead to an increase of $\bra X_\CR \ket$.

The low level of $\bra X_\CR \ket$ in our non-radiative simulations disagrees
with the substantially larger predictions of $\bra X_\CR \ket \sim 20$ per cent
to 50 per cent by \citet{2001ApJ...559...59M}. This could be due to a number of
reasons.  \citet{2000ApJ...542..608M} identified shocks with Mach numbers in
the range $4 \lesssim \M \lesssim 5$ as the most important in thermalizing the
plasma. In contrast, \citet{2003ApJ...593..599R} and
\citet{2006MNRAS.367..113P} found that the Mach number distribution peaks in
the range $1 \lesssim \M \lesssim 3$.  Since diffusive shock acceleration of
CRs depends sensitively on the Mach number (cf.~Fig.~\ref{fig:zeta}), this
implies a more efficient CR injection in the simulations by
\citet{2001ApJ...559...59M}. Secondly, CR injection according to the ``thermal
leakage'' model in the implementation of \citet{2001CoPhC.141...17M} has no
injection limiter for the saturated stage of non-linear CR acceleration which
might have led to an overestimate of the CR injection for strong shocks.
Finally, the grid-based cosmological simulations have been performed in a
cosmological box of side-length $50\,h^{-1}$~Mpc with a spatial resolution of
$200\,h^{-1}$~kpc, assuming an Einstein-de Sitter cosmological model
\citep{2001ApJ...559...59M}.  The lack of resolution inside the clusters in the
grid-based approach does not allow one to properly resolve the adiabatic
compression of the composite of CRs and thermal gas. As a result, on one hand
the disfavouring of the CR pressure relative to the thermal pressure upon
compression is not accounted for, and on the other hand, this underestimates CR
loss processes at the dense cluster centres due the lack of resolution there.

While our simulations have much higher spatial resolution inside
clusters and can therefore be expected to be more accurate, one has to be
aware of resolution limitations in them as well. This applies
especially to the radiative simulations, which are not able to resolve
faint cluster galaxies, and are not guaranteed to give already a
converged value for the total amount of stars forming in the
clusters. Also, we know that simulated clusters exhibit cooling flows
that are much stronger than observed in nature, which can amplify
effects related to the CR compression and slow infall of ICM gas in
the simulated clusters. However, our main purpose in this paper is to
study the principle changes in cluster structure brought about by the
inclusion of CRs, and our resolution is high enough to answer this
question reliably.

Taking into account that the two radiative CR simulations are chosen to bracket
the realistic case, we roughly predict a ratio of the CR-to-thermal energy of
$E_\CR / E_\th = 9$, 20, and 50 per cent for small CC, large CC, and merging
clusters.  These predictions are in good agreement with CR constraints from
gamma-ray and radio observations \citep{2003A&A...407L..73P,
2004A&A...413...17P}, considering that these authors assumed a constant CR
energy density relative to the thermal gas.

\subsection{Profiles of non-radiative simulations}
\label{sec:non-rad_profiles}

\begin{figure*}
\begin{center}
\resizebox{0.5\hsize}{!}{\includegraphics{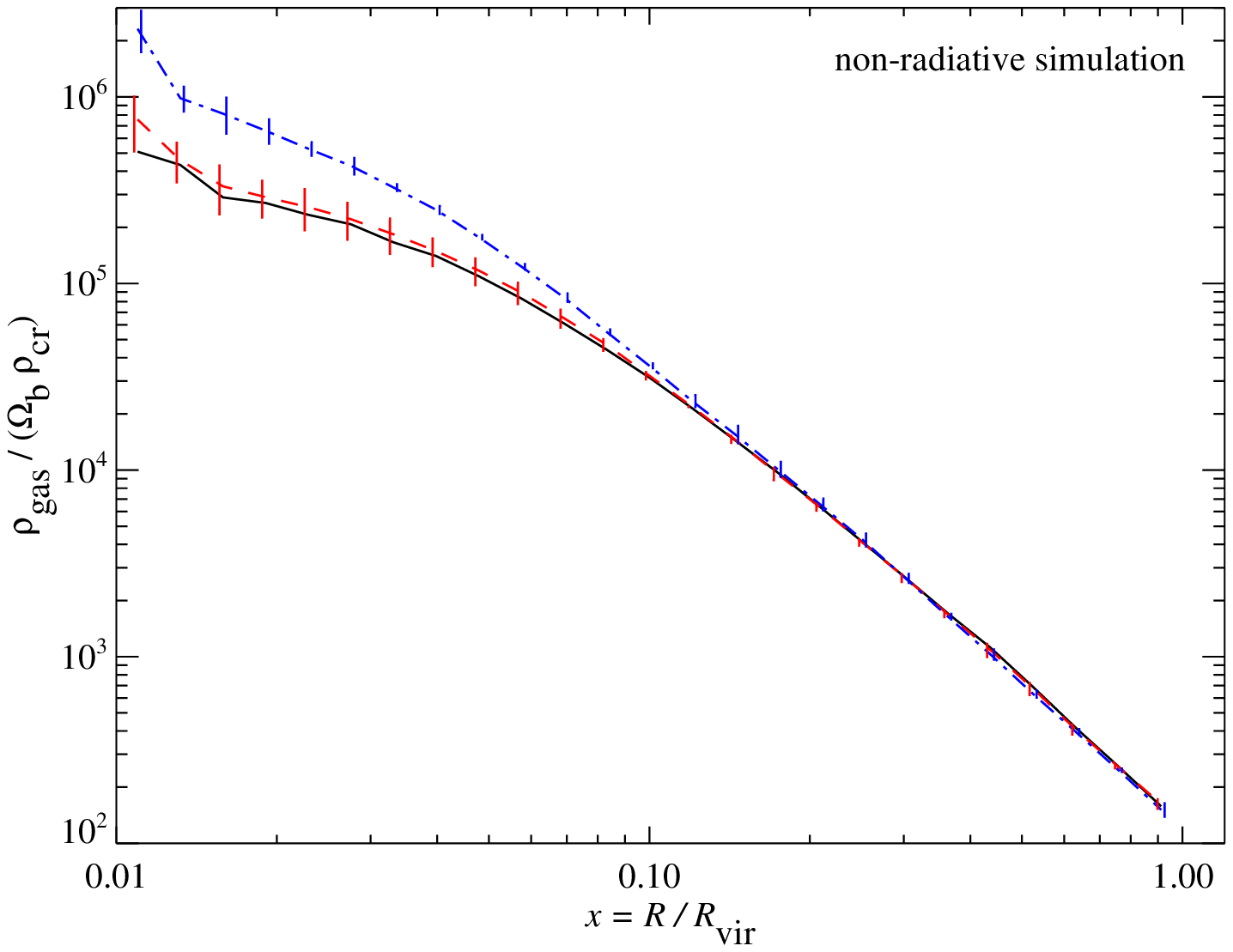}}%
\resizebox{0.5\hsize}{!}{\includegraphics{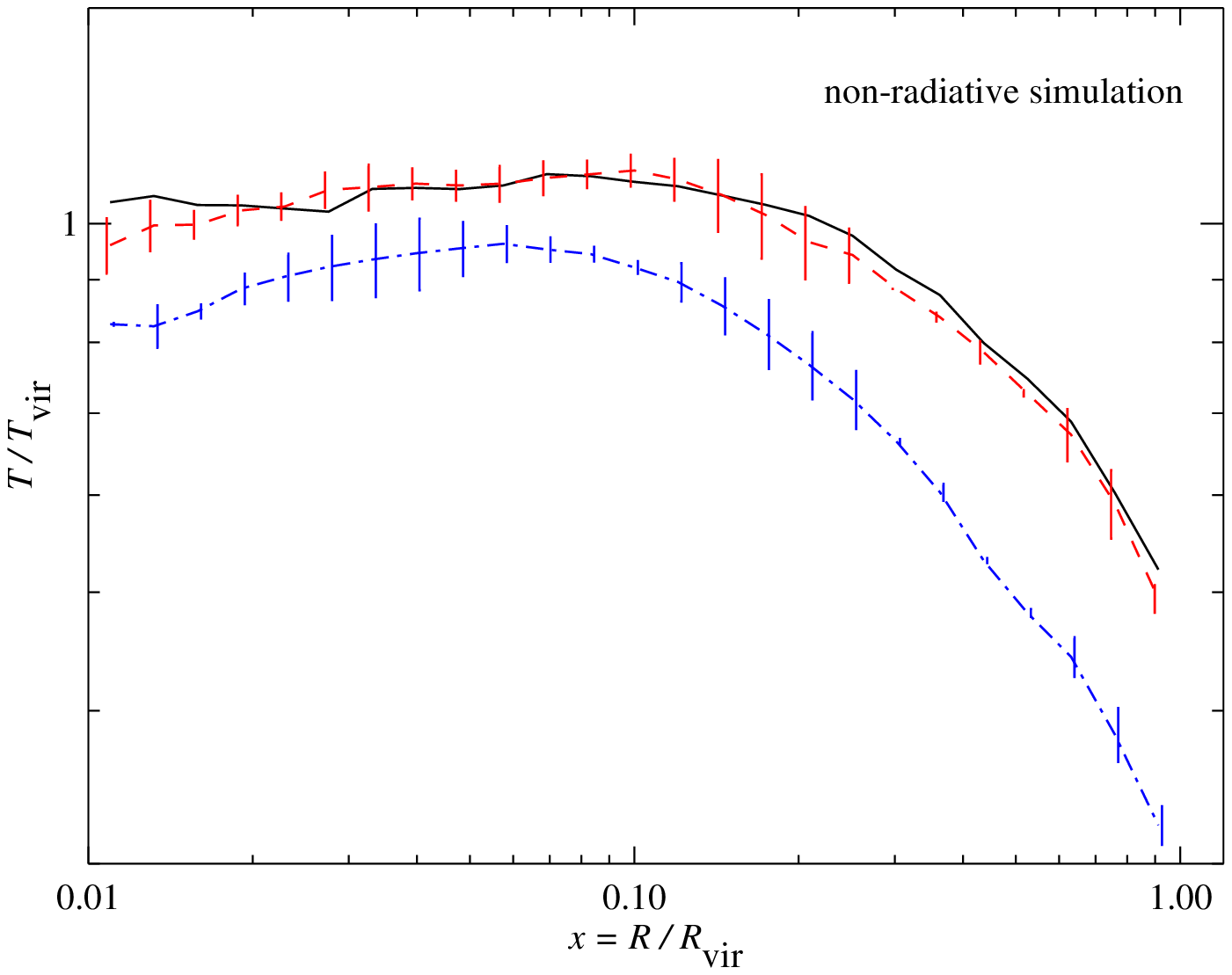}}\\
\resizebox{0.5\hsize}{!}{\includegraphics{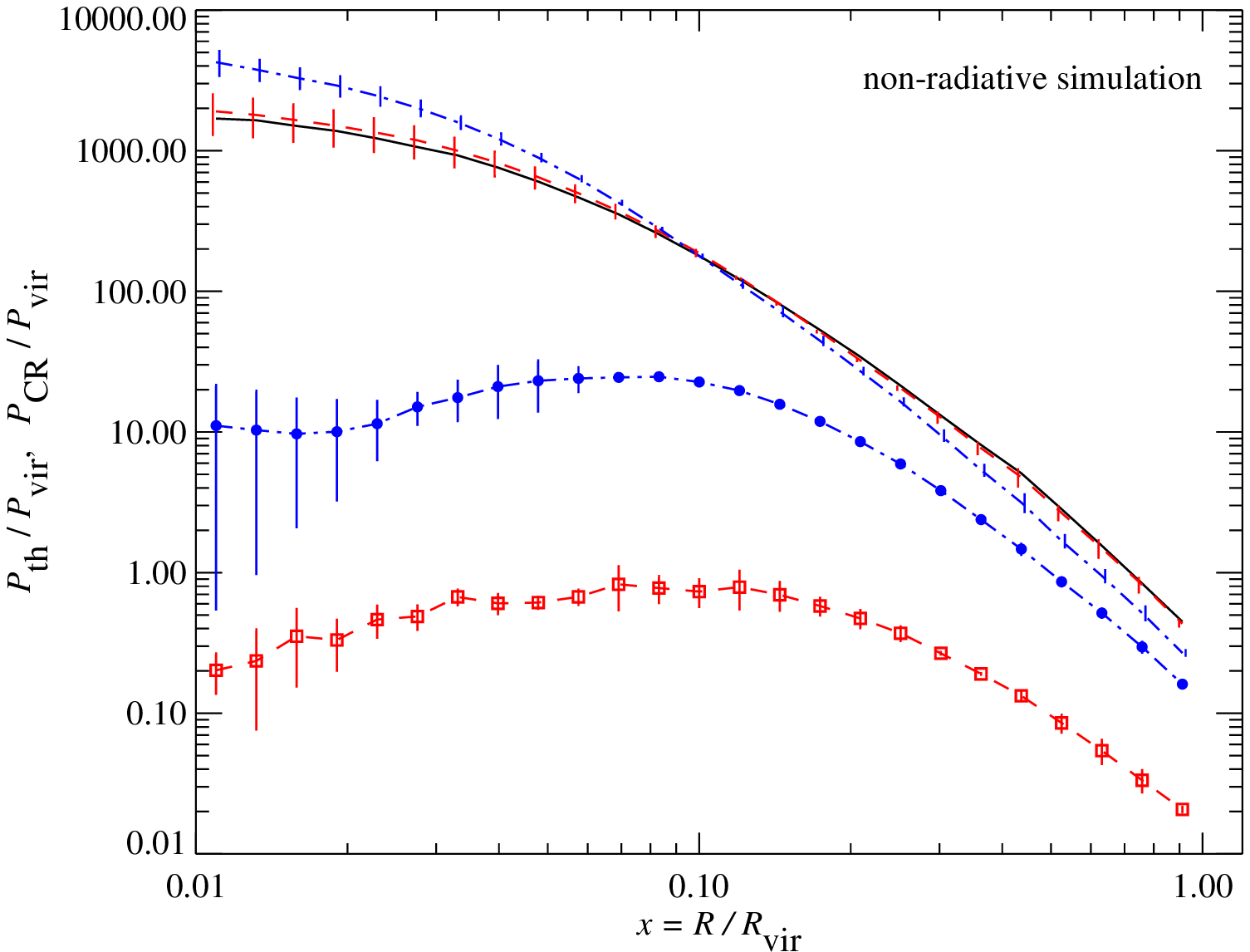}}%
\resizebox{0.5\hsize}{!}{\includegraphics{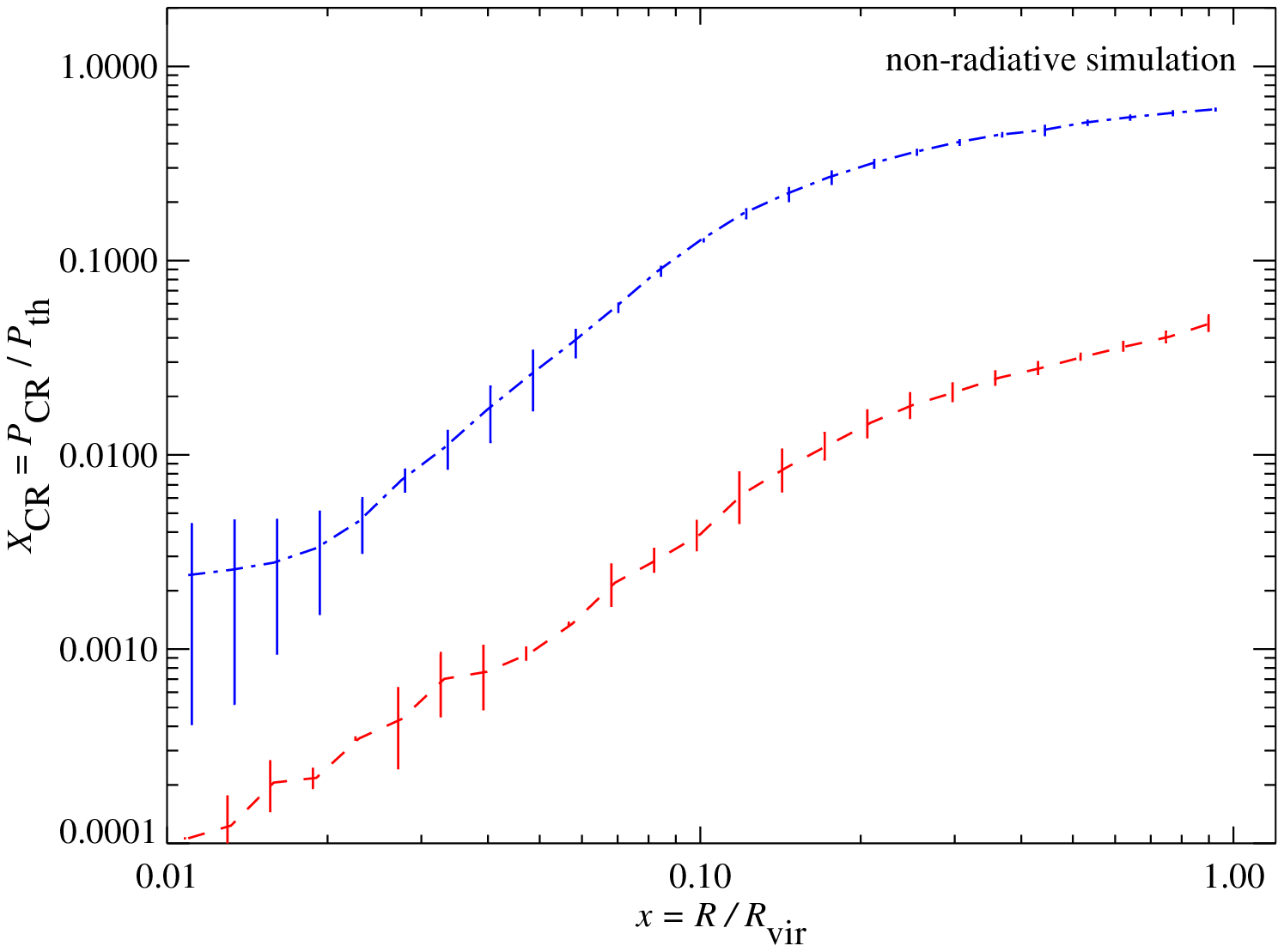}}\\
\resizebox{0.5\hsize}{!}{\includegraphics{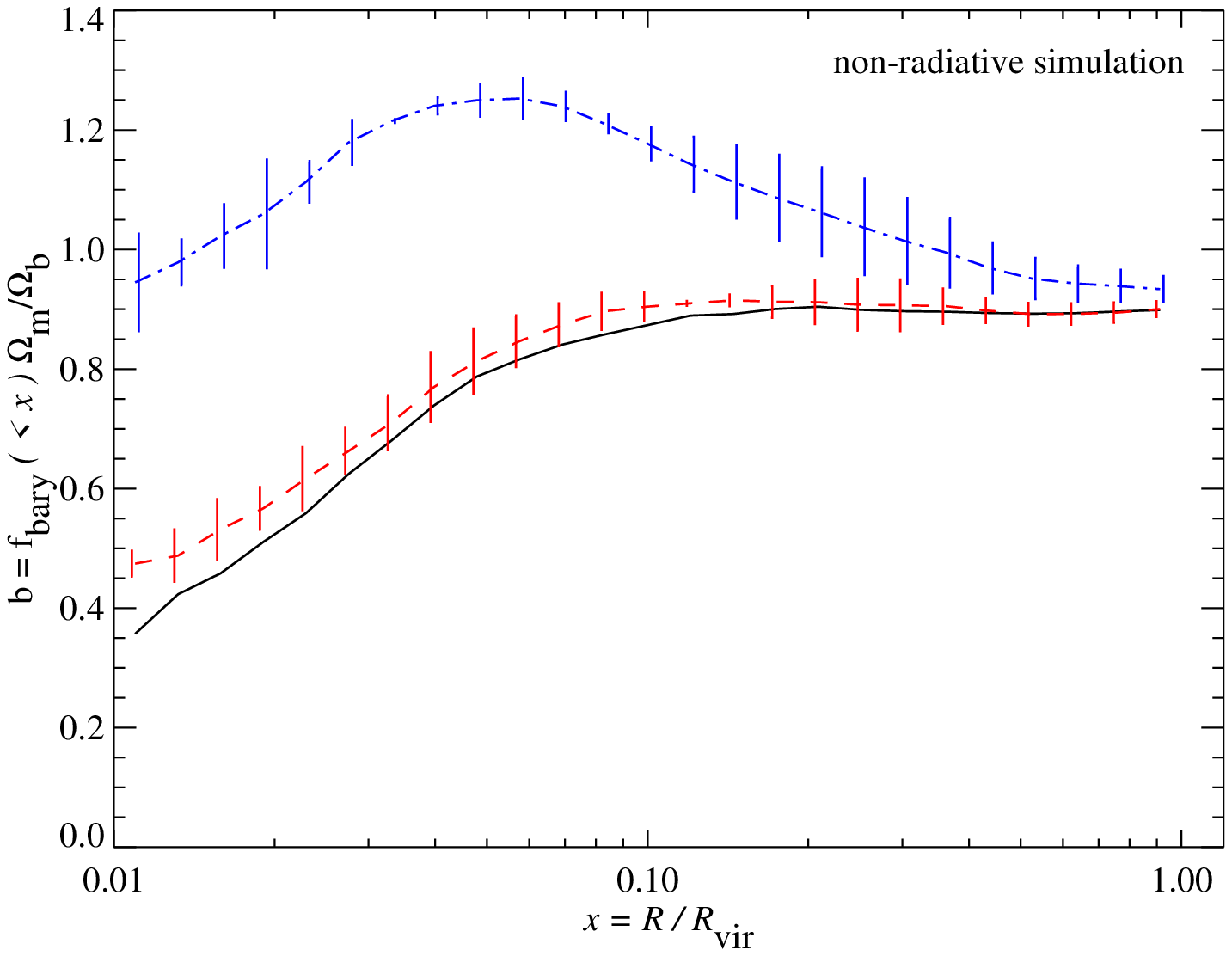}}%
\resizebox{0.5\hsize}{!}{\includegraphics{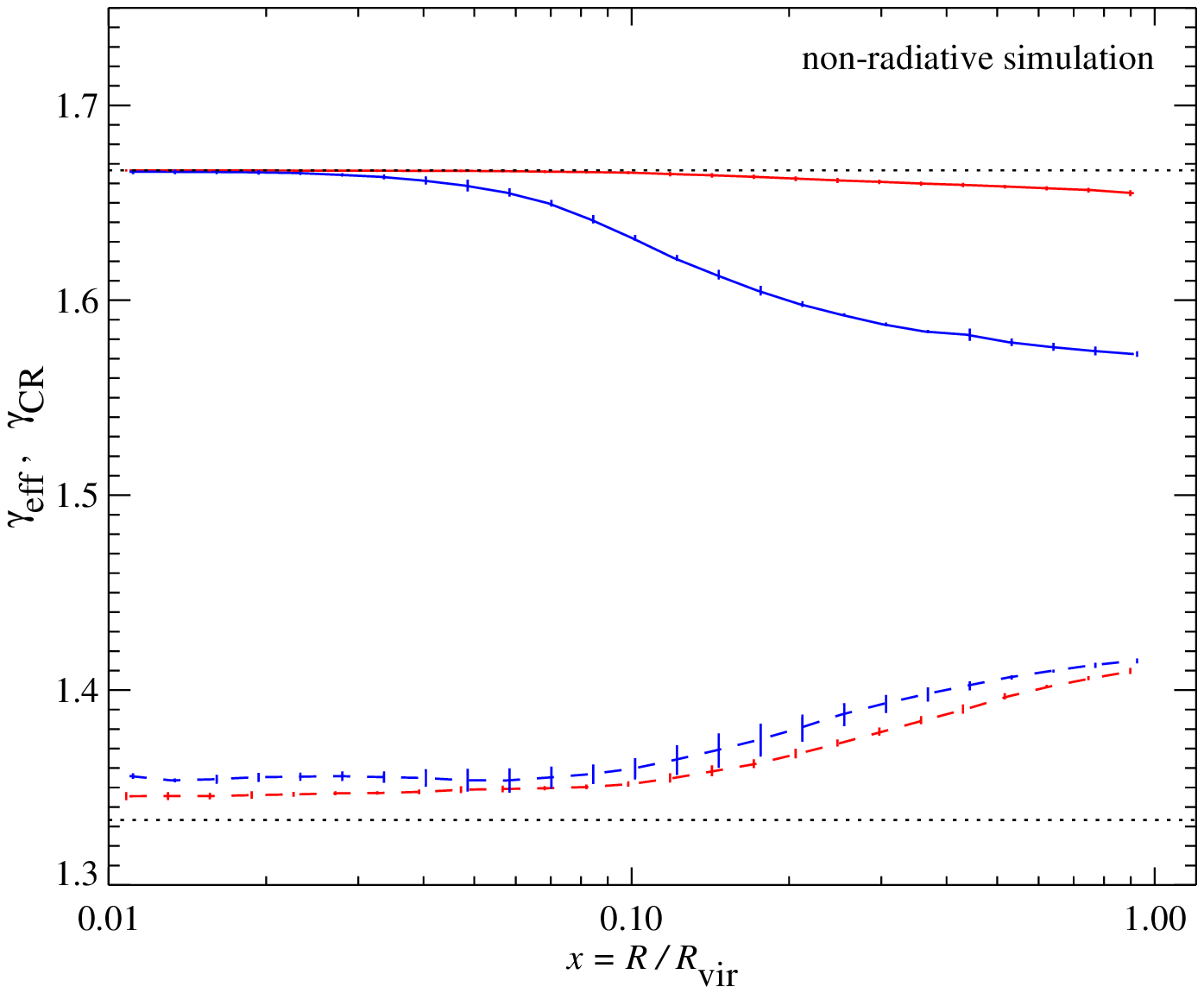}}\\
\end{center}
  \caption{Average profiles of our sample of small cool core clusters (clusters
  10, 11) in our non-radiative simulations at redshift $z=0$. Here and in the
  following plot, the error bars represent the standard deviation from the
  sample mean.  Shown are the appropriately scaled profiles of the gas density
  $\rho_\rmn{gas}$, temperature $T$, thermal and CR pressure ($P_\th$ and
  $P_\CR$, the latter with symbols), relative CR pressure $X_\CR = P_\CR /
  P_\th$, baryon fraction $b$, and the effective adiabatic indices
  $\gamma_\rmn{eff}$ (solid) in comparison with the CR adiabatic indices
  $\gamma_\CR$ (dashed).  Colour scheme: black: reference simulations without
  CRs, red: CRs are injected only through shock acceleration using our Mach
  number dependent scheme, blue: CR shock acceleration with constant energy
  injection efficiency.}
  \label{fig:non-rad}
\end{figure*}

Results for the appropriately scaled average profiles of our non-radiative
simulations are shown in Fig.~\ref{fig:non-rad} only for our sample of small
cool core clusters. The main conclusions drawn from this sample are
representative for our complete sample, while including the remaining clusters
would only increase the cluster to cluster variance of the profiles, primarily
due to the non-sphericity (for merging clusters) and the different
concentrations (for large cool core clusters). The radial coordinate has been
scaled with $R_\rmn{vir}$ to allow comparison between differently sized
objects.  In all plots of the profiles, the error bars represent the standard
deviation $\sigma$ from the sample mean of the indicated subsample of clusters.
The colour scheme for the non-radiative profiles is {\em black} for our
reference simulations without CRs, {\em red} for simulations where CRs are
injected only through shock acceleration using Mach number dependent scheme
(realistic CR model), and {\em blue} for simulations where CRs are accelerated
at structure formation shocks with a constant injection efficiency $\zeta=0.5$
(simplified CR model).  One should bare in mind that the latter model
represents an extreme simplification that exaggerates the CR effects for better
visibility.

\subsubsection{Density and temperature profiles}
\label{sec:non-rad_density}
The gas density has been scaled with $\Omega_\rmn{b}\, \rho_\rmn{crit}$ while
the temperature has been scaled by its virial quantity that is defined by
$kT_\rmn{vir} = G M_\rmn{vir} \, \mu\, m_\p / (2 R_\rmn{vir})$, where $\mu = 4
/(3 X_\rmn{H} + 1 + 4\,X_\rmn{H}\,x_\e)$ denotes the mean molecular weight,
$X_\rmn{H}=0.76$ is the primordial hydrogen mass fraction, and $x_\e$ is the
ratio of electron and hydrogen number densities which we dynamically track in
our radiative simulations.  Both the average density and the temperature
profile of our realistic CR model shows only very small differences to our
reference simulations. The tendency of the enhanced density and the reduced
temperature in our CR simulations is clearly enhanced in our simplified CR
model. While the density enhancement is largest in the centre, the decrease of
the cluster temperature is larger at the cluster outskirts for reasons that are
described below (cf.~Sect.~\ref{sec:picture}).

\subsubsection{Pressure and relative CR pressure profiles}
The lines without symbols denote the thermal pressure, the ones with symbols
the CR pressure.  The CR and thermal pressure have been scaled with
$P_\rmn{vir} = kT_\rmn{vir}\, 200\, \Omega_\rmn{b}\, \rho_\rmn{crit} / (\mu\,
m_\p)$. Again, there is only a very small difference visible between the thermal
pressure of our realistic CR model and our reference simulations. The trend of
a reduced thermal pressure at cluster outskirts and of the increased
central pressure is enhanced in our simplified CR model. In the outer regions,
the reduced thermal pressure adds up with the CR pressure to the total
(thermal) pressure in our reference simulation while the central pressure
enhancement reflects the density enhancement. The CR pressure shows a flat
central plateau and decreases at $r> 0.1 R_\rmn{vir}$ outwards less steeply
compared to the thermal pressure.

This yields a rising profile for the relative CR pressure throughout the
cluster $X_\CR = P_\CR / P_\th$ due to a combination of the following
effects: CR acceleration is more efficient at the peripheral strong accretion
shocks compared to weak central flow shocks, adiabatic compression of a
composite of CRs and thermal gas disfavours the CR pressure relative to the
thermal pressure due to the softer equation of state of CRs, and CR loss
processes are more important at the dense centres.

\subsubsection{Baryon fraction}
\label{sec:baryon_frac}
In our self-consistent CR model (with varying $\zeta(\M)$), the central baryon
fraction scaled with the universal value $b = f_\rmn{bary} ({<x})\,
\Omega_\rmn{m} / \Omega_\rmn{b}$ is increased by $\sim 10$ per cent and reaches the
same value at the virial radius of $b = 0.9$ compared to our reference
simulation.  Note that the baryon
fraction is increasing as a function of radius which can be traced back to the
different nature of dark matter and gas. The inner slope of the density profile
of collisionless dark matter does not show any sign of flattening
\citep{2004MNRAS.349.1039N} while the gas profile is cored due to
the isotropic hydrodynamic pressure.  This yields to an decreasing
baryon fraction towards the centre, an effect that has been observed in many
works \citep[e.g.][ and references therein]{2006MNRAS.365.1021E}.

Remarkably, the baryon fraction in our simplified CR model (with
$\zeta= 0.5$) even exceeds unity at intermediate radii and reaches a
value at the virial radius of $b = 0.95$. This is mostly due to the
enhanced gas compressibility which boosts the density within the
central regions.

\subsubsection{Effective adiabatic index}
When studying the hydrodynamics of a composite of thermal and CR gas, it is
appropriate to define an effective microscopic adiabatic index by
\begin{equation}
\label{eq:gammaeff}
\gamma_\rmn{eff} \equiv \left.\frac{\dd \log (P_\rmn{th} + P_\CR)}
{\dd \log \rho}\right|_S = 
\frac{\gamma_\rmn{th} + \gamma_\CR\, X_\CR}
{1 + X_\CR},
\end{equation}
where $\gamma_\rmn{th}=5/3$ is the ideal gas value of the ratio of specific
heats and the adiabatic exponent of the CR population is defined by
\begin{equation}
\label{eq:gammaCR}
\gamma_\CR \equiv \left.\frac{\dd \log P_\CR}{\dd \log \rho}\right|_S,
\end{equation}
while the derivative has to be taken at constant entropy $S$
\citep{2006...Ensslin}. Depending on the CR momentum distribution function,
$\gamma_\CR$ can vary between $\gamma_\CR \to 4/3$ (in the ultra-relativistic
limit) to $\gamma_\CR \to 5/3$ (in the non-relativistic limit). One should not
confuse this effective {\em microscopic} adiabatic index of our composite fluid
that enters the hydrodynamic equations and governs the compressibility of the
medium with the effective {\em macroscopic} adiabatic index one can derive from
cluster density and pressure profiles by means of $\gamma = \dd \log P / (\dd
\log r)\, [\dd \log \rho / (\dd \log r)]^{-1}$. If one compressed gas with such
an effective {\em macroscopic} adiabatic index into a spherical cluster that
has the appropriate density profile (without considering shocks and cooling),
one would obtain the simulated pressure profiles.

The effective adiabatic index $\gamma_\rmn{eff}$ is shown with a solid
line while $\gamma_\CR$ is shown with a dashed line. As expected,
$\gamma_\rmn{eff}$ reflects the profile of $X_\CR$ and is only
different from the adiabatic index $\gamma_\th=5/3$ of the thermal gas
in the external cluster regions where CRs make up a larger fraction of
the pressure support (in particular in our simplified
model). Interestingly, $\gamma_\CR$ shows a different behaviour with a
slightly softer value in the central cluster regions. This is due to
enhanced Coulomb cooling in denser core region which effectively
re-thermalizes low-energetic CRs and increases the low momentum cutoff
of the CR distribution function. This leads to a more relativistic CR
population with a lower CR pressure and thus to the observed behaviour
of the adiabatic indices.

\subsubsection{Discussion of CRs profiles in non-radiative simulations}
\label{sec:picture}
Structure formation shock waves propagate through the cosmic tenuous gas, which
is compressed at the transition layer of the shock while a part of the kinetic
energy of the incoming gas is dissipated into internal energy of the post-shock
gas according to the Rankine-Hugoniot jump conditions.  Considering CR
injection at shocks through diffusive shock acceleration, a part of the energy
that has been otherwise completely thermalized is now transferred to the CR
energy reservoir while employing energy conservation.  This effect lowers the
amount of thermalized energy and thus the temperature which is observed at the
periphery in clusters (cf.~Sect.~\ref{sec:non-rad_density}) where it is most
pronounced due to strong shock waves that take place there and inject
efficiently CRs \citep{2000ApJ...542..608M, 2003ApJ...593..599R,
2006MNRAS.367..113P}.

\begin{figure}
\begin{center}
  \resizebox{7.7cm}{!}{\includegraphics{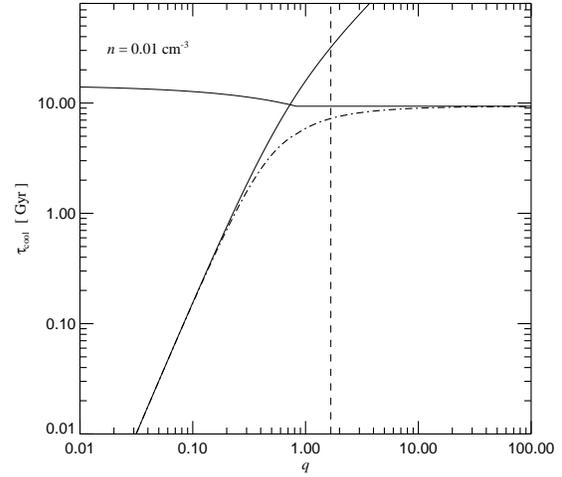}}
  \resizebox{7.7cm}{!}{\includegraphics{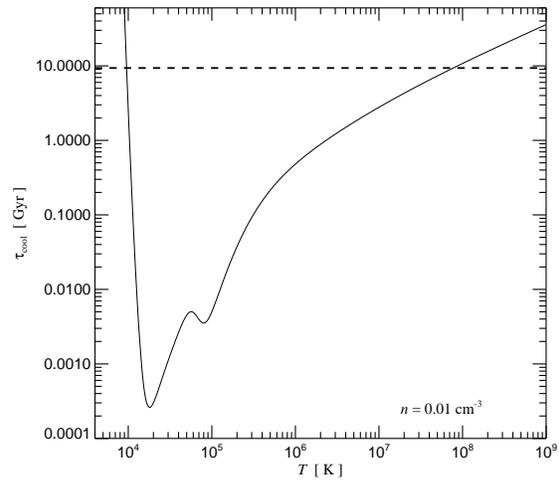}}\vspace*{-.4cm}
\end{center}
  \caption{The top panel shows the cooling times due to Coulomb losses (rising
    solid line) and hadronic dissipation (nearly horizontal line) as a function
    of the spectral cut-off $q = p_\rmn{min} / (m_\p c)$. The dot-dashed line
    gives the total cooling time, while the vertical dashed lines marks the
    asymptotic equilibrium cut-off reached by the CR spectrum when no sources
    are present.  The bottom panel shows the cooling time of ordinary thermal
    gas due to radiative cooling (for primordial metallicity), as a function of
    temperature. The horizontal dashed line marks the cooling time of CRs with
    a high momentum cut-off ($q\gg 1$), for comparison. In both panels, the
    times have been computed for a number density of $n = 0.01\,{\rm cm^{-3}}$,
    a typical number density in the centres of galaxy clusters. Note however
    that the cooling times all scale as $\tau\propto 1/n$, i.e.~for
    different densities, only the vertical scale would change but the relative
    position of the lines would remain unaltered.
 \label{fig:CoolTimes}}
\end{figure}

The successive build-up of a cluster profile of thermodynamic variables (such
as entropy or pressure) through mergers and mass accretion can be thought of as
a combination of two effects. (1) Merger shock waves violently increase the
cluster's entropy and (2) permanent accretion of new shells of mass at the
peripheral cluster regions exert a pressure that adiabatically compresses the
central gas. These two effects have important consequences for the resulting
thermal profiles that are modulated by CR pressure.  Since CR acceleration is
more efficient at the peripheral strong accretion shocks compared to weak
central flow shocks, this yields a stratification of the relative CR
pressure profile which is increasing outwards.  Adiabatic compression of a
composite of CRs and thermal gas disfavours the CR pressure relative to the
thermal pressure due to the softer equation of state of CRs. Roughly, one
expects the ratio of $X_\CR = P_\CR / P_\th$ after adiabatic compression
(denoted with a prime) to that before compression to scale as
\begin{equation}
  \label{eq:ad_compression}
  \frac{X_\CR'}{X_\CR} = \left(\frac{\rho'}{\rho}
  \right)^{\gamma_\CR - \gamma_\th} \sim 0.05
\end{equation}
for a compression ratio of $\rho' / \rho \sim 10^4$. Additionally, CR loss
processes are more important at the dense cluster centres.  Thus, the central
density has to increase dynamically in our CR simulation compared to our
reference simulations in order to balance the gravitational pressure and make
up for the initially `missing' thermal pressure. Shock heating and adiabatic
compression heats the thermal gas in the cluster centre with a higher rate
compared to the external regions. Thus, the thermal pressure $P_\th = n_\th\,
kT$ is increased in the cluster centre in our simulations with CRs compared to
our reference simulations without CRs.

\subsection{Profiles of radiative simulations}
\label{sec:rad_profiles}

The larger CR pressure support in our radiative simulations makes CR
related phenomena more pronounced compared to the non-radiative
simulations. This crucially depends on the small ratio of the thermal
to the comparatively long CR cooling time in regions where the
temperature drops below $T \lesssim 10^7$~K. Here, the CR cooling time
of an aged CR population with an equipartition value for the lower
momentum cutoff $q \sim 1$ remains almost constant, but it decreases
for the thermal gas as it is cooling (cf.~Fig.~\ref{fig:CoolTimes})
and becomes progressively shorter than the CR cooling time. This
equipartition value for $q$ results from a balance of Coulomb cooling
that increases $q$ and leaves the normalisation of the CR distribution
function almost unchanged and CR cooling by hadronic interactions
which decreases the lower momentum cutoff $q$ and the normalisation of
the CR distribution function \citep{2006...Jubelgas}.  Current
cosmological radiative simulations that do not include feedback from
AGN bubbles \citep{2006MNRAS.366..397S} have an over-cooling problem
in the cluster centre where they overproduce the amount of stars. This
problem is larger in cool core clusters which did not recently
experience a major merger. Compared to X-ray observations, the
simulated clusters show enhanced central gas densities, too small
central temperatures, too strong central entropy plateaus, and an
enhanced baryon fraction that is driven by the large stellar mass
fraction towards the cluster centre.  These are all manifestations of
the cooling flow problem of simulated clusters, an issue that
universally occurs in $\Lambda$CDM simulations, independent of the
employed numerical method \citep[e.g.,][]{2004MNRAS.348.1078B,
2005ApJ...625..588K, 2006MNRAS.365.1021E}.

In our plots of cluster profiles, the error bars represent the standard
deviation $\sigma$ from the sample mean of the indicated subsample of
clusters. The error on the mean can easily be obtained by means of $\sigma /
\sqrt{N-1}$, where $N$ is the number of clusters in the sample, i.e.~$N=2$
(small CC), $N=3$ (large CC), $N=5$ (all CC), $N=9$ (merging cluster).  The
colour scheme for the profiles of our radiative simulations is {\em black} for
our reference simulations without CRs, {\em red} for simulations where CRs are
injected only at structure formation shocks using Mach number dependent scheme
({\em CR-shock model}), and {\em blue} for the {\em complete model} that
accounts for CR injection from structure formation shocks on large scales (with
the Mach number dependent scheme) and supernova shocks within the ISM of
individual galaxies with a constant energy injection efficiency of
$\zeta_\rmn{SN} = 0.3$. Although our complete model might seem to be more
relevant astrophysically, both CR models are chosen to represent extreme cases
that are bracketing the realistic scenario in between for the following
reasons. The CR-shock model approaches the realistic case from below because it
does not take into account CRs from supernova shock waves. Our complete CR
model brackets the model from above because it assumes a high CR injection
efficiency from supernovae in addition to the too large star formation rate of
current radiative simulations. This possibly overestimates the CR energy
injection into the ICM in the complete model.  After describing the simulated
profiles, we synthesise our results in Sect.~\ref{sec:rad_picture} and provide
further support for our interpretation by considering CR related physical
processes in more detail in Sect.~\ref{sec:physical_processes}.

\begin{figure*}
\begin{center}
\resizebox{0.5\hsize}{!}{\includegraphics%
{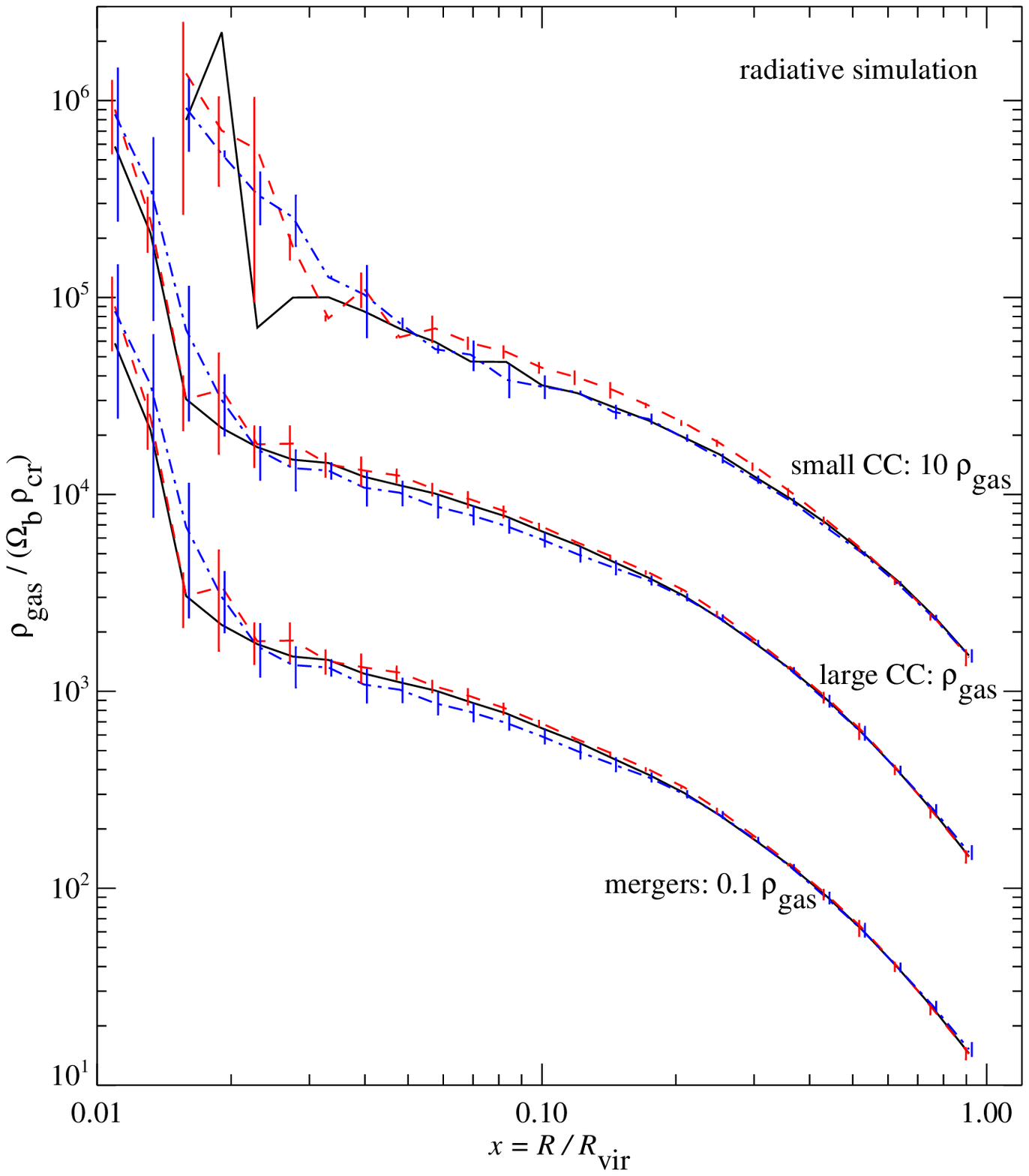}}%
\resizebox{0.5\hsize}{!}{\includegraphics%
{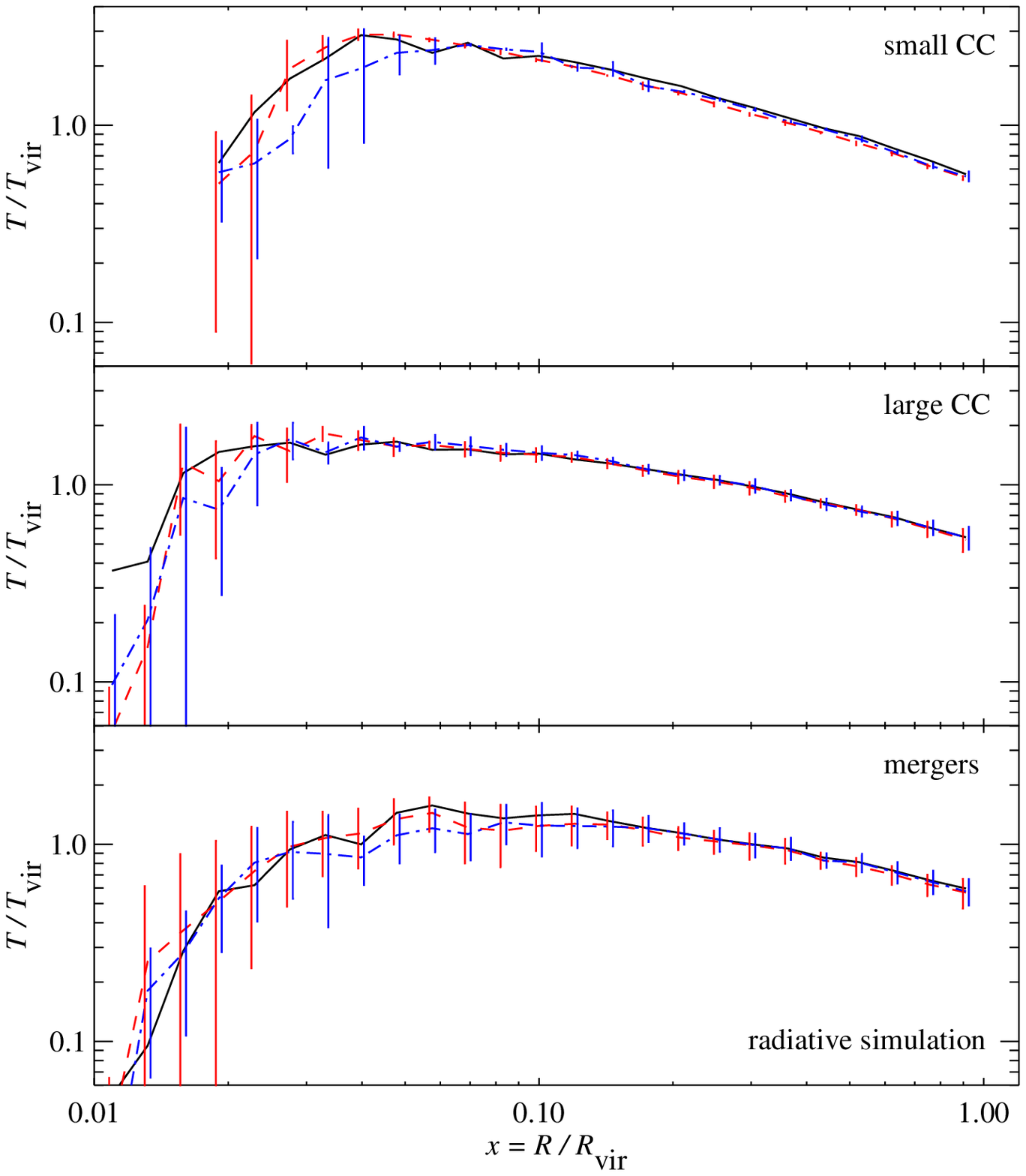}}\\
\end{center}
  \caption{Average profiles of our sample of clusters in our radiative
  simulations at redshift $z=0$. Shown are the appropriately scaled profiles of
  the gas density $\rho_\rmn{gas}$ (left-hand side) and temperature $T$
  (right-hand side).  Colour scheme: black: reference simulations without CRs,
  red: CRs are injected only through shock acceleration using Mach number
  dependent scheme, blue: complete CR-model including CR injection from
  supernovae.}
  \label{fig:density}
\end{figure*}

\subsubsection{Density and temperature profiles}
Results for the appropriately scaled average density and temperature
profiles of our radiative simulations are shown in
Fig.~\ref{fig:density}. Comparing both CR simulations to our reference
simulation, the most striking feature is the density enhancement in
the central regions at $r< 0.04\, R_\rmn{vir}$. Our CR-shock model
shows a density enhancement at larger radii which is strongest in
small cool core clusters and visible up to radii as large as $r =
0.5\, R_\rmn{vir}$. In contrast, our complete model shows a density
decrement at intermediate radii that almost disappears in small cool
core clusters.

Similarly, the temperature in the central regions is reduced in our CR
simulations while it almost coincides at larger radii with our reference
simulation. The cluster-to-cluster variance is largest in our merger
simulation due to ongoing virialisation processes mediated by structure
formation shock waves.

\begin{figure*}
\begin{center}
  \begin{minipage}[t]{0.495\textwidth}
    \centering{\it \Large Cool core clusters:}
  \end{minipage}
  \hfill
  \begin{minipage}[t]{0.495\textwidth}
    \centering{\it \Large Merging clusters:}
  \end{minipage}
\resizebox{0.5\hsize}{!}{\includegraphics{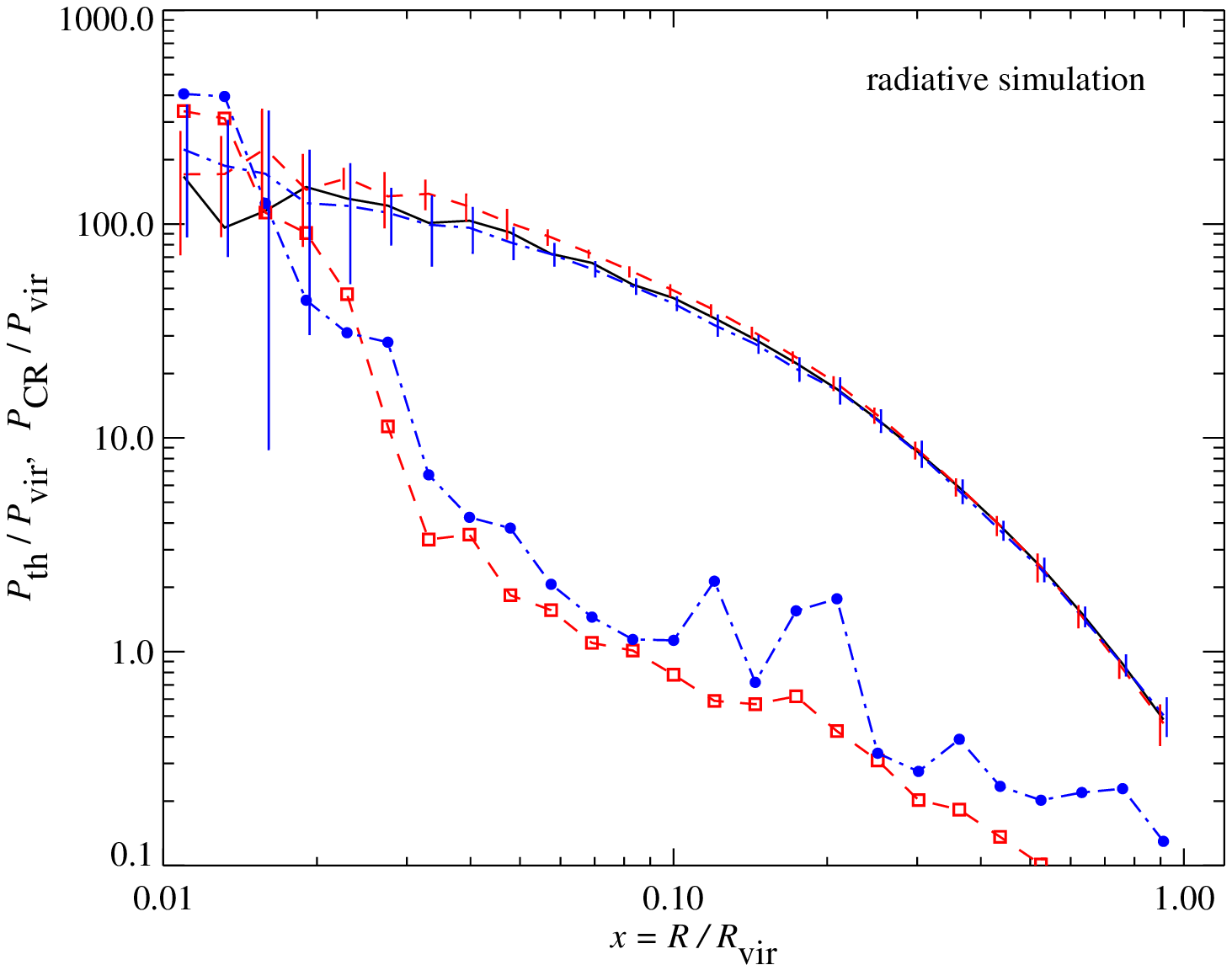}}%
\resizebox{0.5\hsize}{!}{\includegraphics{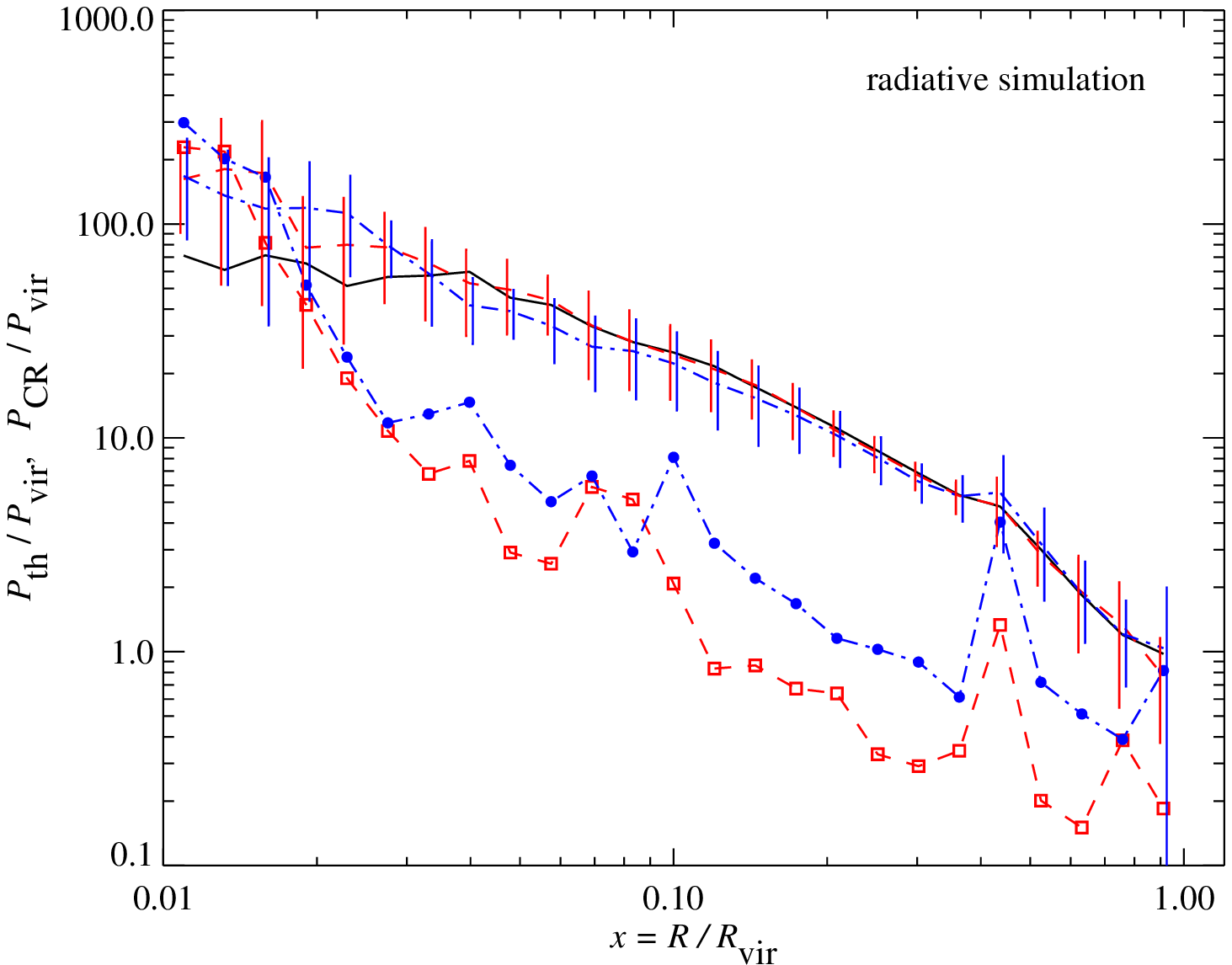}}\\
\resizebox{0.5\hsize}{!}{\includegraphics{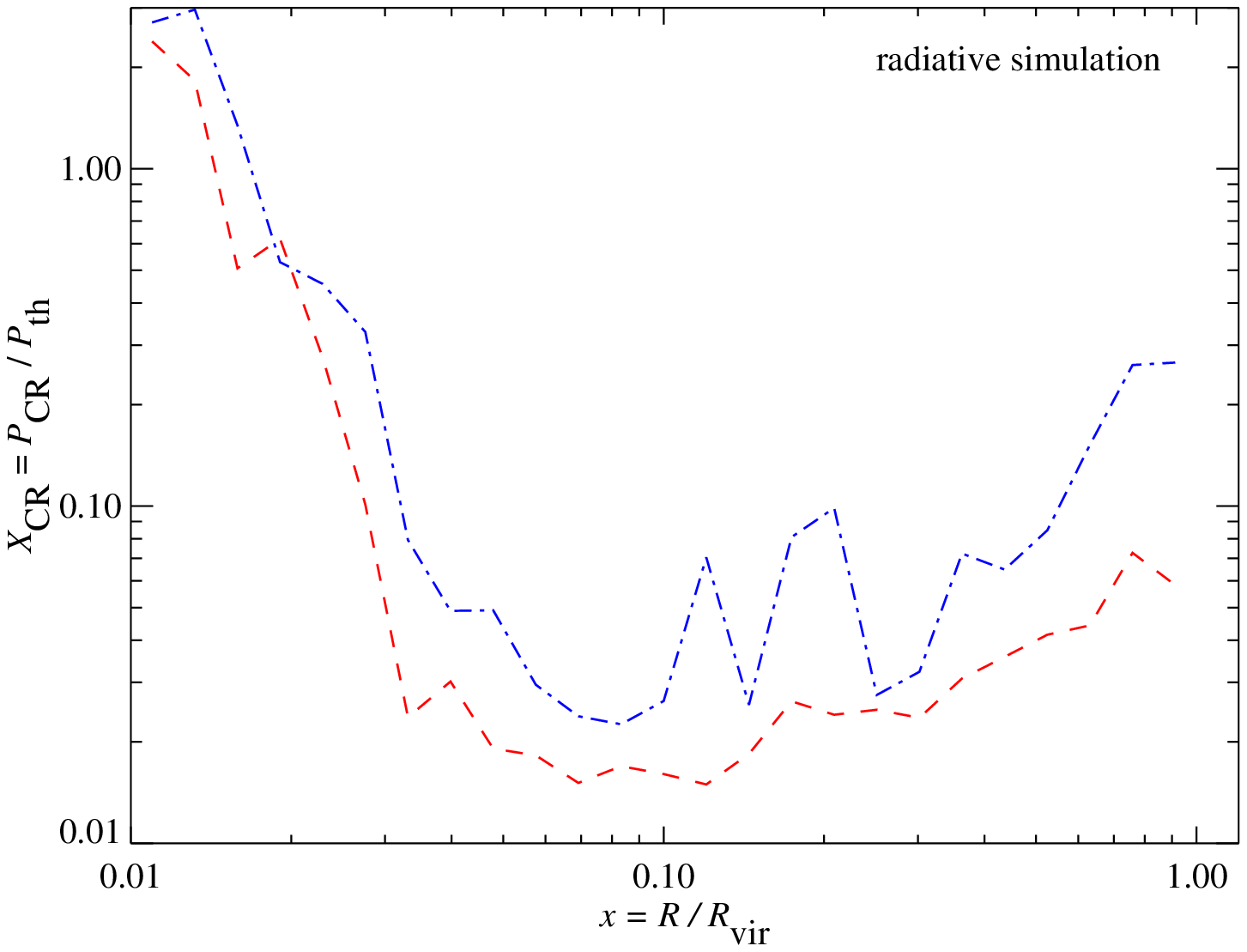}}%
\resizebox{0.5\hsize}{!}{\includegraphics{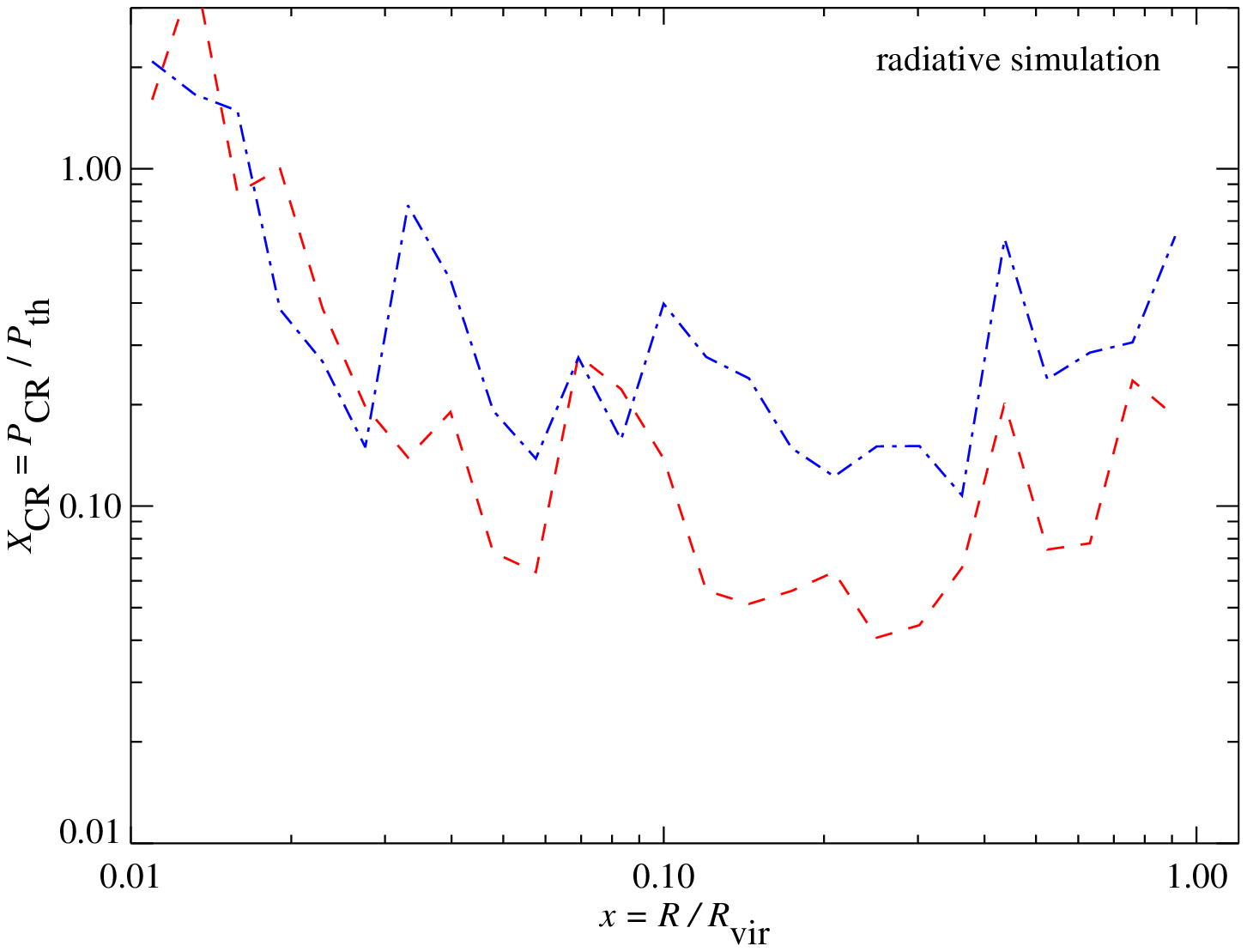}}\\
\resizebox{0.5\hsize}{!}{\includegraphics{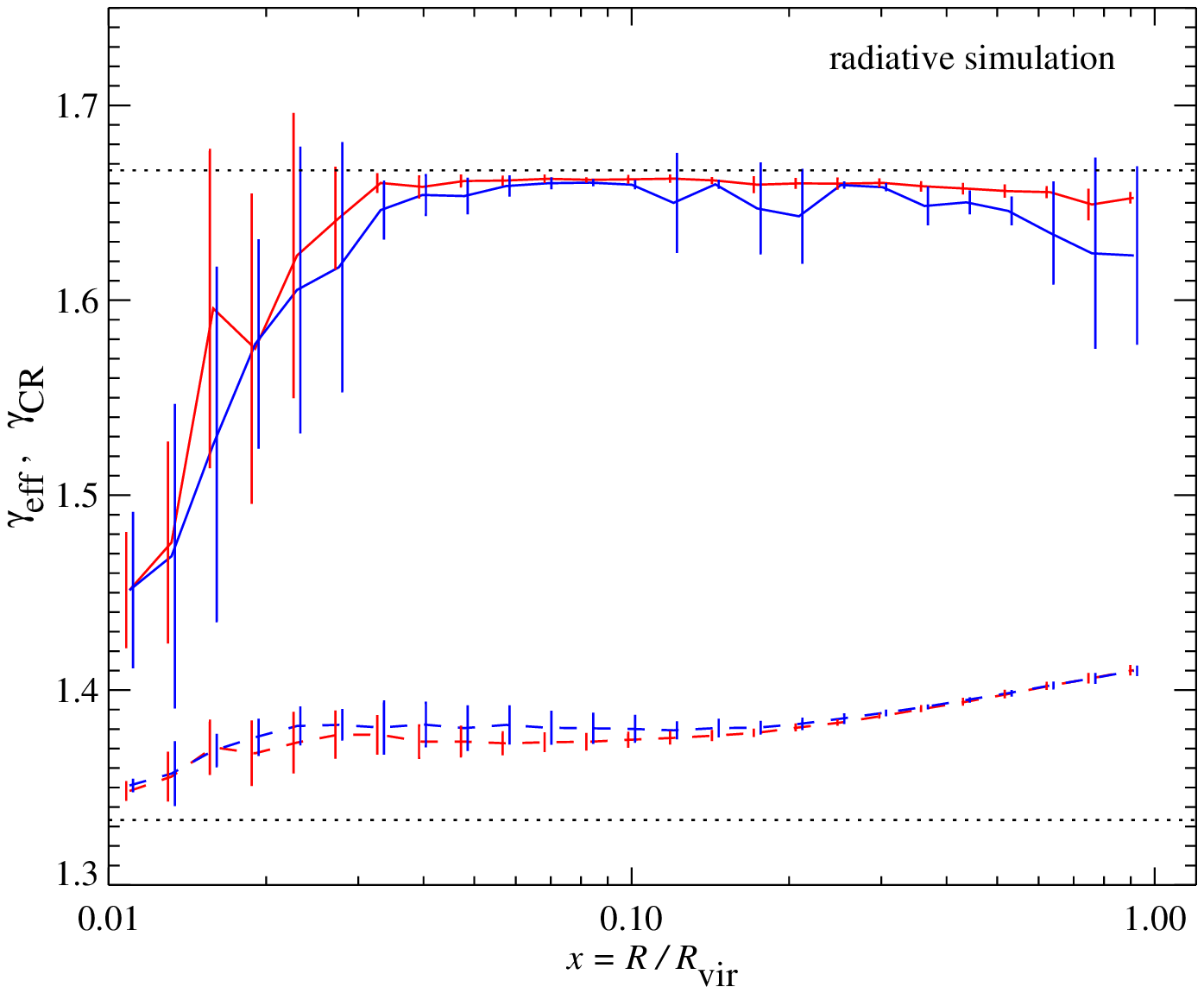}}%
\resizebox{0.5\hsize}{!}{\includegraphics{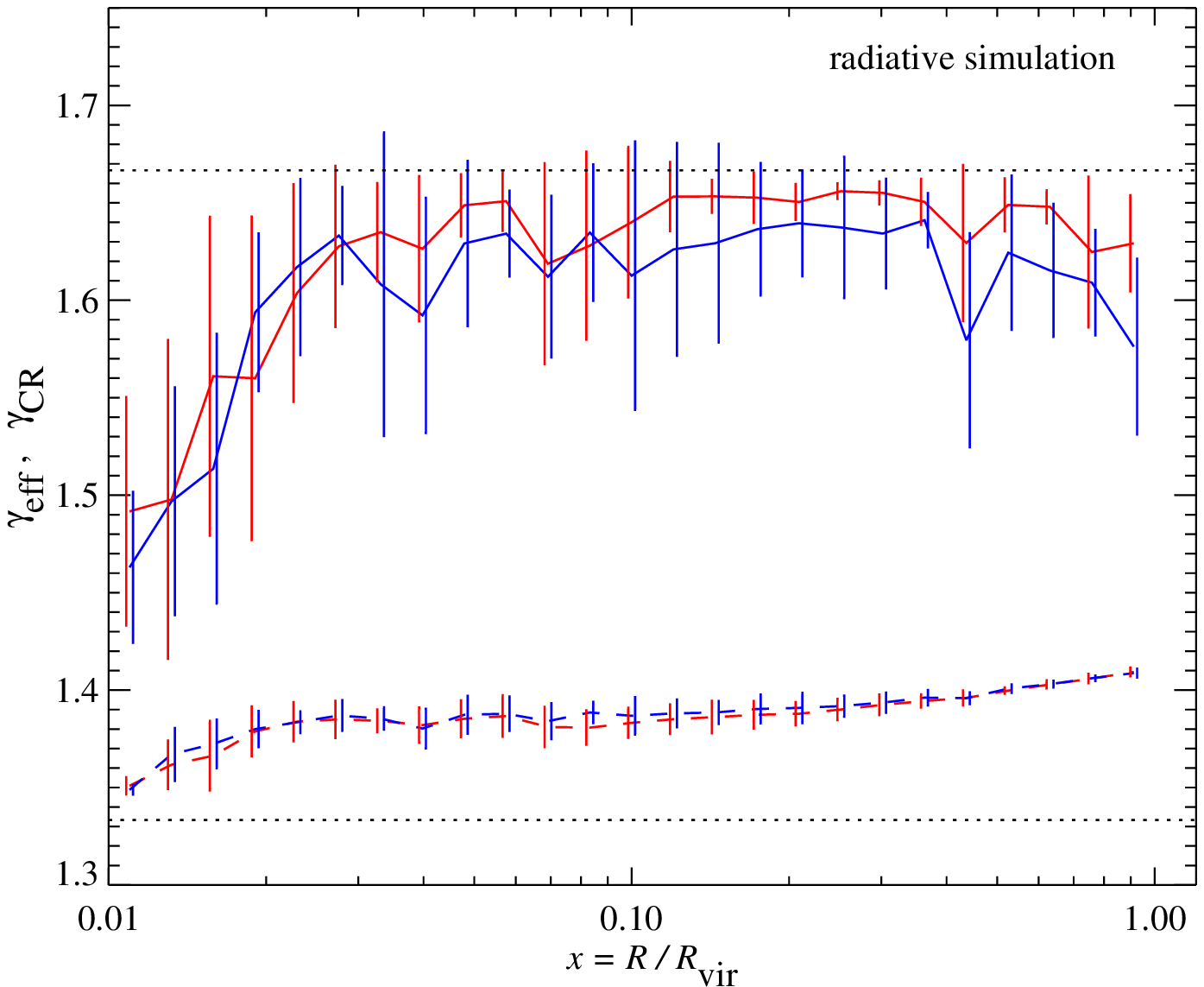}}\\
\end{center}
  \caption{Average profiles of our sample of all cool core clusters (left-hand
    side) and merging clusters (right-hand side) in our radiative simulations
    at redshift $z=0$.  Shown are the appropriately scaled profiles of the
    thermal and CR pressure ($P_\th$ and $P_\CR$, the latter with symbols),
    relative CR pressure $X_\CR = P_\CR / P_\th$, and the effective adiabatic
    indices $\gamma_\rmn{eff}$ (solid) in comparison with the CR adiabatic
    indices $\gamma_\CR$ (dashed).  Colour scheme: black: reference simulations
    without CRs, red: CRs are injected only through shock acceleration using
    Mach number dependent scheme, blue: complete CR-model including CR
    injection from supernovae.}
  \label{fig:pressure}
\end{figure*}

\begin{figure*}
\begin{center}
  \begin{minipage}[t]{0.495\textwidth}
    \centering{\it \Large Cool core clusters:}
  \end{minipage}
  \hfill
  \begin{minipage}[t]{0.495\textwidth}
    \centering{\it \Large Merging clusters:}
  \end{minipage}
\resizebox{0.5\hsize}{!}{\includegraphics{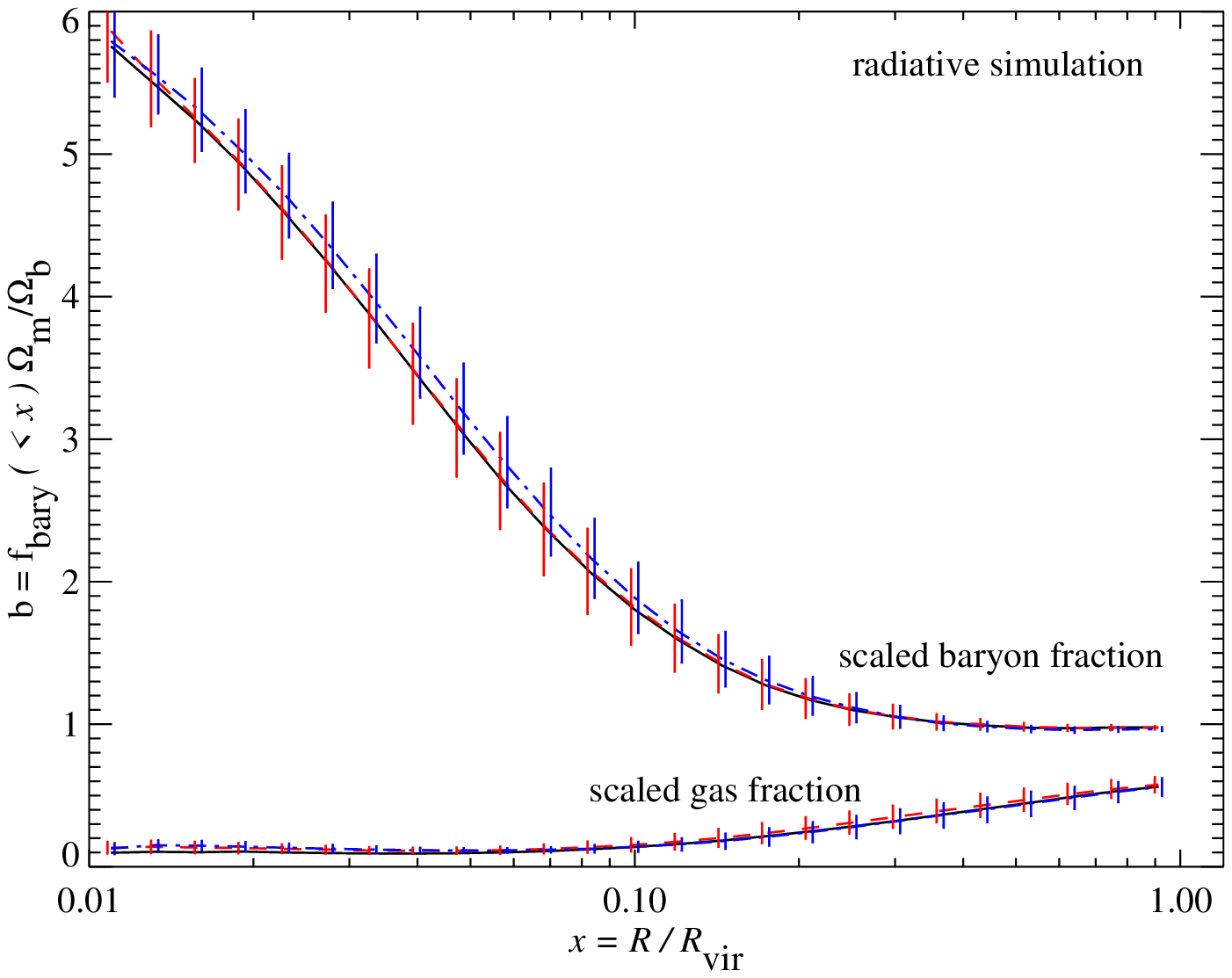}}%
\resizebox{0.5\hsize}{!}{\includegraphics{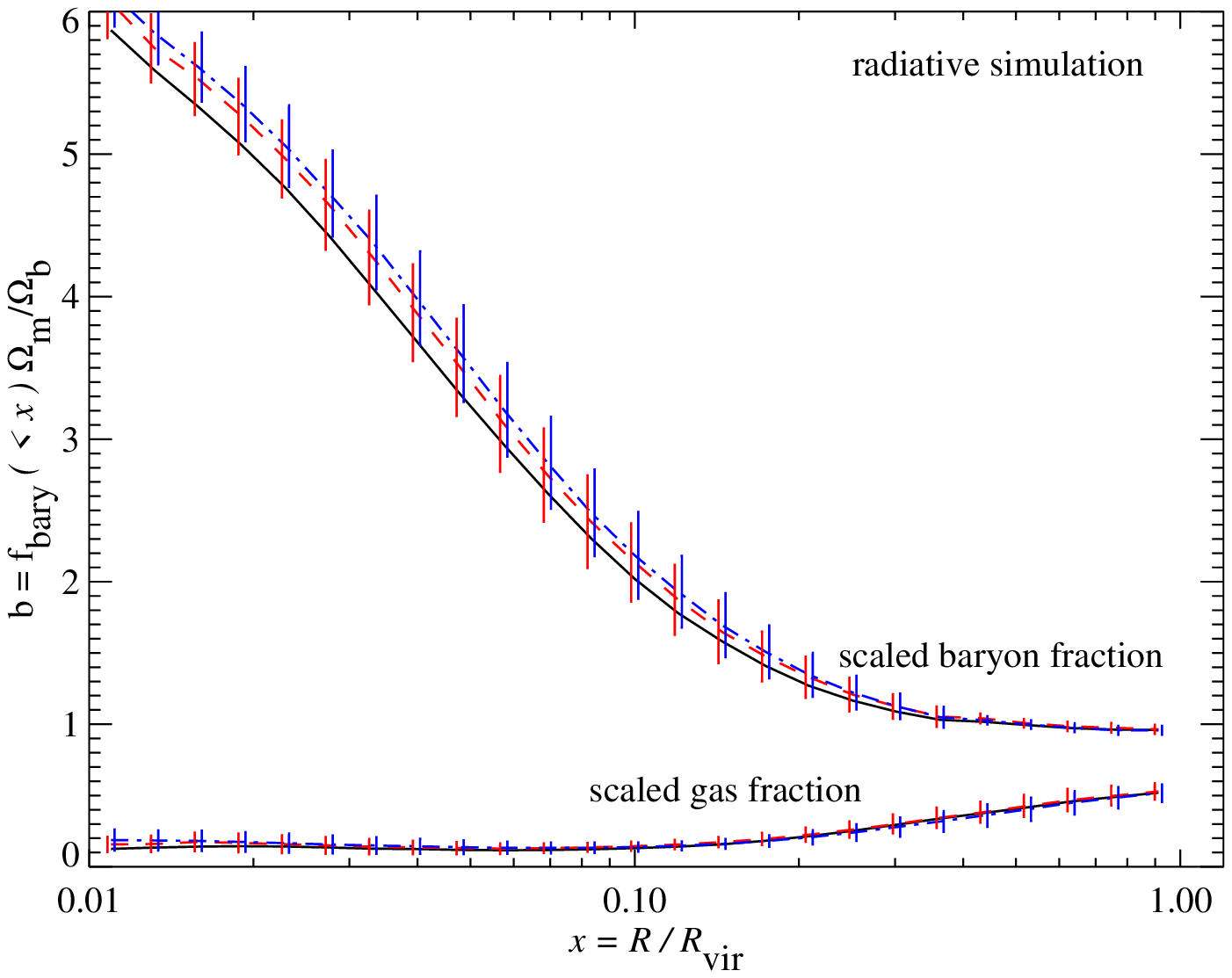}}\\
\resizebox{0.5\hsize}{!}{\includegraphics{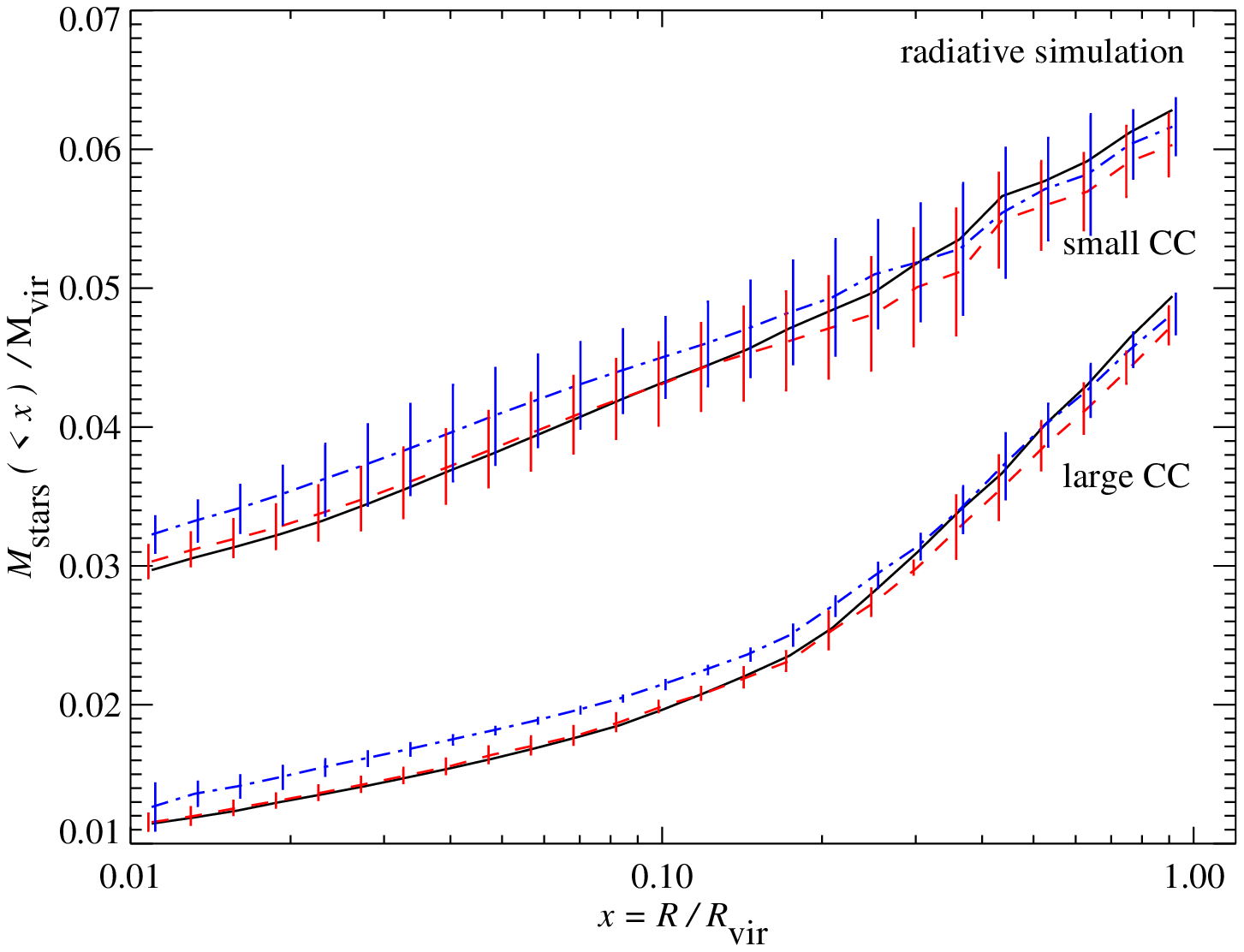}}%
\resizebox{0.5\hsize}{!}{\includegraphics{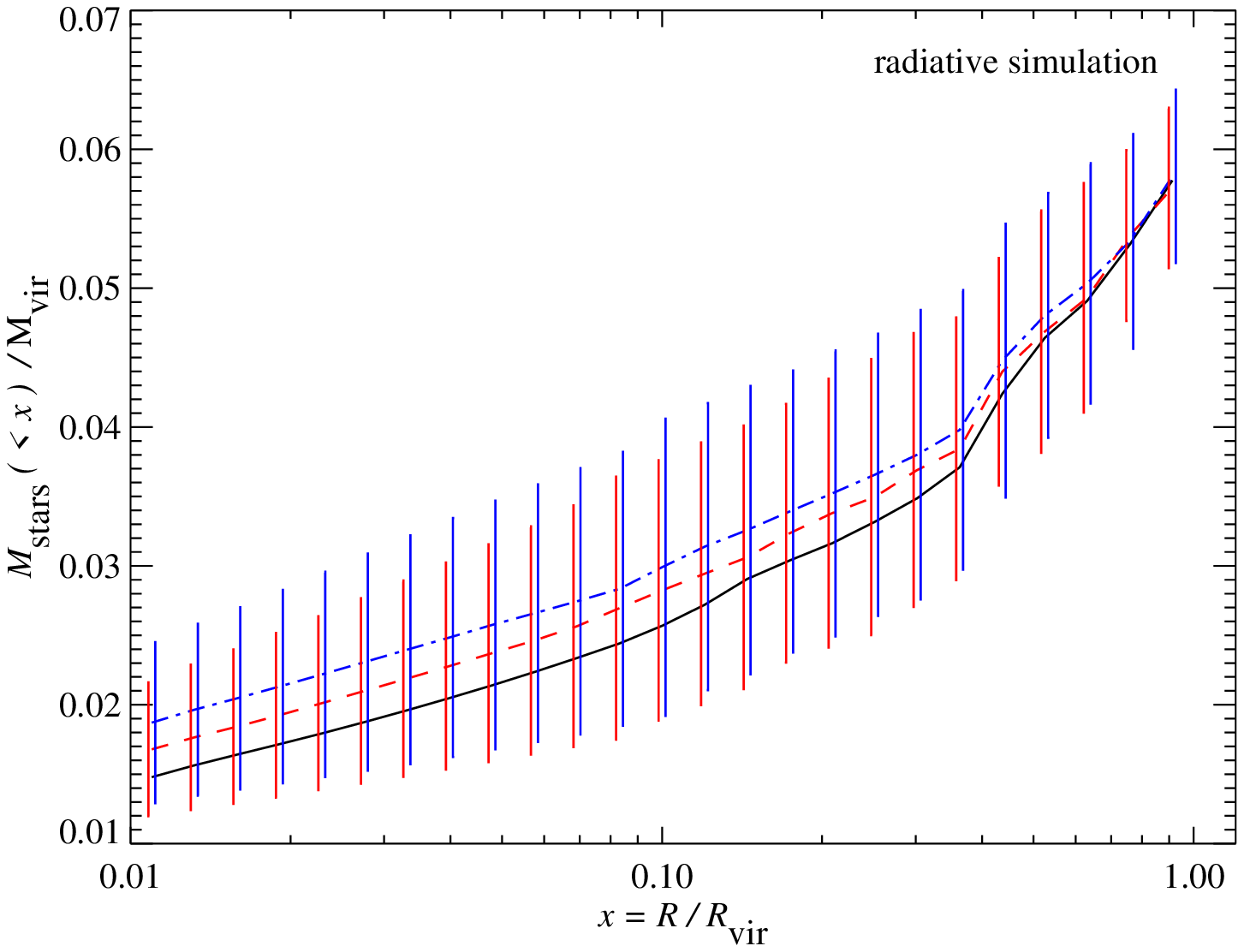}}\\
\end{center}
  \caption{Average profiles of our sample of all cool core clusters (left-hand
    side) and merging clusters (right-hand side) in our radiative simulations
    at redshift $z=0$.  Shown are the appropriately scaled profiles of the
    baryon fraction $b$ (top panels) and the stellar mass $M_\rmn{stars} (<x)$
    (bottom panels).  Colour scheme: black: reference simulations without CRs,
    red: CRs are injected only through shock acceleration using Mach number
    dependent scheme, blue: complete CR-model including CR injection from
    supernovae.}
  \label{fig:mstar}
\end{figure*}

\subsubsection{Pressure profiles}
Figures~\ref{fig:pressure} and \ref{fig:mstar} compare profiles of our sample
of all cool core clusters (CC clusters, on the left-hand side) to our sample of
merging clusters (on the right-hand side). Our samples of small and large CC
clusters differ only marginally for these profiles.  While the thermal pressure
profile of the CC clusters follows a cored $\beta$-profile, the pressure
profile of merging clusters almost shows a flat single power-law behaviour that
is due to the ongoing merging activities (cf.~Fig.~\ref{fig:pressure}, top
panels). The thermal pressure of the CC clusters in our CR-shock model exceeds
the one in our reference simulation, while the thermal pressure in the merging
clusters shows no significant difference among our different models apart from
the innermost central region where we observe an excess in both of our CR models.

In contrast, the CR pressure profiles do not show a core; their  profile
increases inwards roughly as $P_\CR \propto r^{-1}$ up to the external regions
of the ISM of the cD galaxy at $r \propto 0.04\, R_\rmn{vir}$, where their
profile steepens considerably and scales as $P_\CR \propto r^{-2}$ in our CC
clusters. The error bars of the CR pressure profiles have been suppressed for
clarity.  The CR pressure profiles in both of our CR models are similar
to each other: the profile in the CR-shock model shows a smoother behaviour
while the one in our complete model has a larger radial variation and a higher
pressure level due to the additionally injected CR pressure from supernovae
within galaxies.  The hypothesis that the supernova CRs are irrelevant for the
total pressure budget is however inadmissible because simulations that consider
only CRs from supernovae shocks show very similar pressure profiles
\citep{2006...Jubelgas}.  The reason lies rather in the fact that any CR
population is processed by adiabatic loss and transport processes in a
comparable fashion.

\subsubsection{Relative CR pressure profiles}
It is very instructive to study the relative CR pressure profiles $X_\CR(r) =
P_\CR(r) / P_\th(r)$ in Fig.~\ref{fig:pressure}, middle panels. Generally,
$X_\CR(r)$ is larger in merging clusters compared to CC clusters and does not
adopt values below $X_\CR=0.1$ at any radius for the complete CR model in
merging clusters.  As already mentioned in Sect.~\ref{sec:rel_CR_pressure},
strong merger shock waves efficiently inject CRs at the external cluster
regions and mix the highly CR-enriched intergalactic medium outside clusters
with the ICM, thus boosting the relative CR pressure during a merger. This also
results in large cluster-to-cluster variations at some given radius. Note that
the error bars of the relative CR pressure profiles have been suppressed for
clarity.

Compared to the very smooth, dynamically unimportant profile of our
non-radiative simulations, $X_\CR(r)$ shows a steep increase towards the
cluster centre within the ISM of the cD galaxy and much larger radial variation
in these radiative simulations. This is due to the short thermal cooling time
scales within dense galactic environments compared to the comparatively long CR
cooling time scales (cf.~Fig.~\ref{fig:CoolTimes}). There, the thermal gas
cools on a small timescale of $\tau \simeq 10^7 \mbox{ yr}$, diminishes
pressure support, condenses out of the gaseous phase and eventually forms
stars.  In the process, the CR pressure reaches values that exceed the equipartition with
the thermal pressure. In the outer parts of the cluster, $X_\CR(r)$ is a rising
function with increasing radius due to the same reasons mentioned above in our
non-radiative simulations.

\subsubsection{Effective adiabatic index}
In our CC cluster sample, the effective adiabatic index $\gamma_\eff$ departs
from its canonical value $\gamma_\th=5/3$ (in the absence of CRs) strongly in
the cluster centre inside the ISM of the cD galaxy and at the external regions
of the cluster where the relative CR pressure profile is dynamically important
(cf. Fig.~\ref{fig:pressure}, bottom panels). The composite of CRs and thermal
gas acquires there a higher compressibility that requires a larger overdensity
in order to exert the same pressure that balances the gravitational pressure
from the combination of dark matter and gas. The value of $\gamma_\eff$ in our
sample of merging clusters is everywhere softer compared to the canonical
adiabatic index although there is a large variation among different
clusters. This again reflects the higher CR pressure level in merging clusters
due to the high CR injection efficiency of merger shock waves.

Interestingly, the adiabatic index of the CR population $\gamma_\CR$ in our
radiative simulations is harder compared to our non-radiative simulations.
This can be traced back to the smaller overdensity in radiative simulations and
thus the smaller CR cooling rate due to Coulomb interactions that re-thermalize
the low-energy part of the CR momentum distribution function and eventually
yield an ultra-relativistic CR population. The lower overdensity in our
radiative simulations basically stems from the fact that gas mass from
intermediate cluster scales is removed since it replenished the condensing star
forming material inside galaxies and most importantly of the central cD galaxy.

\subsubsection{Baryon fraction}
Figure~\ref{fig:mstar} shows the baryon fraction scaled with the
universal value, $b = f_\rmn{bary} ({<x})\, \Omega_\rmn{m} /
\Omega_\rmn{b}$, that reaches values of unity at the virial radius
independent of the adopted model and dynamical state of the
cluster. The profile of $b$ rises steeply towards the centre, which is
mostly due to the increase of the stellar mass fraction relative to
the dark matter. Additionally, Fig.~\ref{fig:mstar} shows the gas
fraction scaled with the universal value, $f_\rmn{gas} ({<x})\,
\Omega_\rmn{m} / \Omega_\rmn{b}$. Its profile is decreasing inwards in
contrast to the scaled baryon fraction.  This shows clearly that the
stellar mass fraction as well as the stellar mass-to-gas fraction is
not constant as a function of radius but strongly increasing
inwards. As a word of caution, a fraction of the inward increase of
$b$ is caused by the over-cooling problem in the cluster centre where
we form too many stars that are too blue compared to observations. We
note that this is a result of the well-known cooling flow problem in
simulated clusters, which is independent of the employed numerical
method \citep[e.g.,][ and references therein]{2004MNRAS.348.1078B,
2005ApJ...625..588K, 2006MNRAS.365.1021E}. CR physics with CR source
terms studied in this paper such as structure formation shock waves
and supernovae shocks on galactic scales seems to reinforce this problem
instead of solving it.  Eventually, self-regulating heating sources
related to AGNs at cluster centres might be required to solve this
problem.

The value of $b$ in our sample of merging clusters is increased by roughly 10 per cent
in our CR models compared to our reference simulation while the enhancement is
smaller in the CC cluster sample. The simulations show a decreasing gas mass
fraction towards the centre for reasons that have been explained in
Sect.~\ref{sec:baryon_frac}.

\subsubsection{Cumulative stellar mass}
The profile of the cumulative stellar mass normalised by the virial mass of
the cluster is shown in Fig.~\ref{fig:mstar}.  One notes that there is a
higher fraction of stellar mass formed relative to $M_\vir$ in our small CC
clusters compared to the larger ones. This might be partly due to the above
mentioned over-cooling problem which has a higher relative impact in smaller
systems. There are more stars formed in our complete CR model on scales $r<
0.3 R_\vir$ irrespective of the dynamical state of the cluster while our
CR-shock model shows less formed stellar mass in the outer cluster
regions in our CC clusters. This shows again that the CR physics studied
here aggravates the cooling flow problem in contemporary simulations.

\subsubsection{Discussion of CRs profiles in radiative simulations}
\label{sec:rad_picture}

The CR distribution in the dilute outer cluster regions is dominated by the CR
injection characteristics with its strong structure formation shocks that
efficiently accelerate CRs into the IGM.  The relative CR pressure $X_\CR$
decreases as we move inwards due to the same reasons that apply in the case of
non-radiative simulations: (1) weak central flow shocks are inefficient in
accelerating CRs, (2) adiabatic compression of a composite of CRs and thermal
gas disfavours the CR pressure relative to the thermal pressure due to the
softer equation of state of CRs, and (3) CR loss processes are more important
at the dense centres.

Rather than becoming dynamically unimportant as it is the case in our
non-radiative simulations, the relative CR pressure increases strongly towards
dense substructures if we consider radiative losses of the thermal gas due to
the long CR cooling time scales compared to those of the thermal gas. The
thermal gas cools much faster radiatively and diminishes its pressure support
while it condenses out and forms stars. The high fraction of pressure support
provided by CRs yields a higher compressibility of the composite fluid that
requires a larger overdensity in order to balance the gravitational pressure
from the dark matter and gas. This increases the density of each galactic
substructure as well as in the cluster centre and leads to a pressure
enhancement, provided the hydrostatic equation is applicable, $\dd P /(\dd r) =
\rho_\rmn{gas}\, G\, M(<r)\, r^{-2}$. This higher density leads to a higher
rate of star formation and thus increases the central baryon fraction 
since gas from larger scales has to replenish the condensed gas in order to
maintain hydrostatic equilibrium.

In the case of ongoing merger activity, the relative CR pressure is boosted due
to a combination of two effects: (1) a merger event triggers large random
motions and merger shock waves that steepen considerably while they are running
into the shallower cluster potential where they efficiently inject CRs. (2)
These large random motions mix the highly CR-enriched intergalactic medium
outside clusters with the ICM.  Weak virialisation shocks traversing the
cluster after the merger thermalize random gas motions, thus increase the
thermal energy and yield a decrease of $X_\CR$. CR loss processes and adiabatic
compression furthermore decreases $X_\CR$ in intermediate cluster regions with
a comparatively long thermal cooling time scale.

\subsection{Cosmic ray related physical processes}
\label{sec:physical_processes}
In the following sections we study the physical processes related to
CRs in more detail, which complements the previous sections. In particular,
we will consider projected maps of physical cluster properties, and
analyse the Mach number statistics during cluster formation. Finally,
Sect.~\ref{sec:thermal_obs} studies the consequences of CRs for
thermal observables such as X-ray emission and the Sunyaev-Zel'dovich
effect.

\subsubsection{Maps of physical quantities}

\begin{figure*}
\begin{center}
\resizebox{0.495\hsize}{!}{\includegraphics{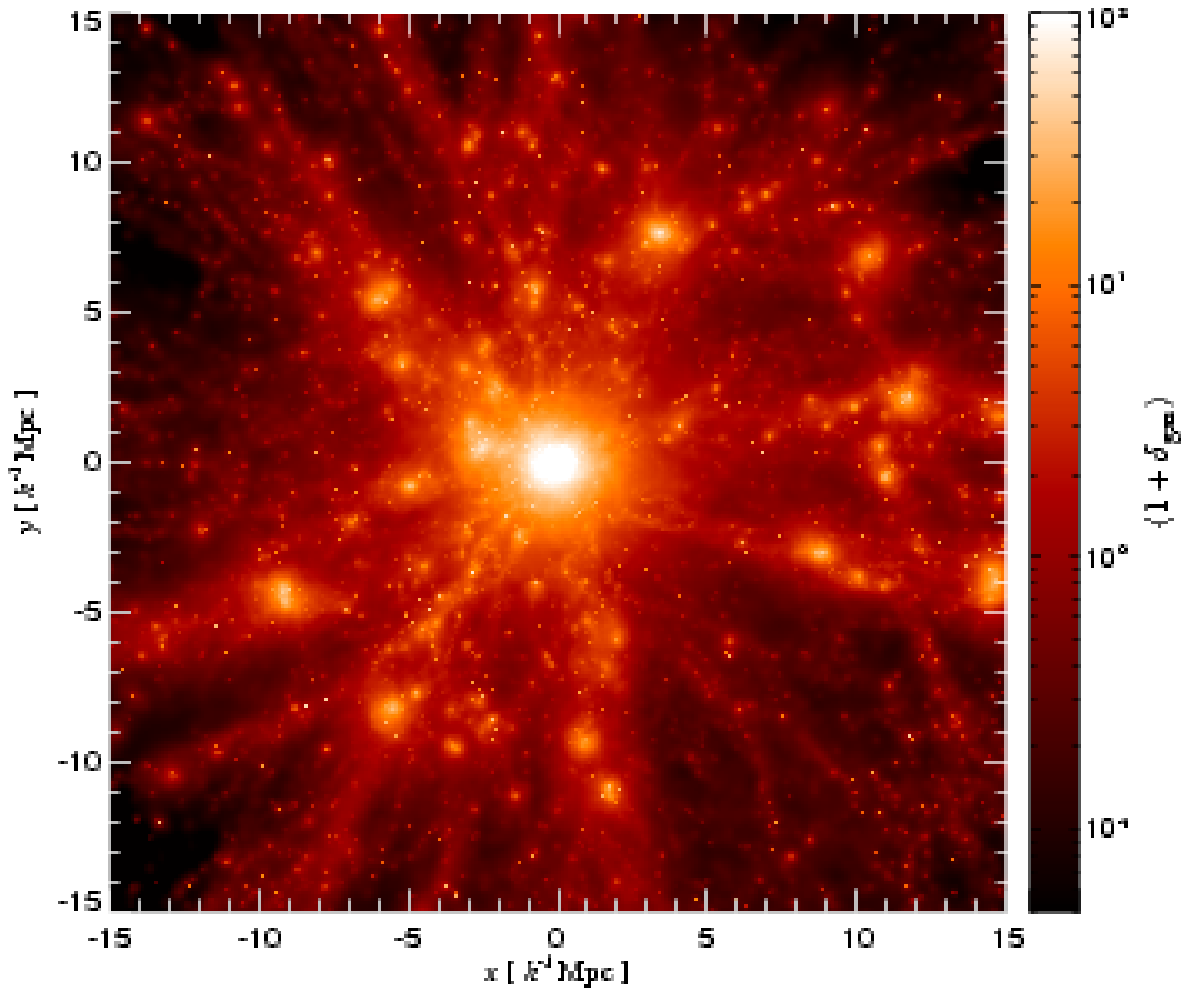}}\hfill%
\resizebox{0.495\hsize}{!}{\includegraphics{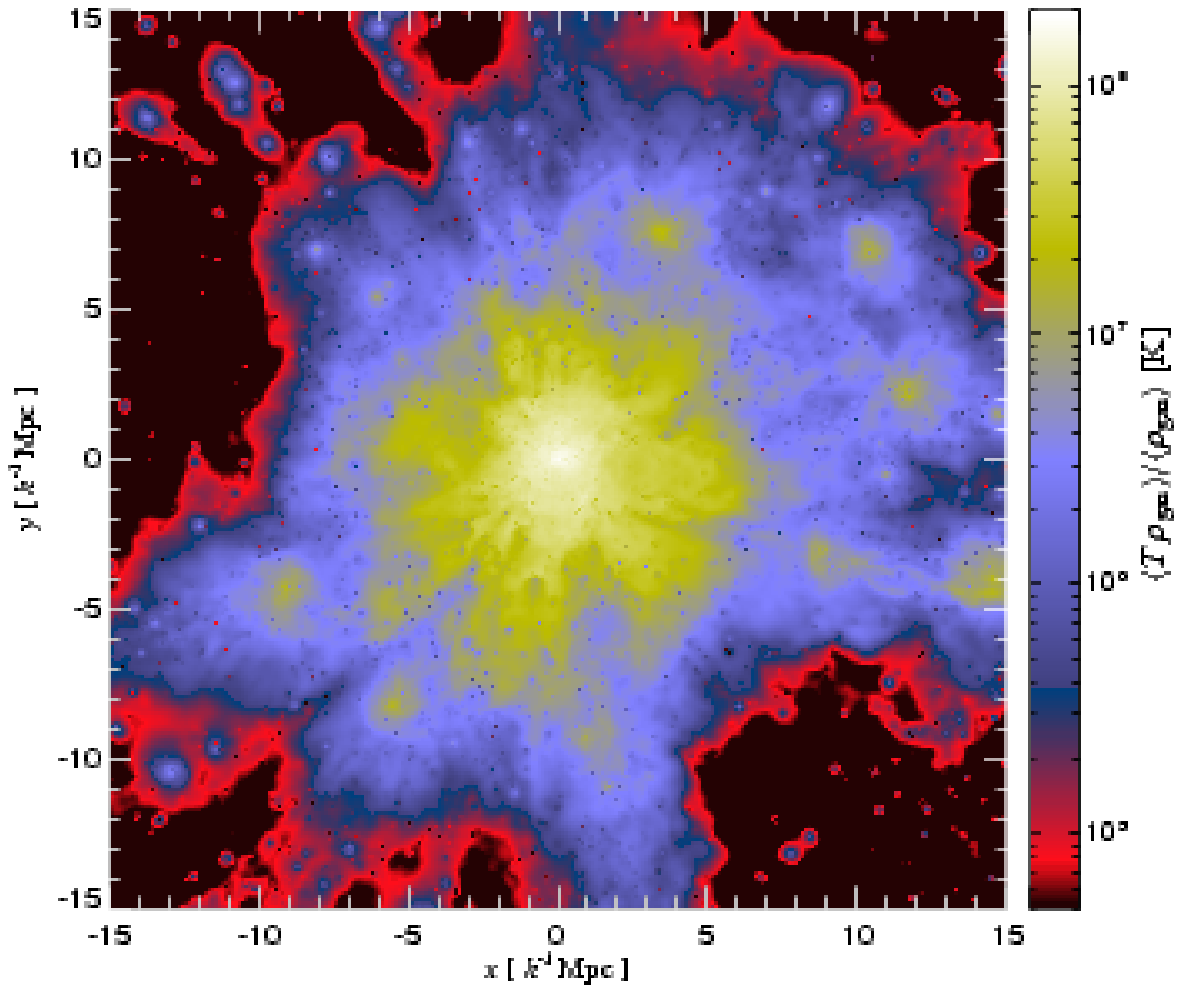}}\\
\resizebox{0.495\hsize}{!}{\includegraphics{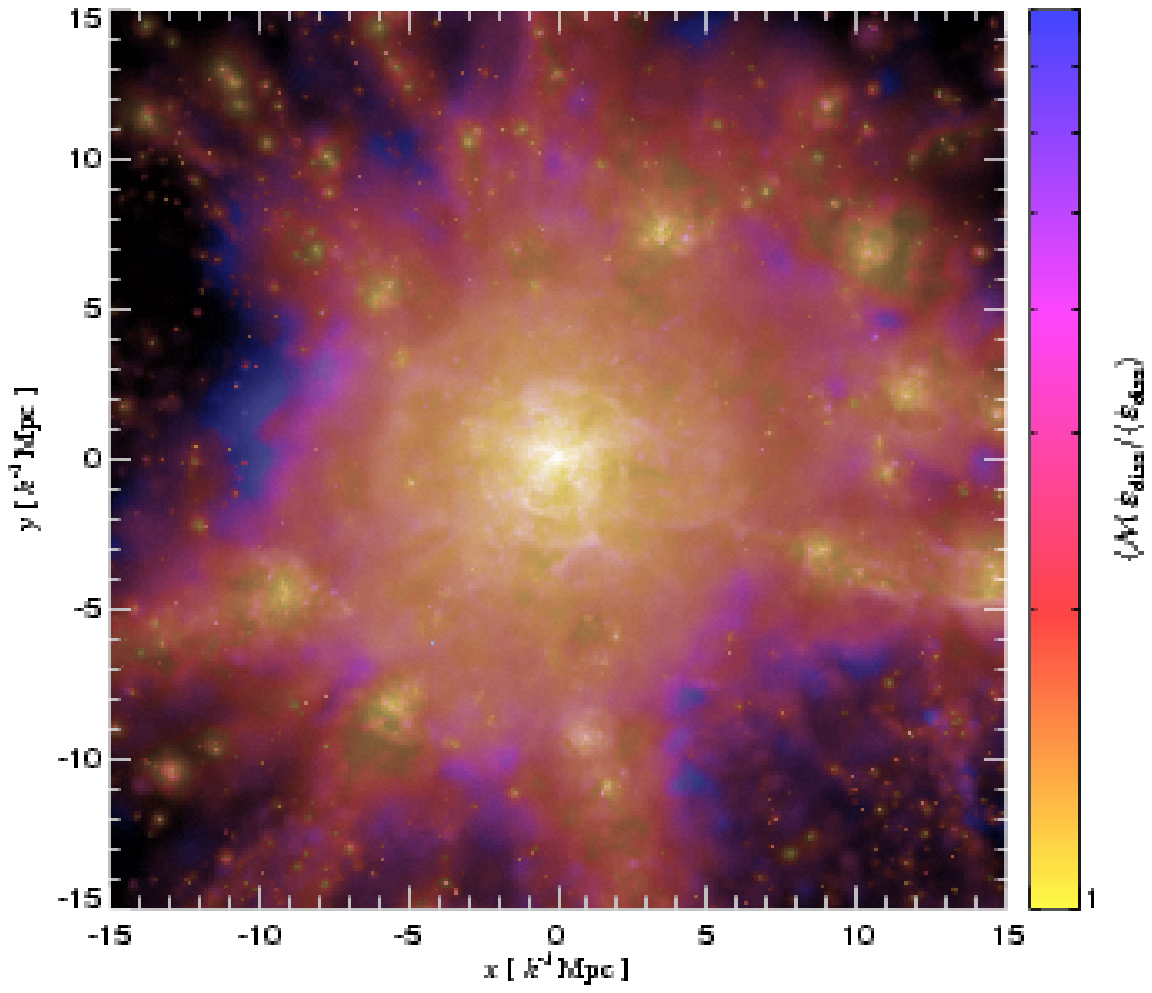}}\hfill%
\resizebox{0.495\hsize}{!}{\includegraphics{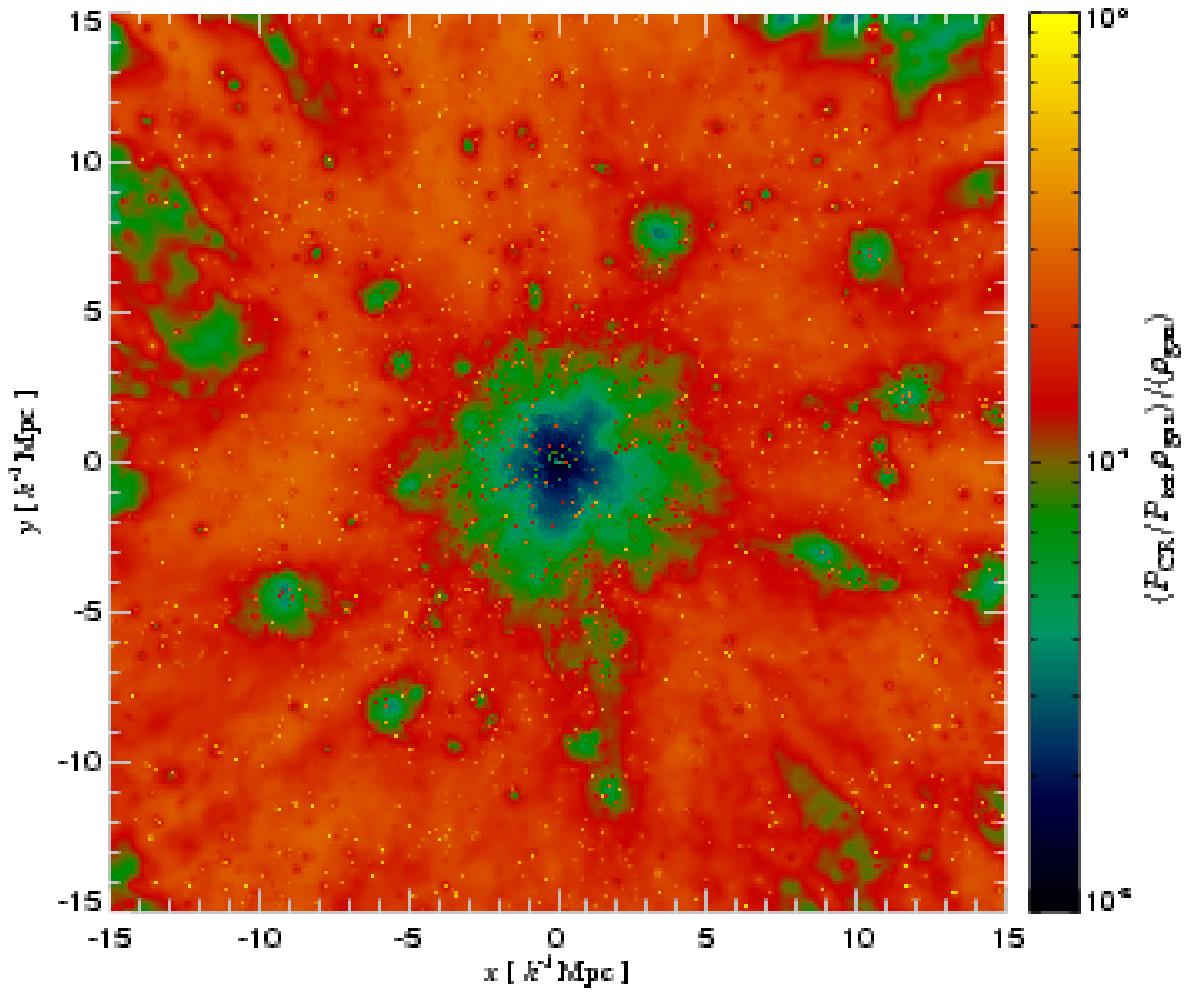}}\\
\end{center}
  \caption{Environment of the largest galaxy cluster in our sample (a cool core
  cluster) in the radiative simulation including CRs from structure formation
  shocks only. Shown are the projected density, mass-weighted temperature, the
  Mach number of shocks weighted by the energy dissipation rate in colour (while
  the brightness displays the logarithm of the dissipation rate), and the mass
  weighted CR pressure relative to the total pressure. We performed
  line-of-sight averages to obtain the projected quantities.}
  \label{fig:g8}
\end{figure*}

\begin{figure*}
\begin{center}
  \begin{minipage}[t]{0.495\textwidth}
    \centering{\it \Large Small cool core cluster:}
  \end{minipage}
  \hfill
  \begin{minipage}[t]{0.495\textwidth}
    \centering{\it \Large Large cool core cluster:}
  \end{minipage}
\resizebox{0.5\hsize}{!}{\includegraphics{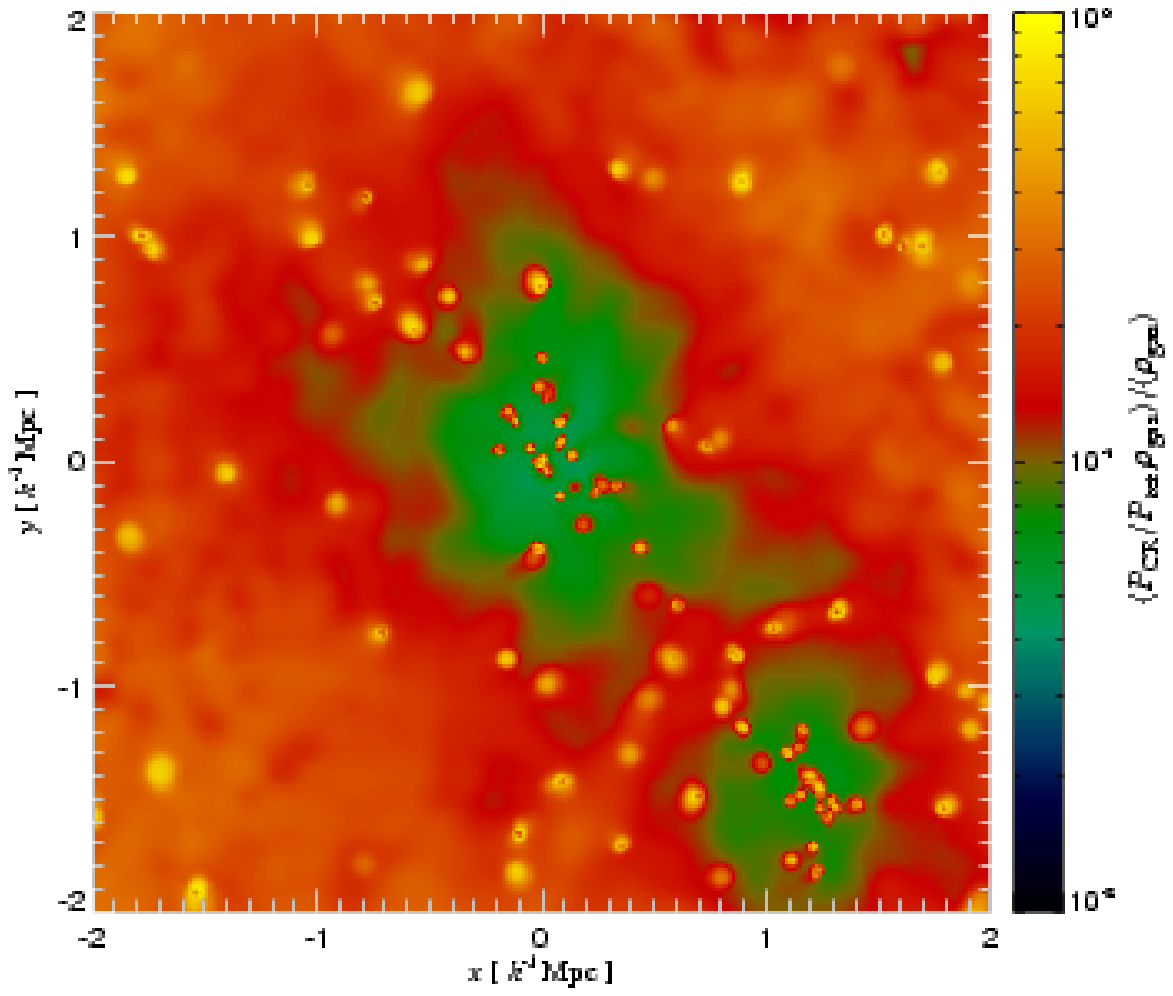}}%
\resizebox{0.5\hsize}{!}{\includegraphics{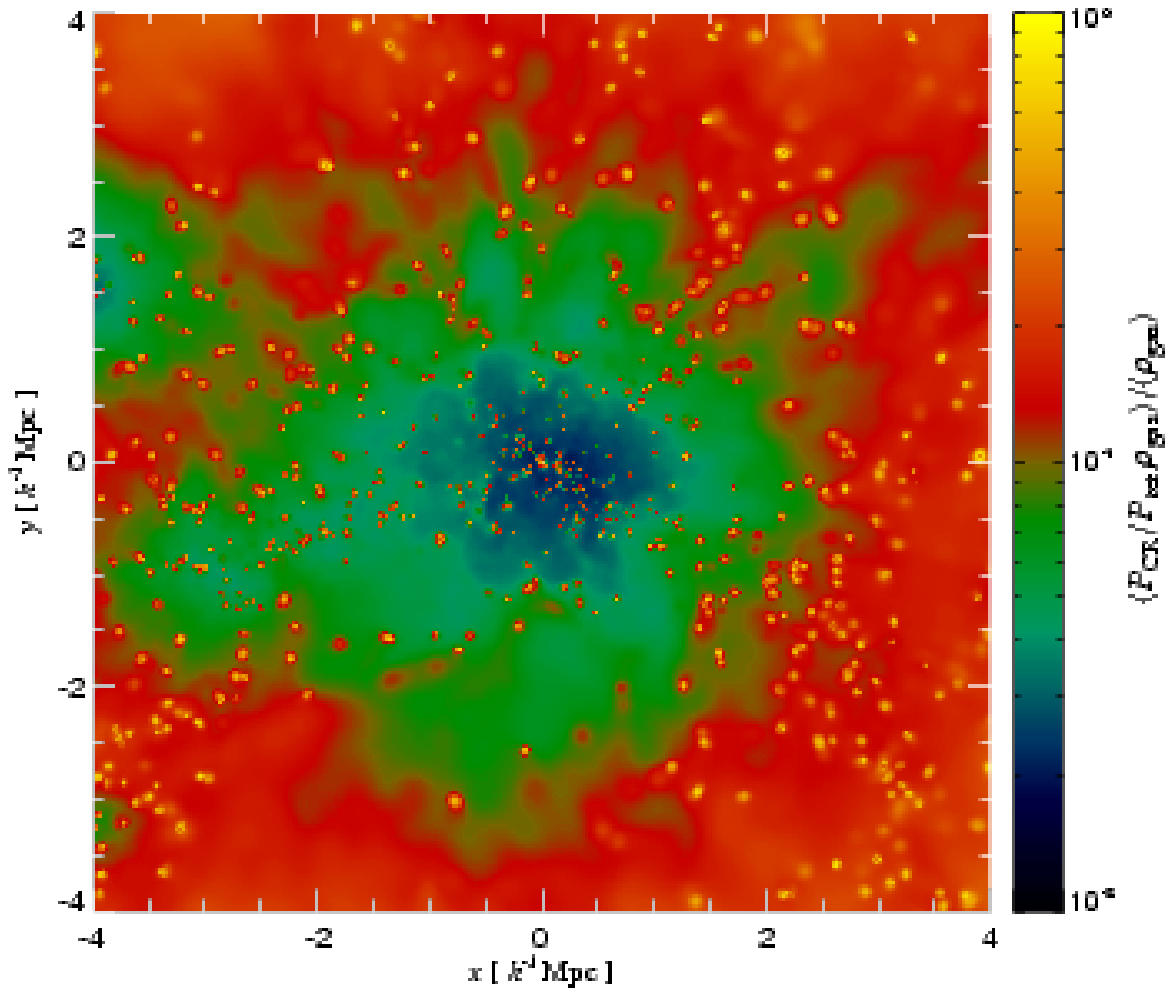}}\\
\resizebox{0.5\hsize}{!}{\includegraphics{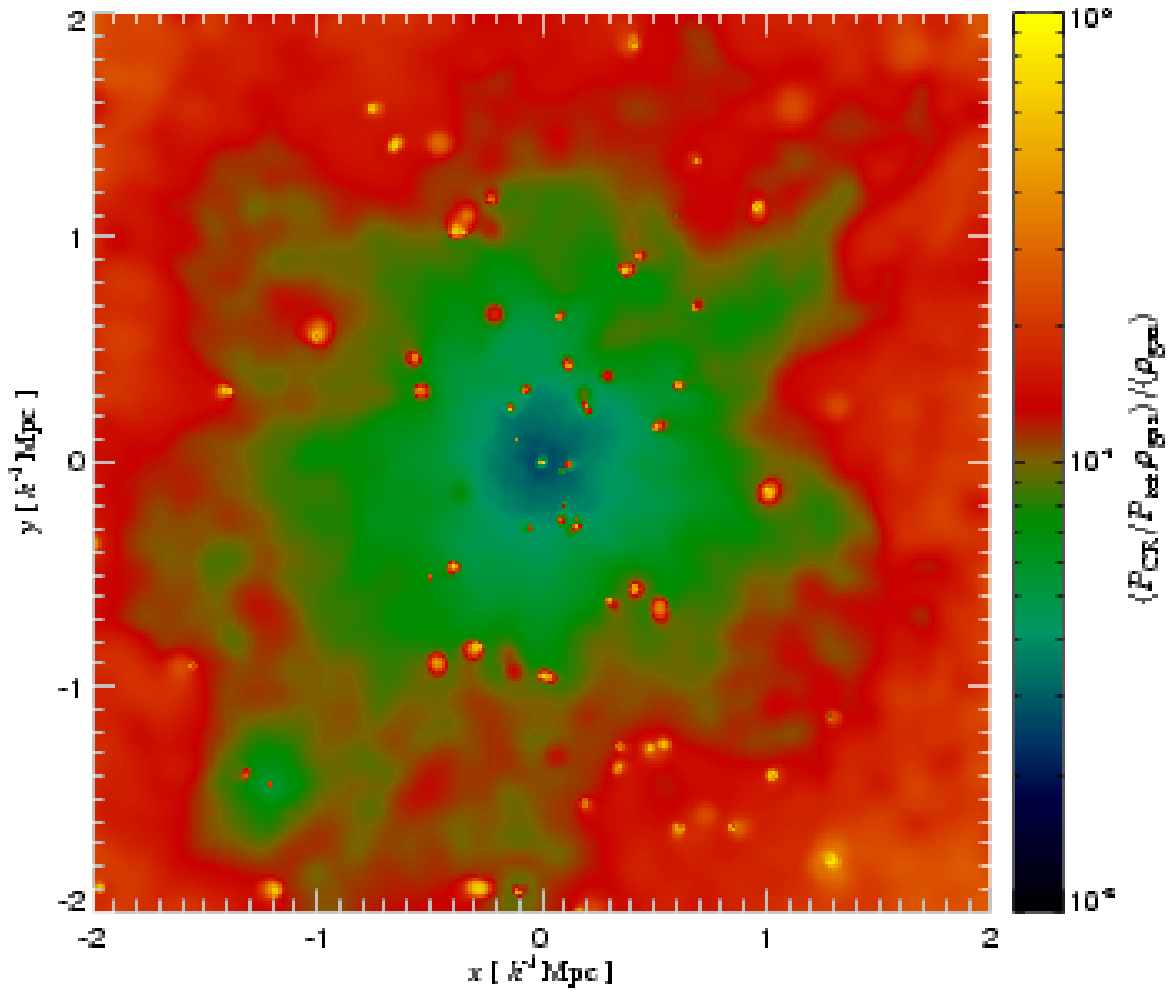}}%
\resizebox{0.5\hsize}{!}{\includegraphics{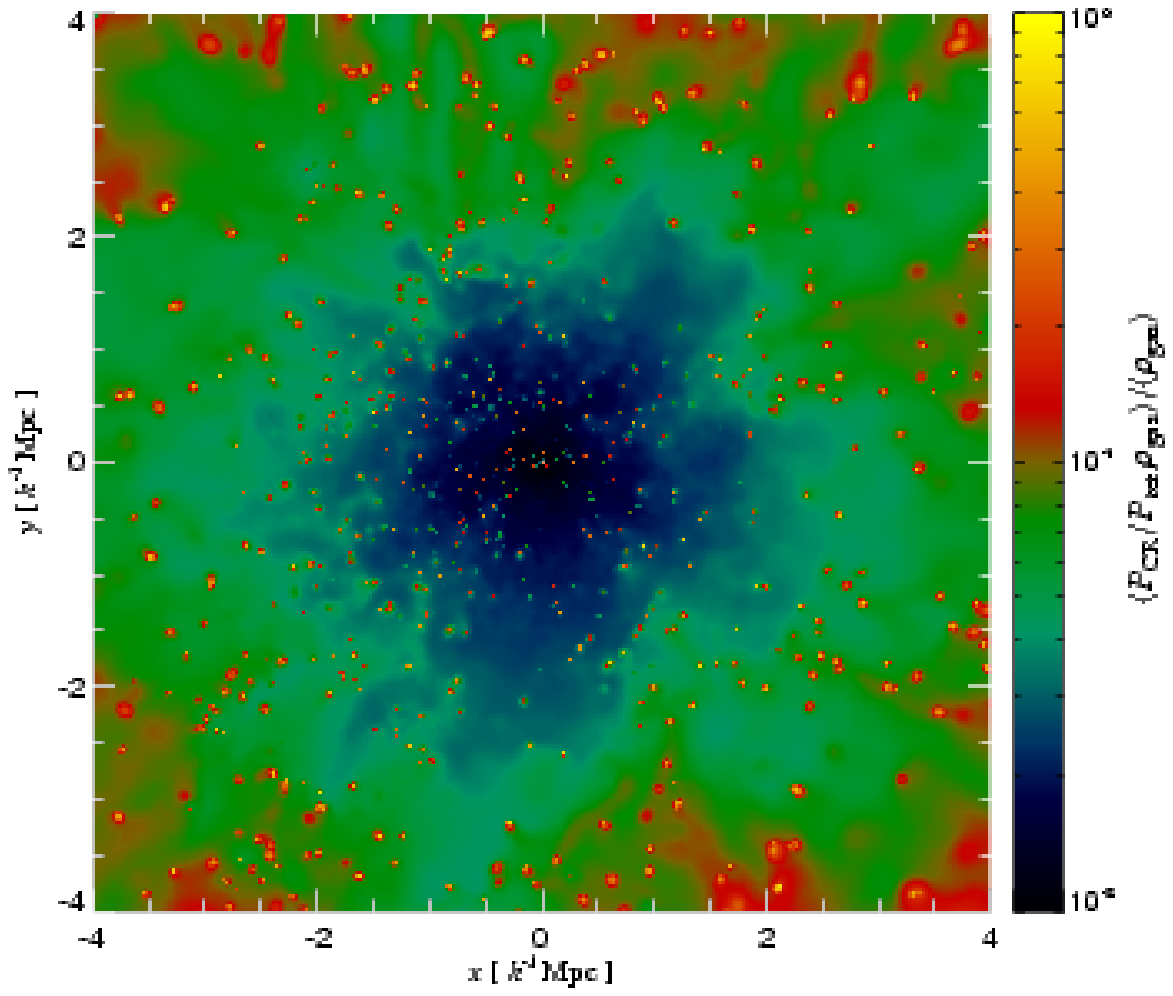}}\\
\end{center}
  \caption{Time evolution of the mass-weighted CR pressure relative to the
  total pressure in the radiative simulation including CRs from structure
  formation shocks. Upper panels show the clusters at $z = 1$, the lower panels
  at $z = 0$.  Left panels: small cool core cluster with a virial mass  of
  $8.8\times 10^{13}\,h^{-1}\,\mbox{M}_\odot$ at $z=0$. Right panels: large cool core cluster with
  a virial mass  of $1.8\times 10^{15}\,h^{-1}\,\mbox{M}_\odot$.}
  \label{fig:XCRtime_evolution}
\end{figure*}

We produced projected maps of the density, the mass-weighted temperature, CR
related quantities, and thermal cluster observables.  Generally, a
three-dimensional scalar field $a(\vecbf{r})$ along any ray was calculated by
distributing the product of $a(\vecbf{r})$ and the specific volume $M_\alpha /
\rho_\alpha$ of the gas particles over a grid comoving with the cosmic
expansion. This yields the projected quantity $A (\vecbf{r}_{\bot})$:
\begin{equation}
  \label{eq:projection}
  A (\vecbf{r}_{\bot,\, ij}) = \frac{1}{L_\rmn{pix}^2}
  \sum_\alpha a_\alpha \frac{M_\alpha}{\rho_\alpha}\,
  W_{\alpha,\,ij}(\vecbf{r}_{\bot,\, ij} - \vecbf{r}_\alpha),
\end{equation}
where $W_{\alpha,\,ij}$ is the value of the projected smoothing kernel
(normalised to unity for the pixels covered) of an SPH particle $\alpha$ at
comoving grid position $\vecbf{r}_{\bot,\, ij}$, and $L_\rmn{pix}^2$ is the
comoving area of the pixel.

The surface mass density map was produced by projecting the quantity $a_\alpha
= \rho_\alpha / (\Omega_\rmn{b}\, \rho_\rmn{cr})$ and the mass-weighted
temperature map by setting $a_\alpha = T_\alpha\, \rho_\alpha$ divided by the
mass projection.  Figure~\ref{fig:g8} shows the radiative simulation of a
super-cluster region (with the identifier g8) which is dominated by a large
cool core cluster and surrounded by smaller satellite clusters.  While the ICM
of the central massive cluster reaches a temperature of the order of its virial
temperature of $kT_\vir = 13.1 \mbox{ keV}$, the surrounding warm-hot
intergalactic medium (WHIM) acquires temperatures of $kT \sim (10^{-2} -
10^{-1}) \mbox{ keV}$.  The spatial distribution of shock strengths can be
studied best by looking at the Mach numbers weighted by the energy dissipation
rate at shocks.  The colour hue of the lower panel on the left-hand side of
Fig.~\ref{fig:g8} has been chosen to represent the projected quantity $a_\alpha
= \M_\alpha \dot{\eps}_{\rmn{diss},\, \alpha}$ divided by the projected energy
dissipation rate $\bra \dot{\eps}_\rmn{diss} \ket_\rmn{los}$, where $\M_\alpha$
denotes the Mach number of the SPH particle. The brightness scales
logarithmically with the projected dissipation rate $\bra \dot{\eps}_\rmn{diss}
\ket_\rmn{los}$. Within this super-cluster region most of the energy is
dissipated in weak internal shocks with Mach numbers $\M\lesssim 2$ which are
predominantly central flow shocks or merger shock waves traversing the cluster
centre. Collapsed cosmological structures are surrounded by several shells of
external shocks with successively higher Mach numbers, but they play only a
minor role in the energy balance of thermalization as can be inferred by its
dim brightness. Clearly visible are spherical shells of shocks at different
radii from the cluster centre. Two distinct accretion shocks at distances of 2
and $3\,h^{-1}\mbox{ Mpc}$ to the cluster centre are visible, followed by
shells of stronger shocks further outwards.

\begin{figure*}
\begin{center}
\resizebox{\hsize}{!}{\includegraphics{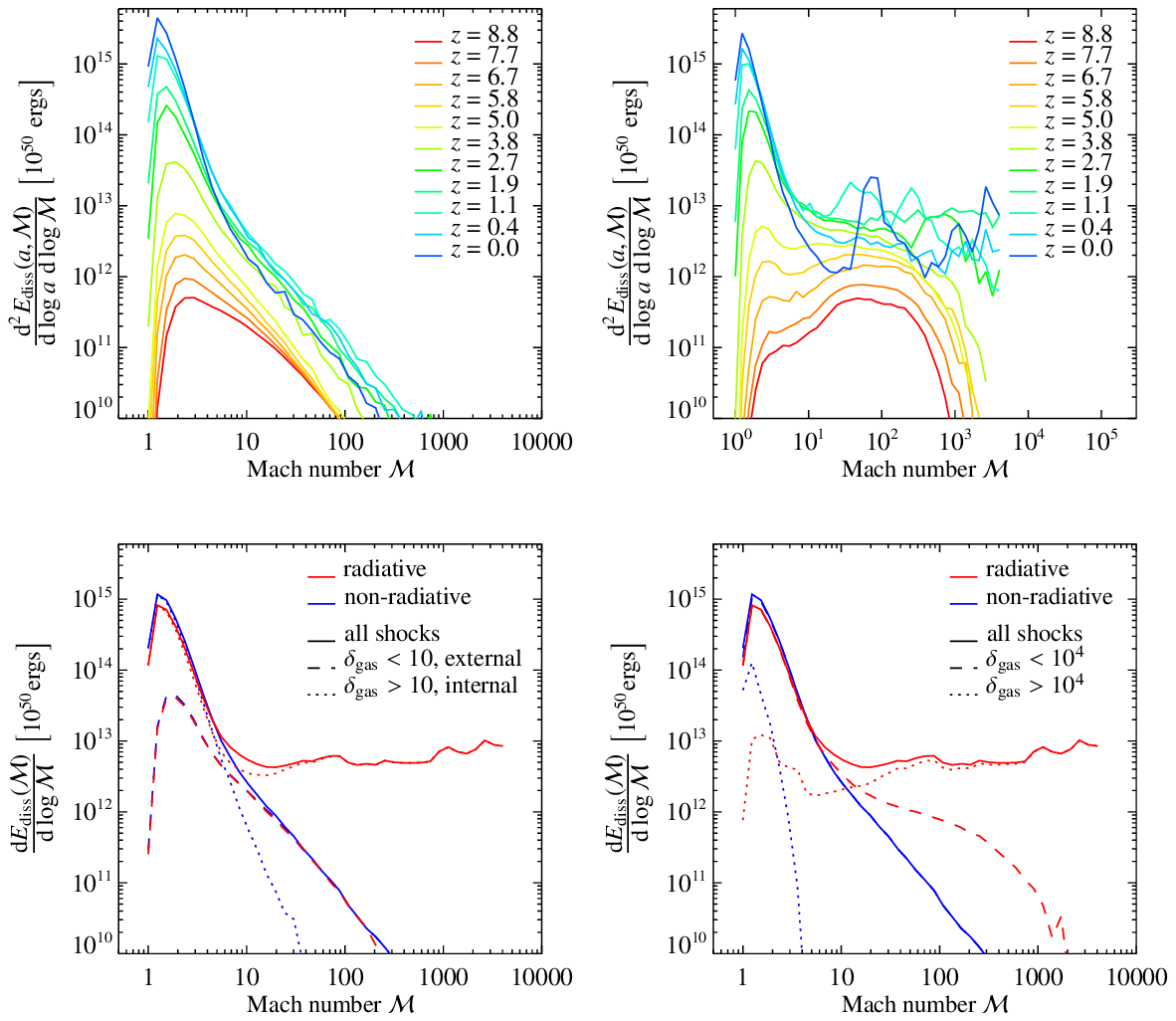}}
\end{center}
  \caption{Influence of radiative cooling and star formation on the Mach number
  statistics of high-resolution cluster simulations (g8). The {\em top
  left-hand panel} shows the differential Mach number distribution $\dd^2
  E_\rmn{diss}(a,\M)/( \dd \log a\, \dd \log\M)$ for our non-radiative
  simulation while the {\em top right-hand panel} shows this distribution for
  the radiative simulation.  The {\em lower left-hand panel} shows both
  distributions integrated over the scale factor, $\dd E_\rmn{diss}(\M)/( \dd
  \log \M)$. External shocks (occurring in overdensities $\delta_\rmn{gas} =
  \rho_\rmn{gas} / (\Omega_\rmn{b}\, \rho_\rmn{cr}\, a^{-3}) - 1< 10$) are shown
  with dashed lines and internal shocks ($\delta_\rmn{gas} > 10$) with dotted
  lines.  The {\em lower right-hand panel} shows again the integrated Mach
  number distribution, $\dd E_\rmn{diss}(\M)/( \dd \log \M)$. Here, the shock
  populations are subdivided into two categories that are separated by the
  overdensity threshold $\delta_\rmn{gas} = 10^4$. This shows unambiguously
  that the flat tail of the Mach number distribution in radiative simulations
  originates from internal shocks within cooled high-density regions that are
  almost absent in non-radiative simulations.}
  \label{fig:MachDistrib}
\end{figure*}

Finally, the mass-weighted CR pressure relative to the total pressure
$\tilde{X}_\CR = P_\CR / P_\rmn{tot}$ was obtained by projecting $a_\alpha =
\rho_\alpha\, \tilde{X}_{\CR,\,\alpha}$ divided by the mass projection (lower
panel on the right-hand side of Fig.~\ref{fig:g8}). The projections of the
relative CR pressure support the picture inferred from our profiles. While
$\tilde{X}_\CR$ acquires comparatively high values within the WHIM that are
hydrodynamically important, their importance diminishes (on average) within
each galaxy cluster for reasons laid out above
(cf.~Sect.~\ref{sec:rad_picture}). Within each individual galaxy, the CR pressure reaches
equipartition or dominates the thermal pressure as can be seen by the numerous
yellow points sprinkled over the map, each corresponding to a galaxy.  The time
evolution of the relative CR pressure from $z=1$ until today and the comparison
of $\tilde{X}_\CR$ for a large and a small CC cluster can be studied in more
detail in Fig.~\ref{fig:XCRtime_evolution}, that shows the central regions of
the galaxy cluster 1 and 10. At $z=0$, the iso-contour of $\tilde{X}_\CR = 0.1$
(the transition from green to red for our colour scale) roughly coincides with
the virial radius in our small cluster $R_\vir \simeq 0.7\, h^{-1}\mbox{ Mpc}$
while it approximately lies at a scale of twice the virial radius in the large
cluster ($R_\vir \simeq 2\, h^{-1} \mbox{ Mpc}$). This already hints that
weaker shocks are responsible for thermal dissipation in high mass halos
compared to low mass halos. We will study this effect in more detail in the
next section. The second striking observation is that at high redshifts the
warmer colours dominate over the cooler ones, indicating a higher level of
relative CR pressure at high redshift. This is because in the hierarchical
model of structure formation, the characteristic halo mass and thus the
associated characteristic temperature grow with cosmic time. The higher
sound speeds of the diffuse gas at later times decreases the characteristic
Mach numbers of the shocks responsible for thermalization and thus decreases
the acceleration efficiency of CRs.

\subsubsection{Mach number statistics}

\begin{figure*}
\begin{center}
  \begin{minipage}[t]{0.495\textwidth}
    \centering{\it \Large Non-radiative simulation:}
  \end{minipage}
  \hfill
  \begin{minipage}[t]{0.495\textwidth}
    \centering{\it \Large Radiative simulation:}
  \end{minipage}
\resizebox{\hsize}{!}{\includegraphics{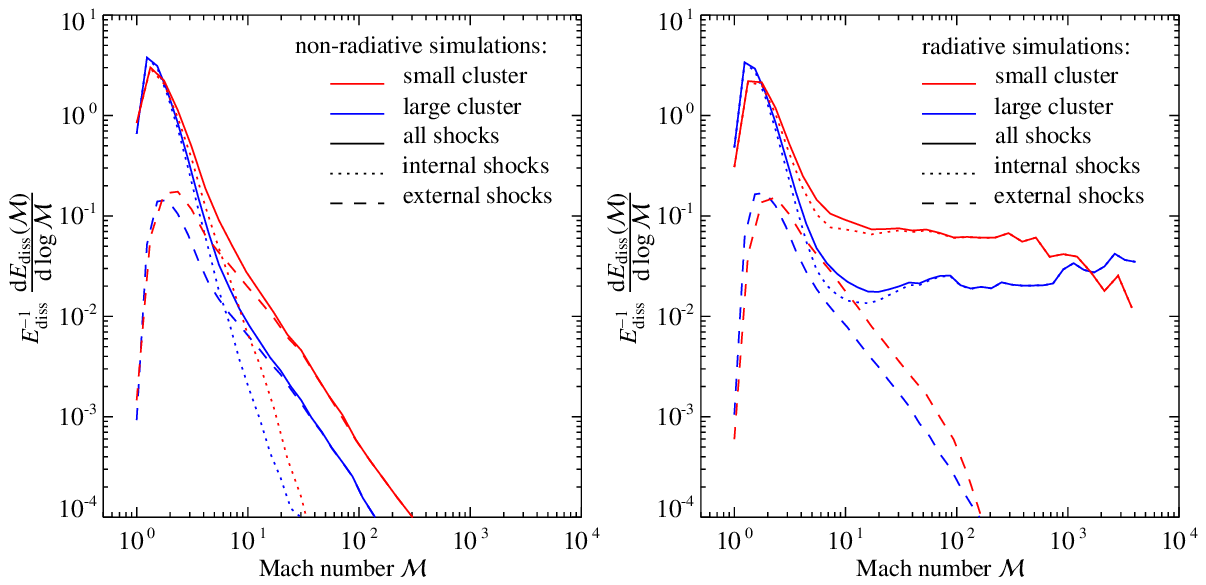}}
\end{center}
  \caption{Influence of the cluster size on the Mach number statistics of
  high-resolution cluster simulations. The {\em left-hand panel} compares the
  appropriately scaled Mach number distribution $\dd E_\rmn{diss}(\M)/( \dd
  \log \M)$ of a large super-cluster region (g8) and a small isolated cluster
  region (g676) in our non-radiative simulation, while the {\em right-hand
  panel} shows the comparison in our simulation with radiative gas physics.
  Internal shocks (occurring at overdensities $\delta_\rmn{gas} =
  \rho_\rmn{gas} / (\Omega_\rmn{b}\, \rho_\rmn{cr}\, a^{-3}) > 10$) are shown with
  dotted lines and external shocks with dashed lines. The total dissipated
  energy integrated over the cosmic formation history in our simulation of our
  isolated small cluster (g676) amounts to $E_\rmn{diss}= 2.4\times
  10^{62}\mbox{ erg}$, in our large super-cluster simulation (g8) the amount of
  dissipated energy was two orders of magnitude larger, $E_\rmn{diss}=
  2.4\times 10^{64}\mbox{ erg}$. The characteristic Mach number which is
  responsible for the energy dissipation decreases with cluster mass due to the
  increasing sound speed in those systems.}
  \label{fig:MachDistrib_small-large}
\end{figure*}

Earlier work on this subject \citep{2000ApJ...542..608M, 2003ApJ...593..599R,
2006MNRAS.367..113P} classified structure formation shocks into two broad
populations which are labelled as {\em internal} and {\em external} shocks.
These categories have been distinguished on whether or not the associated
pre-shock gas was previously shocked in the cosmic history. Since we are also
interested in assessing the properties of shock waves in radiative simulations,
it turns out to be favourable not to use a thermodynamical criterion such as
the temperature but a criterion based on the overdensity $\delta_\rmn{gas} =
\rho_\rmn{gas} / (\Omega_\rmn{b}\, \rho_\rmn{cr}\, a^{-3}) - 1$ in order not to
confuse the shock definition in the presence of radiatively cooling gas in
galaxies (here, $\rho_\rmn{gas}$ denotes the physical gas mass density and $a$
is the cosmic scale factor). In practice, we use the criterion of a critical
pre-shock overdensity $\delta_\rmn{gas}>10$ or the physical gas density
$\rho_\rmn{gas} > 7.5\times 10^{-24} \mbox{ g cm}^{-3}$ for the classification
of an internal shock.  With this criterion, {\em external} shock surfaces
surround filaments, sheets, and haloes where the pristine adiabatically cooling
gas is shocked for the first time while {\em internal} shocks are located
within the regions bound by external shocks and are created by flow motions
accompanying hierarchical structure formation. Their population includes
accretion shocks produced by infalling material along the filaments into
clusters, merger shocks resulting from infalling haloes, and flow shocks inside
nonlinear structures which are excited by supersonic motions of subclumps.  The
thermal gas of the warm-hot intergalactic medium has a low sound velocity $c =
\sqrt{\gamma_\th P/\rho}=\sqrt{\gamma_\th k T / (\mu m_\p)}$ owing to the low
temperature, so once the diffuse gas breaks on mildly nonlinear structures,
strong shock waves develop that are characterised by high Mach numbers
$\M=\vel_\rmn{s}/ c$.  Nevertheless, the energy dissipation rate of internal
shocks is always higher compared to external shocks within the WHIM because the
mean shock speed and pre-shock gas densities are significantly larger for
internal shocks.

To quantify these considerations, we compute the differential Mach number
distribution weighted by the shock-dissipated energy $\dd^2 E_\rmn{diss}
(a,\M)/( \dd \log a\, \dd \log\M)$ at different redshifts characterised by
scale factor $a$. The top left-hand panel of Fig.~\ref{fig:MachDistrib} shows
this Mach number distribution in our non-radiative simulation of a
super-cluster region (g8), while the top right-hand panel shows this
distribution for the radiative simulation using the same initial conditions.
The lower left-hand panel shows both distributions integrated over the scale
factor, $\dd E_\rmn{diss}(\M)/( \dd \log \M)$.  Internal shocks are shown with
dotted lines and external shocks with dashed lines.  The lower right-hand panel
shows again the integrated Mach number distribution. This time, however,
the shock population is subdivided into two categories that are separated by
the overdensity threshold $\delta_\rmn{gas} > 10^4$ to study the origin of the
flat horizontal tail of the Mach number distribution extending to very high
Mach numbers in our simulation with radiative gas physics.

\begin{figure*}
\begin{center}
  \begin{minipage}[t]{0.495\textwidth}
    \centering{\it \Large Non-radiative simulation:}
  \end{minipage}
  \hfill
  \begin{minipage}[t]{0.495\textwidth}
    \centering{\it \Large Radiative simulation:}
  \end{minipage}
\resizebox{0.5\hsize}{!}{\includegraphics{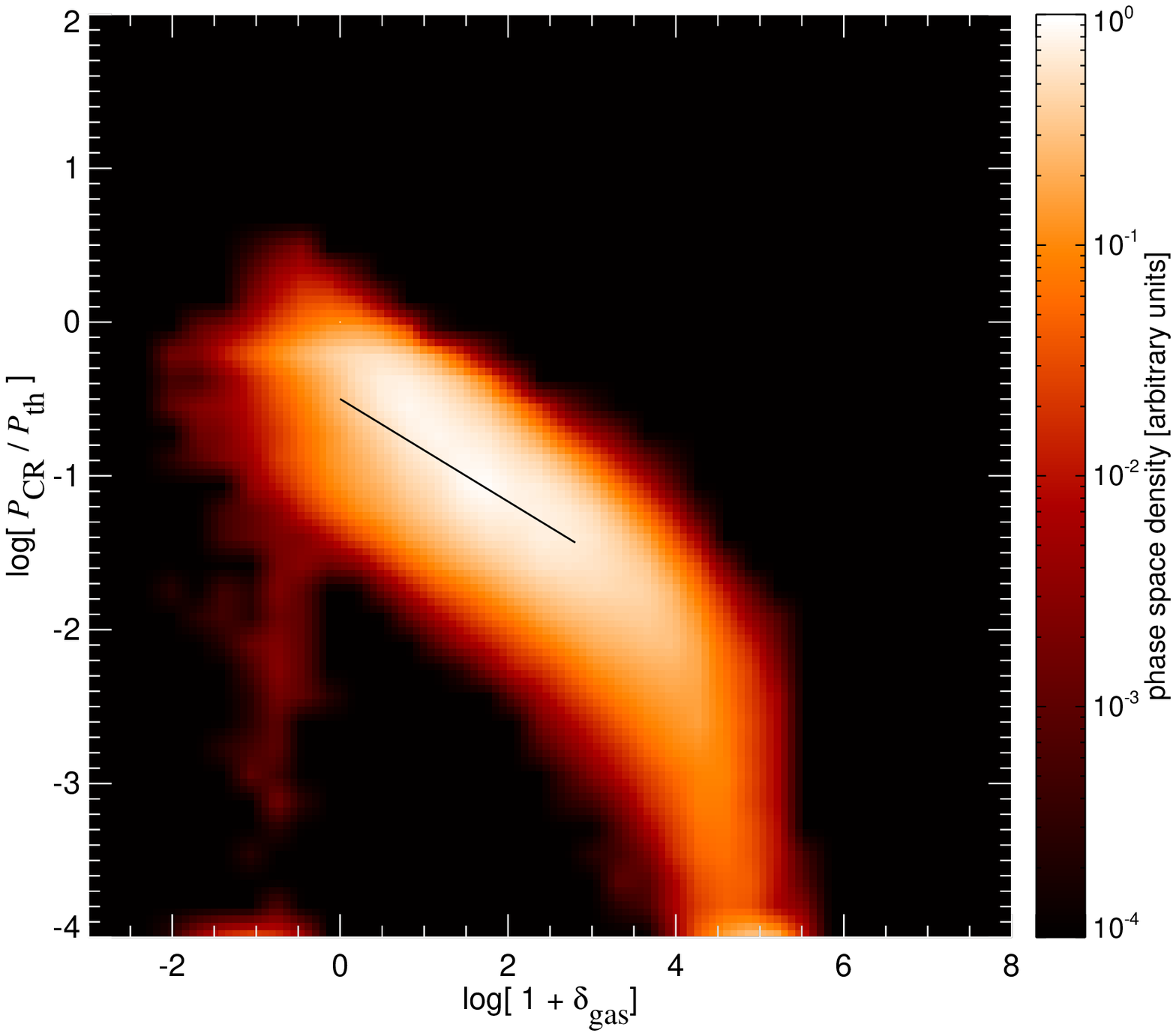}}%
\resizebox{0.5\hsize}{!}{\includegraphics{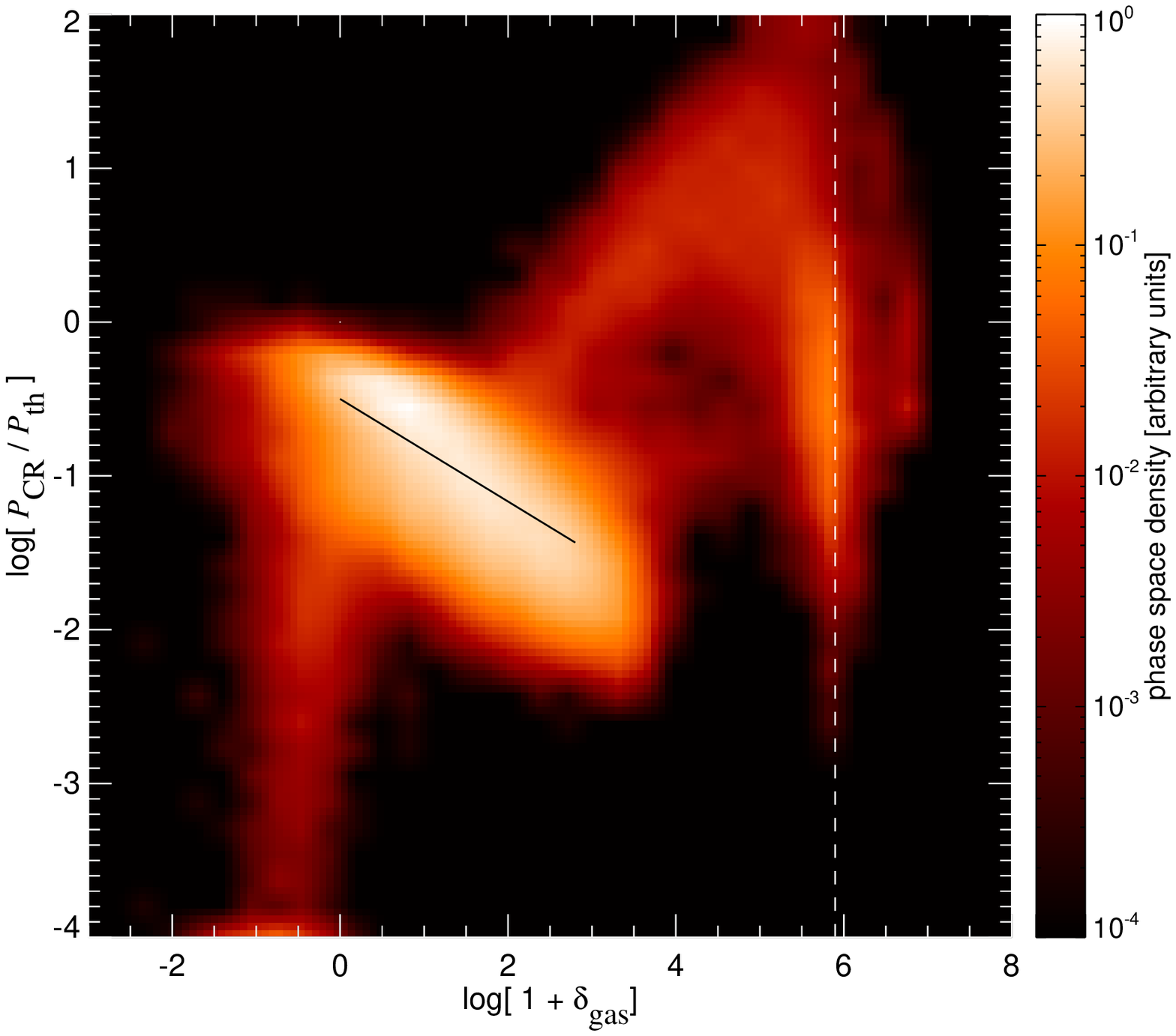}}\\
\end{center}
  \caption{Phase space distribution of the CR pressure relative to the thermal
    pressure at $z = 0$. Left-hand side: radiative simulation, right-hand side:
    non-radiative simulation of our small cluster 10. The black line shows a track
    of constant entropy of a composite of CRs and thermal gas during adiabatic
    compression (using the same normalisation in both panels to guide the
    eye). Dashed white line: star formation threshold of the simulation. Shown
    is the distribution of SPH particles with $r < 10 \, R_\rmn{vir}$ that are
    identified by being in regions not contaminated by boundary dark matter
    particles in the simulation of cluster 10.}
  \label{fig:phase_space}
\end{figure*}

Several important points are apparent. (1) The non-radiative simulation shows a
smooth distribution that is skewed towards lower values of the Mach number.
The mean of the Mach number distribution weighted by the dissipated energy
decreases as cosmic time evolves, i.e.{\ }the average shock becomes weaker at
later times. (2) In contrast, the radiative simulation shows a broad symmetric
distribution in $\log \M$ at early times and develops later on a peak at low
Mach numbers that resembles that in the non-radiative case followed by a
significant disturbed tail towards higher Mach numbers of $\M \gtrsim 10$.  Due
to this tail, the mean of the distribution is not a monotonic function of time any
more and achieves high values of $\bra \M\, \dd \dot{E}_\rmn{diss} /( \dd \log
\M) \ket / \bra \dd \dot{E}_\rmn{diss} /( \dd \log \M) \ket \simeq 100$ already
at $z = 10$.  This reflects the merging history of the clusters forming in that
simulation in combination with the ongoing development of dense cooling
regions that permit the formation of strong shocks during merger events.  (3)
The time integrated Mach number distribution weighted by the dissipated energy,
$\dd E_\rmn{diss}(\M)/( \dd \log \M)$, peaks at Mach numbers $\M\lesssim
2$. Compared to full-size cosmological simulations, the peak position remains
unchanged while the smooth tail in our non-radiative super-cluster simulation
is shifted towards lower values \citep[cf.][ Fig.~6]{2006MNRAS.367..113P}. In
non-radiative simulations, the peak of the Mach number distribution originates
from weak flow shocks internal to dense halo regions, while the smooth tail
towards strong shocks results from external shocks where the diffuse gas breaks
on mildly nonlinear structures. These under-dense regions are under-represented
in our super-cluster simulation and thus we miss a subsample of this shock
population compared to the cosmic mean.  (4) The horizontal tail extending
towards high Mach numbers in radiative simulations can unambiguously be
attributed to internal shocks within cooled high-density regions of
$\delta_\rmn{gas}>10^4$. Part of this result is due to the over-cooling of
these dense regions in our radiative simulation, which yields an overestimate
of the Mach numbers due to the too small sound speeds.  However, the main
contribution in terms of energy dissipation originates from internal shocks
that occur in regions with a pre-shock overdensity $10< \delta_\rmn{gas}
<10^4$. This is because most of the ICM that is being virialised lies simply at
overdensities $\delta_\rmn{gas} < 10^4$.  (5) In both types of simulations,
there is an increasing amount of energy dissipated at shock waves per
logarithmic interval in the scale factor as the universe evolves because the
mean shock speed is significantly growing when the characteristic mass becomes
larger with time.  This trend starts to level off at redshift $z\simeq 1$ when
galaxy clusters are in place.

\begin{figure*}
\begin{center}
  \begin{minipage}[t]{0.495\textwidth}
    \centering{\it \Large Non-radiative simulation:}
  \end{minipage}
  \hfill
  \begin{minipage}[t]{0.495\textwidth}
    \centering{\it \Large Radiative simulation:}
  \end{minipage}
\resizebox{0.5\hsize}{!}{\includegraphics{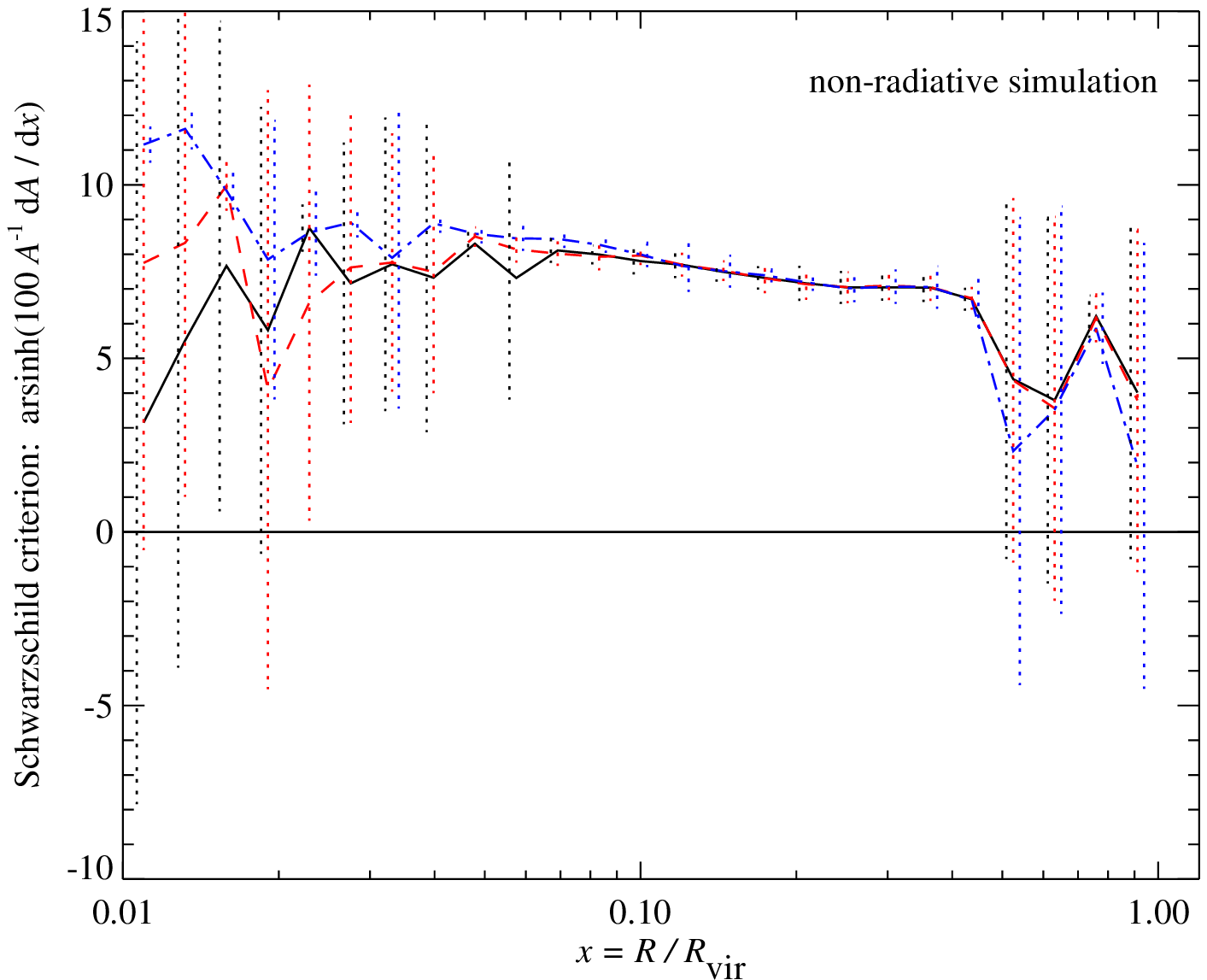}}%
\resizebox{0.5\hsize}{!}{\includegraphics{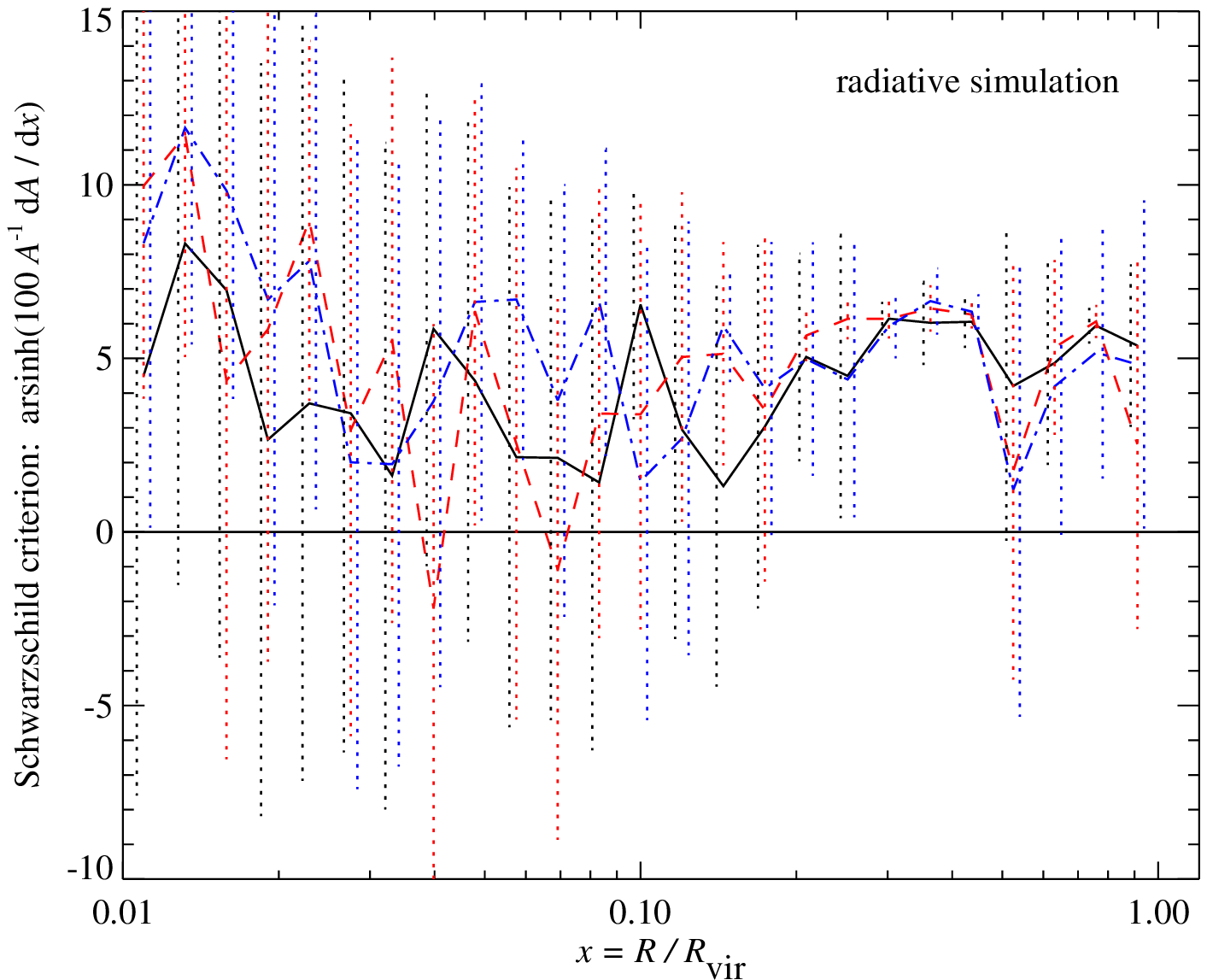}}\\
\end{center}
  \caption{Average profiles of the Schwarzschild criterion for convective
  instability of all clusters. Left-hand side: non-radiative simulations,
  right-hand side: radiative simulations. Positive values of this criterion
  indicate convective stability. Colour scheme: black: reference simulations
  without CRs, red: CR are injected only through shock acceleration using Mach
  number dependent scheme, blue radiative: full model including CRs from
  supernovae, blue non-radiative: CR shock acceleration with constant energy
  injection efficiency.}
  \label{fig:Schwarzschild}
\end{figure*}

Figure~\ref{fig:MachDistrib_small-large} compares the shock strengths of our
super-cluster region (g8) with that of an isolated small cluster region
(g676). While the dissipated energy integrated over the cosmic formation
history in our simulation of the isolated small cluster (g676) amounts to
$E_\rmn{diss}= 2.4\times 10^{62}\mbox{ erg}$, in our large super-cluster
simulation (g8) the amount of dissipated energy was two orders of magnitude
larger, $E_\rmn{diss}= 2.4\times 10^{64}\mbox{ erg}$. Normalising the Mach
number distribution $\dd E_\rmn{diss}(\M)/( \dd \log \M)$ with $E_\rmn{diss}$,
we observe in our radiative as well as in our non-radiative simulation that the
characteristic Mach number which is responsible for the energy dissipation
decreases with cluster mass due to the increasing sound speed in more massive
systems. In other words, weaker shocks are responsible for the energy
dissipation in larger systems.  Apparently, the characteristic shock speed in
these large systems does not increase accordingly to make up for the increase
in sound speed. This confirms the suggestion of \citet{2006...Jubelgas} that
virialisation processes in more massive systems is mediated by shocks with
smaller Mach numbers.  Since strong shocks are more effective in accelerating
CRs, the relative CR pressure level is increased in low mass halos as shown in
Table~\ref{tab:XCR} and Fig.~\ref{fig:XCRtime_evolution}.

\subsubsection{Cosmic ray phase space distributions}

Figure~\ref{fig:phase_space} compares the phase space distribution of the CR
population at $z = 0$ of a radiative simulation to a non-radiative simulation.
Specifically, we present the CR pressure relative to the thermal pressure,
$X_\CR$, versus the overdensity of the gas $\delta_\rmn{gas} + 1=
\rho_\rmn{gas} / (\Omega_\rmn{b}\,\rho_\rmn{cr}\, a^{-3})$. Thus, this figure
shows the variance in $X_\CR$ at a given overdensity $\delta_\rmn{gas}$ in our
simulations of cluster 10.  At low densities, both phase space distributions
resemble each other. Low density regions are characterised by strong shocks
that efficiently accelerate CRs and yield a high value of $X_\CR$. As we move
inwards the cluster, the density increases and the characteristic shock
strength decreases which makes CR acceleration less efficient; an effect that
reduces $X_\CR$. Adiabatic compression of a composite of CRs and thermal gas
disfavours the CR pressure relative to the thermal pressure due to the softer
equation of state of CRs. This is visualised by the black line which shows a
track of constant entropy $X_\CR \propto \delta_\rmn{gas}^{\gamma_\CR -
\gamma_\th}$ of a composite of CRs and thermal gas during adiabatic
compression. Finally, CR loss processes are more important in denser regions, a
third process that reduces $X_\CR$. While the story ends here for the
non-radiative simulations, it goes an important step beyond in radiative
simulations.  There, the thermal gas cools on a small timescale, diminishes
pressure support, condenses out of the gaseous phase and eventually forms stars
at the characteristic density threshold as indicated by the white dashed
line. Meanwhile, the relative CR pressure increases due to the comparatively
long CR cooling time leading to a CR pressure that dominates over the thermal
one at densities $\delta_\rmn{gas}>10^4$. Finally, at the star formation
threshold, we obtain a very broad distribution of $X_\CR$.

\subsubsection{CR buoyancy -- the generalised Schwarzschild criterion}

Figure~\ref{fig:Schwarzschild} shows the generalised Schwarzschild criterion
for convective instability of our sample of all clusters.  The criterion for
convective stability is given by 
\begin{equation}
\label{eq:convective_stability_main}
\frac{\Delta \ln A}{\Delta r} > 0,
\end{equation}
where the effective entropic function is denoted $A = (P_\rmn{th} + P_\CR)\,
\rho^{-\gamma_\eff}$ (cf.~Eqn.~(\ref{eq:convective_stability})). The ICM of our
non-radiative and radiative simulations is overall convectively stable. This is
no surprise since in any convective unstable region, the dynamics of the
instability will quickly restore a stable configuration, thus instabilities are
self-erasing.  The radiative simulations show a larger cluster-to-cluster
variation with respect to individual radial bins, preferably in merging
clusters that are formally convectively unstable. Infalling clumps of matter
distort the spherically averaged profile such that the Schwarzschild criterion
classifies them as convectively unstable.  The simulations including CR physics
do not show significantly different behaviour compared to our reference
simulations. CRs from supernovae shocks that are injected near the cluster
centre could have triggered convective instabilities as proposed by
\citet{2004ApJ...616..169C}, but this is not really seen in our simulations.
The situation might be different and more unstable if one included an
additional central source in the form of CRs that are diffusing out of AGN
bubbles at cluster centres.

\subsection{Thermal cluster observables}
\label{sec:thermal_obs}

\subsubsection{Enhanced X-ray emission}

\begin{figure*}
\begin{center}
  \begin{minipage}[t]{0.495\textwidth}
    \centering{\it \Large Small cool core cluster:}
  \end{minipage}
  \hfill
  \begin{minipage}[t]{0.495\textwidth}
    \centering{\it \Large Large cool core cluster:}
  \end{minipage}
\resizebox{0.5\hsize}{!}{\includegraphics{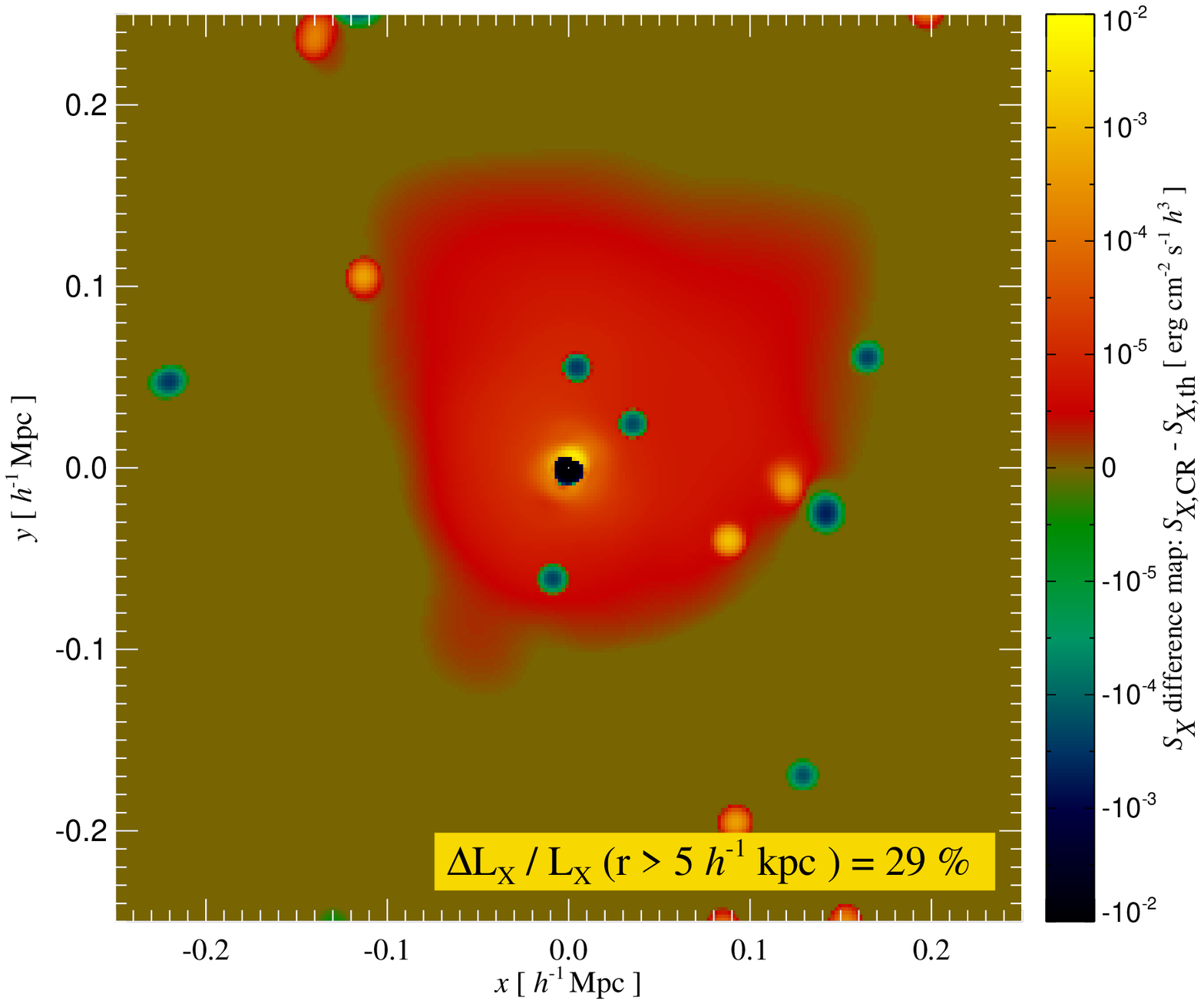}}%
\resizebox{0.5\hsize}{!}{\includegraphics{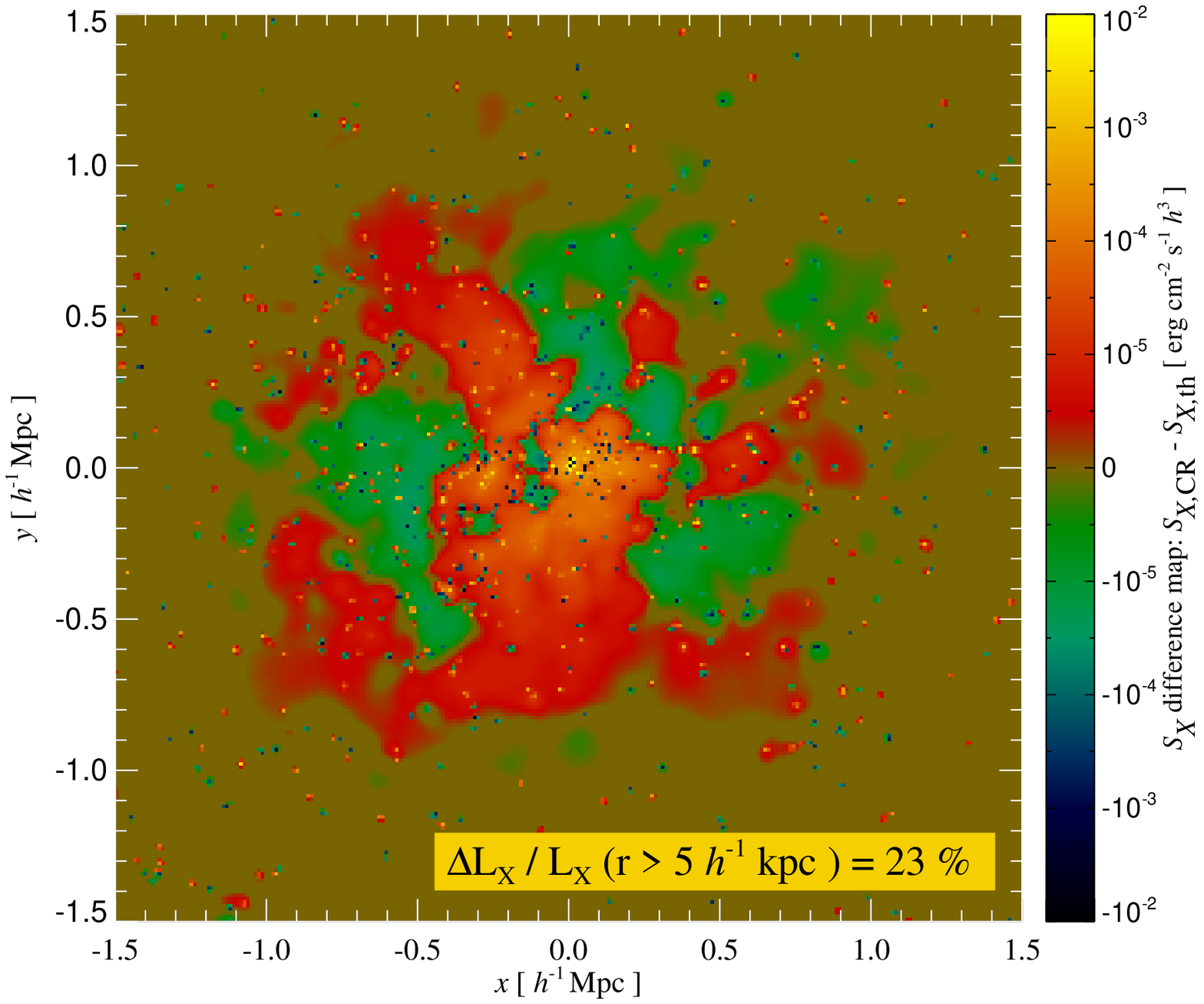}}\\
\resizebox{0.5\hsize}{!}{\includegraphics{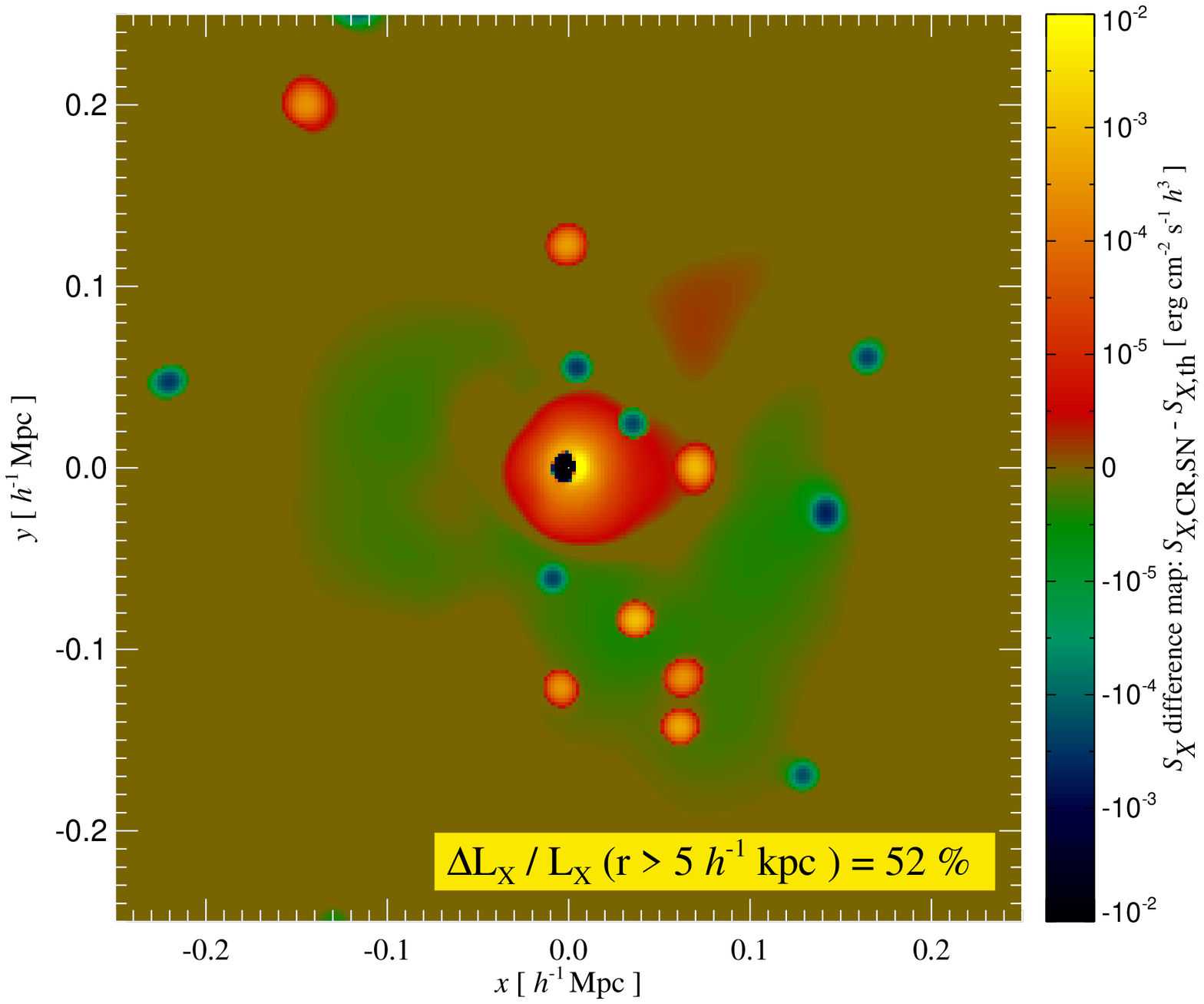}}%
\resizebox{0.5\hsize}{!}{\includegraphics{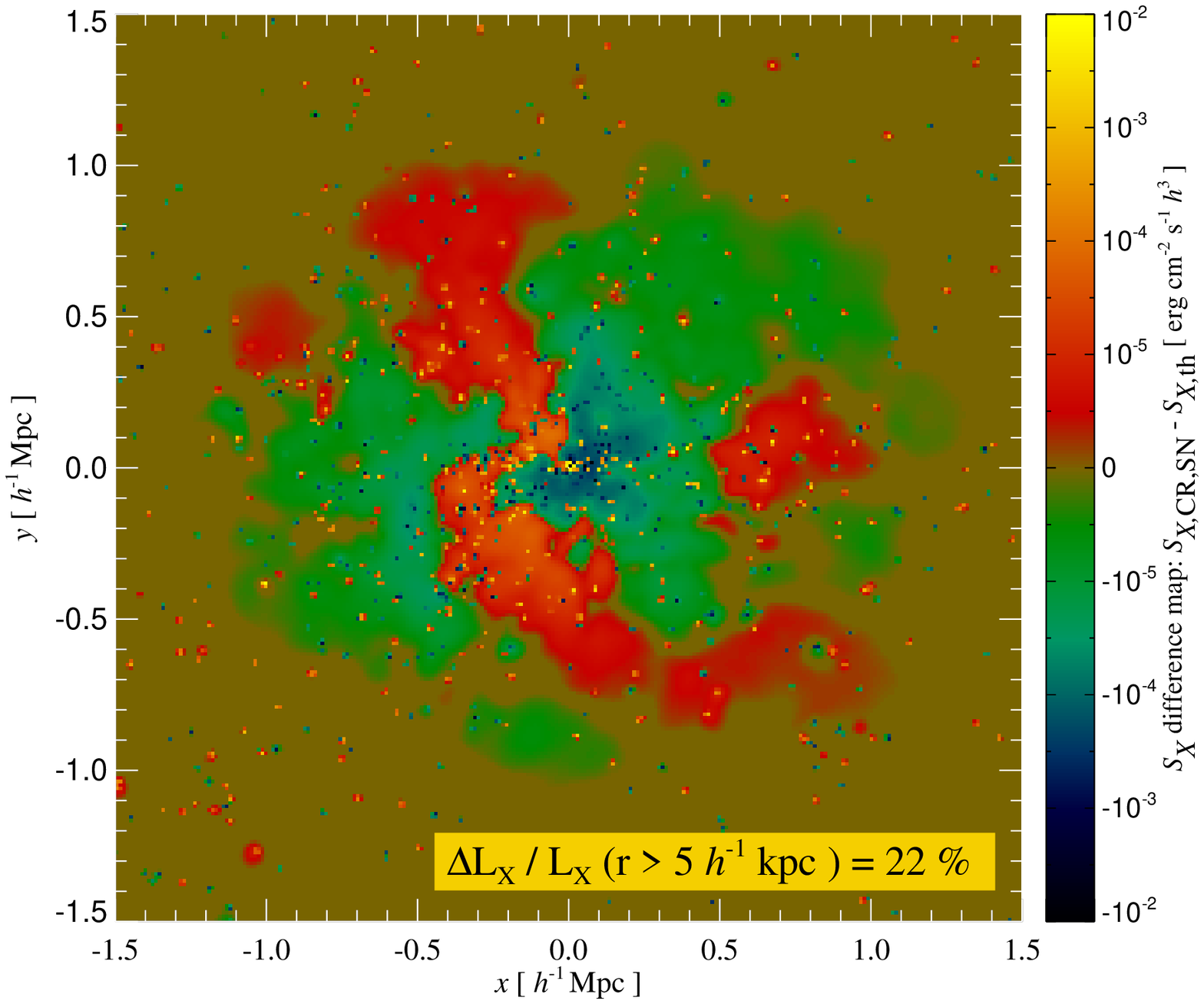}}\\
\end{center}
  \caption{Difference of the X-ray surface brightness in a radiative simulation
  with CRs and the corresponding reference simulation without CRs.  Left-hand
  side: small cool core cluster with a virial mass of $8.8\times
  10^{13}\,h^{-1}\,$Mpc at $z=0$, right-hand side: large cool core cluster with a virial
  mass of $1.8\times 10^{15}\,h^{-1}\,$Mpc. Upper panels: simulations only
  account for CR acceleration at structure formation shocks, lower panels: CRs
  from supernovae are also considered. The relative difference of the
  integrated X-ray luminosity (excluding the central over-cooling region of $r <
  5\,h^{-1}\,$kpc) is given in the inlay.}
  \label{fig:SXdiff}
\end{figure*}

Our map of the bolometric X-ray surface brightness in the bremsstrahlung regime
was produced by means of Eqn.~(\ref{eq:projection}), projecting the quantity
\begin{eqnarray}
  \label{eq:X-ray_projection}
  a_\alpha &=& \Lambda_0\, \left(\frac{k T_\alpha}{\mbox{keV}}\right)^{1/2}\,
  \left(\frac{\rho_\alpha}{\mu\,m_\p}\right)^2, \\
  \label{eq:Lambda0}
  \Lambda_0 &=& \left(\frac{2\pi\,\mbox{keV}}{3\, m_\e}\right)^{1/2}
  \frac{2^5\,\pi\,e^6}{3\, h\, m_\e c^3}\,
  \bar{g}_\rmn{B}(T_\alpha) \,\mu^2 \frac{1+X_\rmn{H}}{2} \\
  &=& 1.78 \times 10^{-24} \mbox{ erg cm}^{-3}\mbox{ s}^{-1}.
\end{eqnarray}
Here, $T_\alpha$ denotes the temperature of the SPH particle (labelled with
$\alpha$), $h$ denotes the Planck constant, $e$ the elementary charge,
$X_\rmn{H}=0.76$ is the primordial hydrogen mass fraction, $\mu = 4 /(5
X_\rmn{H} + 3) = 0.588$ is the mean molecular weight of fully ionised gas, and
$\bar{g}_\mathrm{B}\simeq 1.2$ is the frequency- and velocity-averaged Gaunt
factor \citep{1979rpa..book.....R, 1978ppim.book.....S}.  To be self-consistent
within the bremsstrahlung assumption, we assume a fully ionised gas of
primordial element composition which consists of hydrogen and helium only.

\begin{table*}
\caption{\scshape: CR induced relative differences of the X-ray luminosity and the
  integrated Compton-$y$ parameter in a representative cross section of our
  cluster sample.} 
\begin{tabular}{l l c c c c c c c c c}
\hline
\hline
Cl. & sim.'s & dyn.~state$^{(1)}$ & $T_{200}^{(1)}$ 
& $\Delta L_X/L_X^{(2)}$ & $\Delta L_X/L_X^{(2)}$ & $\Delta Y/Y$ & $\Delta Y/Y$ 
& $\Delta y_0/y_0$ & $\Delta y_0/y_0$ & $y_0$ \\
& & & [keV] & shock-CRs$^{(3)}$  & compl. CRs$^{(3)}$
            & shock-CRs$^{(3)}$  & compl. CRs$^{(3)}$
            & shock-CRs$^{(3)}$  & compl. CRs$^{(3)}$ & [$10^{-5}$]\\
\hline
1  & g8a  & CC    & 13.1 & $+23\%$ & $+22\%$ & $+0.1\%$ & $-0.6\%$ & $+~8\%$ & $\pm~0\%$ & ~26.7\\
2  & g1a  & CC    & 10.6 & $+18\%$ & $+14\%$ & $+1.1\%$ & $-0.1\%$ & $+~8\%$ & $-~4\%$   & ~15.4\\
                                                               	      		      		       		       
3  & g72a & PostM & 9.4  & $+~9\%$ & $+15\%$ & $-1.6\%$ & $-0.1\%$ & $-~7\%$ & $-~7\%$   & ~14.6\\
5  & g1b  & M     & 4.7  & $+16\%$ & $+69\%$ & $-1.5\%$ & $+1.1\%$ & $+~7\%$ & $-~4\%$   & ~2.73\\
9  & g1d  & M     & 1.7  & $-10\%$ & $+~8\%$ & $-4.5\%$ & $-2.2\%$ & $-12\%$ & $-~1\%$   & 0.475\\
                                                               	      		      		       		       
10 & g676 & CC    & 1.7  & $+29\%$ & $+52\%$ & $-0.8\%$ & $-4.5\%$ & $+23\%$ & $+41\%$   & 0.960\\
11 & g914 & CC    & 1.6  & $+36\%$ & $+23\%$ & $-0.1\%$ & $-3.7\%$ & $+25\%$ & $+~5\%$   & ~1.01\\
\hline
\end{tabular}   
\begin{quote} 
  {\scshape Notes:}\\
  (1) The definitions for the dynamical state of the cluster and the virial
  temperature is given in Table~\ref{tab:sample}. \\
  (2) The relative difference of the integrated X-ray luminosity has been
  computed without taking into account the central over-cooling region of $r <
  5\,h^{-1}\,$kpc. Positive values for the relative difference indicate an
  enhancement of the thermal cluster observable when taking CRs into
  account.\\ 
  (3) The shock-CR model is represented by our radiative simulations that account
  only for structure formation CRs while the complete model is based on our
  radiative simulations that take additionally supernova CRs into account.\\
\end{quote}
\label{tab:CReffect}
\end{table*} 

Figure~\ref{fig:SXdiff} shows the difference of the X-ray surface brightness
in a radiative simulation with CRs and the corresponding reference simulation
without CRs.  The small blue and yellow circles represent individual galaxies
in each of the simulations whose orbits did not match up when one includes
different physical CR processes such that small changes in the energy budget
modify the galactic orbits, accumulate and eventually lead to 
macroscopic separation of equivalent galaxies in the final spatial
distribution.  The left-hand side shows a small cool core cluster with a
virial mass today of $8.8\times 10^{13}\,h^{-1}\,$Mpc (cluster~10) and the
right-hand side a large cool core cluster with a virial mass today of
$1.8\times 10^{15}\,h^{-1}\,$Mpc (cluster~1). While the upper panels show the
simulation that only accounts for CR acceleration at structure formation
shocks, the simulations displayed in the lower panels take also CRs from
supernovae into account. The relative difference of the integrated X-ray
luminosity (without the central over-cooling region of $r < 5\,h^{-1}\,$kpc)
is given in the inlay, appropriately colour-coded such that the integrated
CR effect on the X-ray luminosity can be easily read off from the colour bar. All
these simulations show an enhancement of the central and integrated X-ray
luminosity in our CR models compared to our reference models without CRs. This
effect derives from the softer effective adiabatic index in the CR simulations
due to the large ratio of the CR-to-thermal cooling time in the cluster centre,
leading to a more compressible gas with an associated density increase that is
required to balance the gravitational pressure of the dark matter. The
resulting enhancement of the X-ray surface brightness is more pronounced in
small CC clusters due to the higher level of relative CR pressure in
combination with a relaxed spherically symmetric structure.

Our previous considerations are confirmed by analysing a
representative subsample of our cluster simulations with respect to
mass and dynamical state in Table~\ref{tab:CReffect}. Positive values
for the relative difference indicate an enhancement of the thermal
cluster observable when taking CRs into account. While the differences
of the cool core cluster simulations yield a consistent picture in our
runs which account for CRs from structure formation shocks only, the
sign and the magnitude of the CR-induced effects in merging systems do
not follow a uniform trend due to the merger induced inhomogeneities
in the ICM. In our complete CR model, the cluster-to-cluster variation
of the CR enhancement of equally sized clusters is larger due the CR
injection by supernovae, which traces the current star formation rate
rather than being an integrated measure over the past cluster/core
activity like in our CR-shock runs.

We can get a quantitative idea of the approximate size of this CR effect by
employing a toy model of cluster formation in the presence of a CR component
which we compare to a model without CRs. Consider at some early stage around
$z \sim 2-3$ a composite of thermal gas and CRs that has been shock heated for
the first time while it was collapsing into mildly non-linear structures.
Today's analogue for such a state of the gas might be given by the WHIM which
is characterised by a comparatively high relative CR pressure contribution of
roughly $X_\CR \sim 0.3-0.5$. Let us assume that the initial thermal pressure in
our reference model without CRs is equivalent to the initial total pressure in
our composite model with CRs, $P_{\th,0}^* = P_{\th,0} + P_{\CR,0} = P_{\th,0}\, (1 +
X_{\CR,0})$, where our reference quantities are denoted by a star.  Neglecting
radiative cooling and CR diffusion for simplicity, further accretion of new
shells of mass onto the dark matter halo exerts a pressure that adiabatically
compresses the gas internal to the halo, yielding
\begin{equation}
\label{eq:adiabatic_comp}
P_\th = P_{\th,0}\, C^{\gamma_\th}
\qquad\mbox{and}\qquad
P_\CR = P_{\CR,0}\, C^{\gamma_\CR}.
\end{equation}
While the gas in our reference model is compressed by the factor $C^*$, the
compression factor in the CR model is given by $C$. Requiring that the final
pressure after adiabatic compression in both models is equivalent to each
other, we obtain the condition
\begin{equation}
\label{eq:compression_model}
\left(\frac{C}{C^*}\right)^{\gamma_\th} + 
X_{\CR,0}\, \left(\frac{C}{C^*}\right)^{\gamma_\CR}\,
C^{*\,\gamma_\CR - \gamma_\th} - (1 + X_{\CR,0}) = 0.
\end{equation}
This equation can be solved for the combination $C / C^*$. Assuming an initial
relative CR pressure of $X_{\CR,0} \simeq 0.3$, a compression factor in the
reference model of $C^* \simeq 10^3$, and equivalent initial densities in both
models, we obtain a density enhancement for our toy cluster in our CR model of
$\Delta \rho / \rho^* \simeq 15$ per cent relative to the reference model. The associated
X-ray emission would be enhanced in the presence of CRs by a factor of $\Delta L_X /
L_X^* \simeq 32$ per cent compared to the model without CRs. This result should be
taken with a grain of salt since we did not consider shock heating, radiative
cooling, nor any CR transport and loss processes. It merely illustrates the
important effect of the softer equation of state of CRs.

\subsubsection{Modified Sunyaev-Zel'dovich effect}

\begin{figure*}
\begin{center}
  \begin{minipage}[t]{0.495\textwidth}
    \centering{\it \Large Small cool core cluster:}
  \end{minipage}
  \hfill
  \begin{minipage}[t]{0.495\textwidth}
    \centering{\it \Large Large cool core cluster:}
  \end{minipage}
\resizebox{0.5\hsize}{!}{\includegraphics{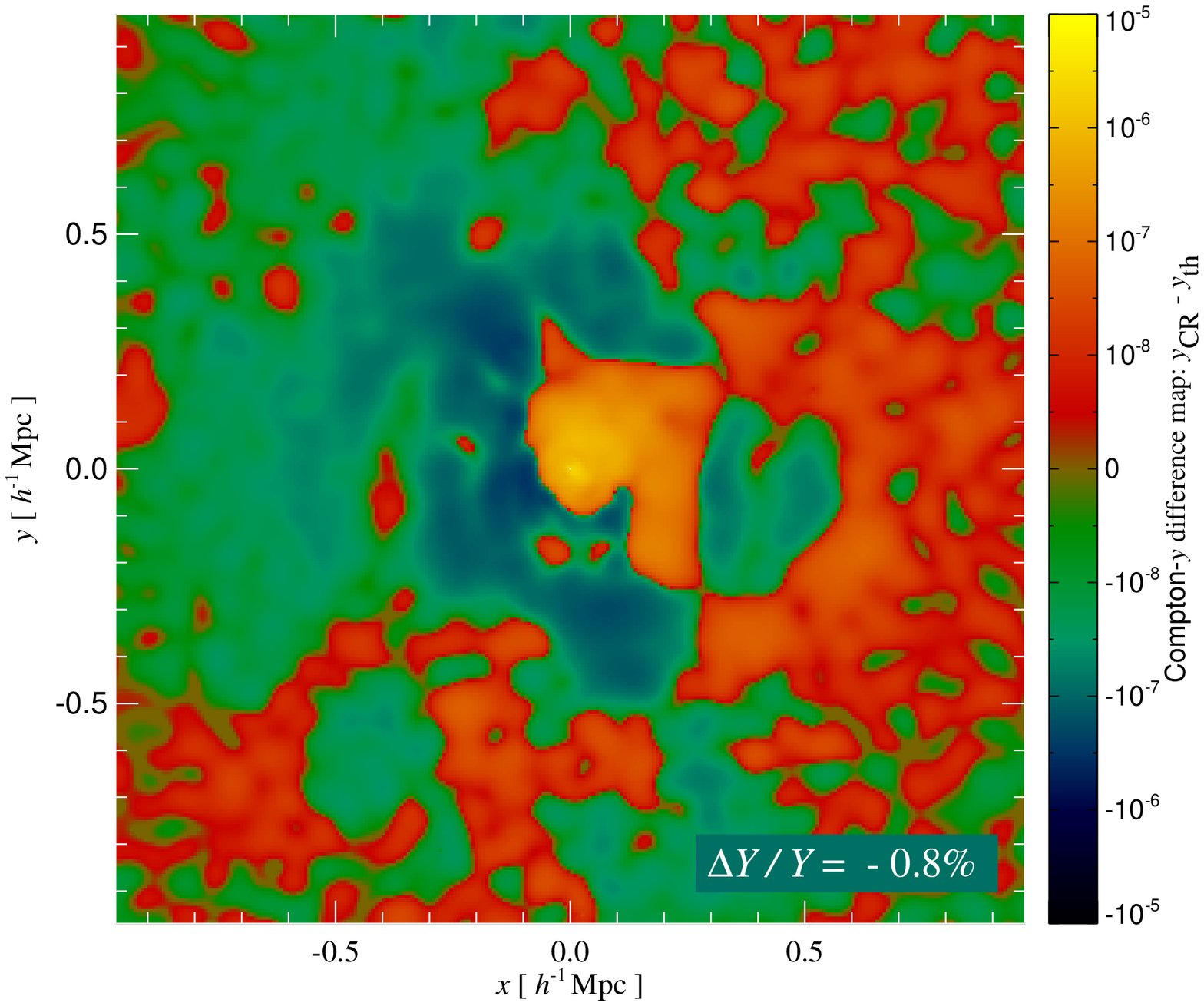}}%
\resizebox{0.5\hsize}{!}{\includegraphics{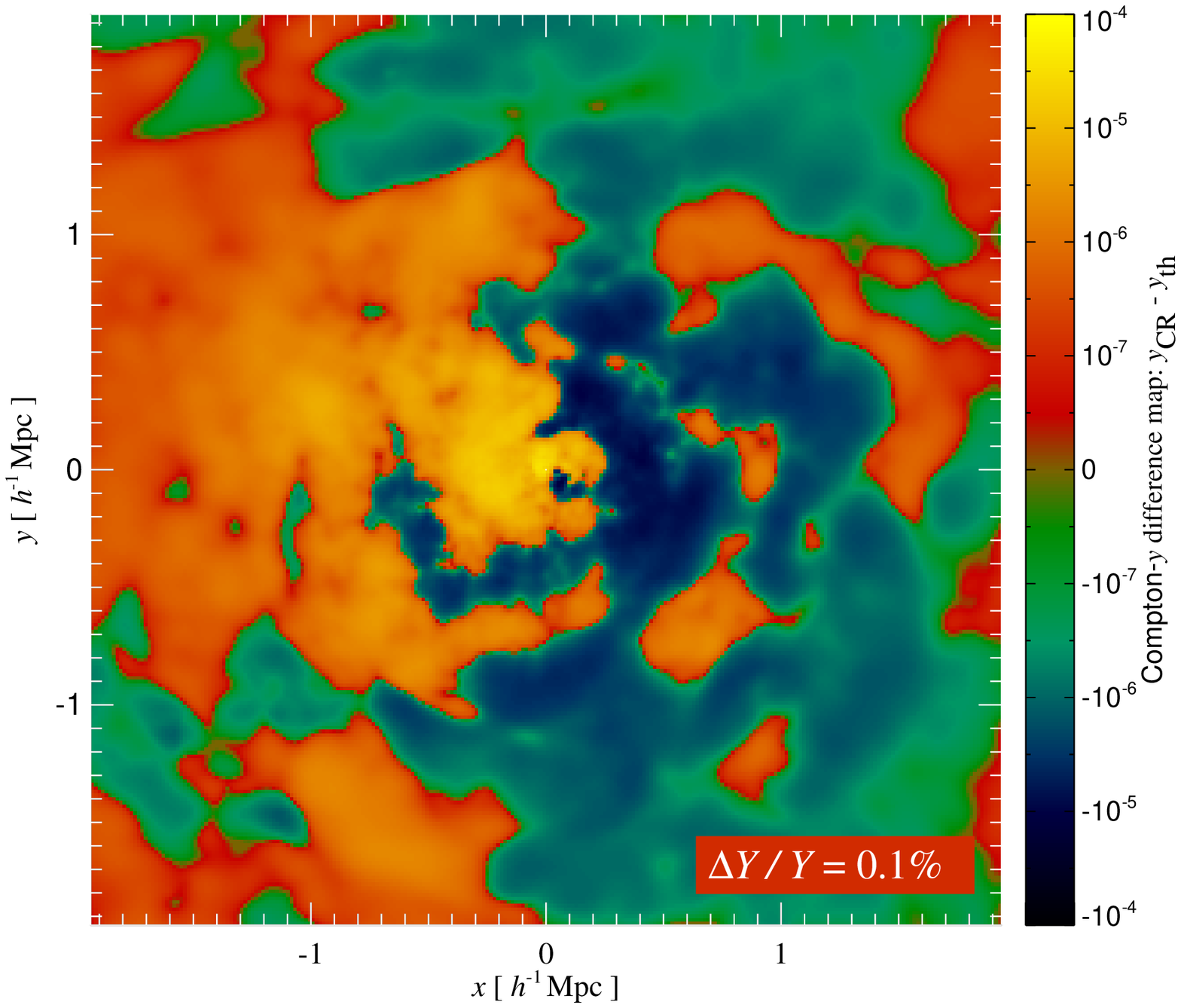}}\\
\resizebox{0.5\hsize}{!}{\includegraphics{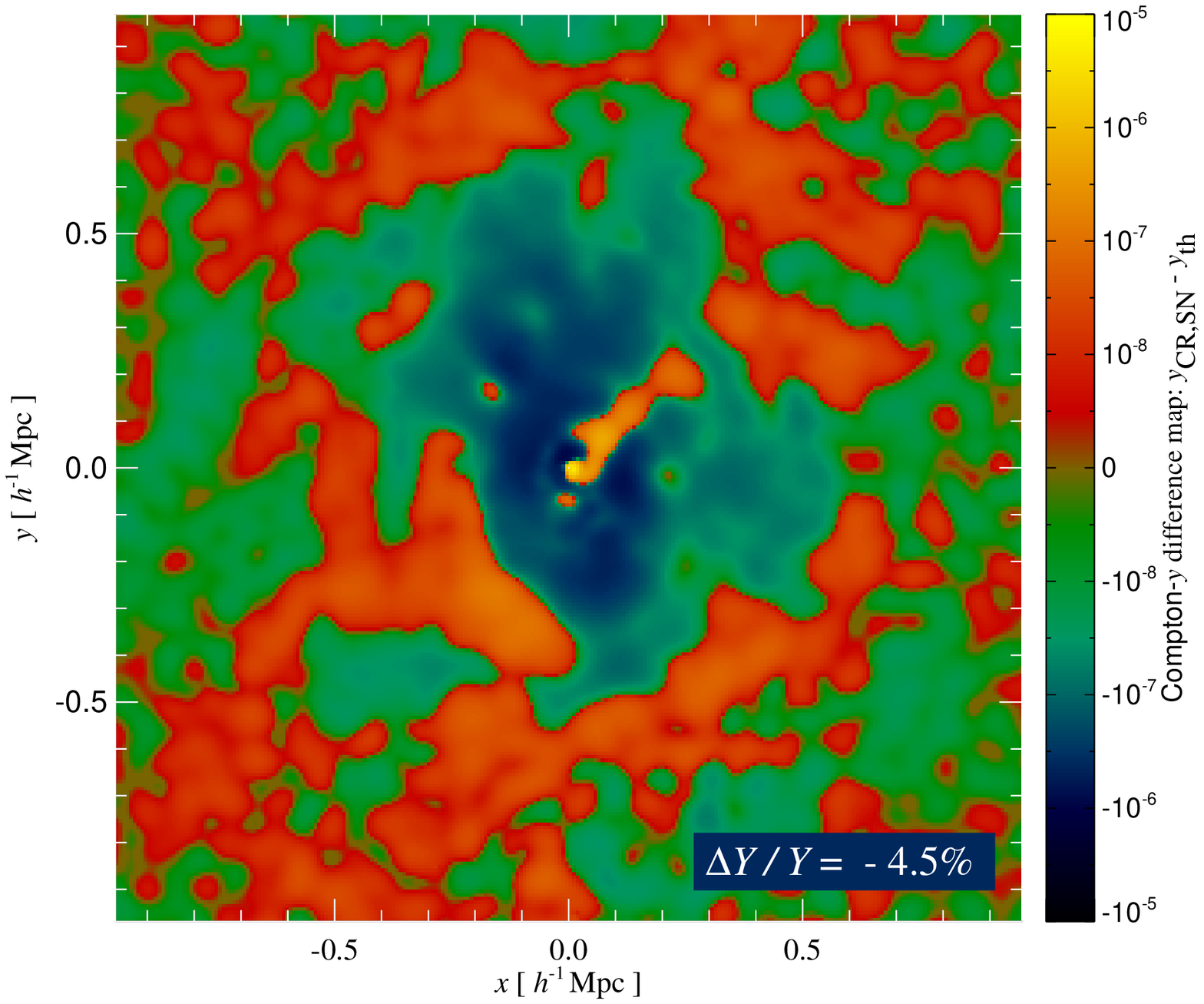}}%
\resizebox{0.5\hsize}{!}{\includegraphics{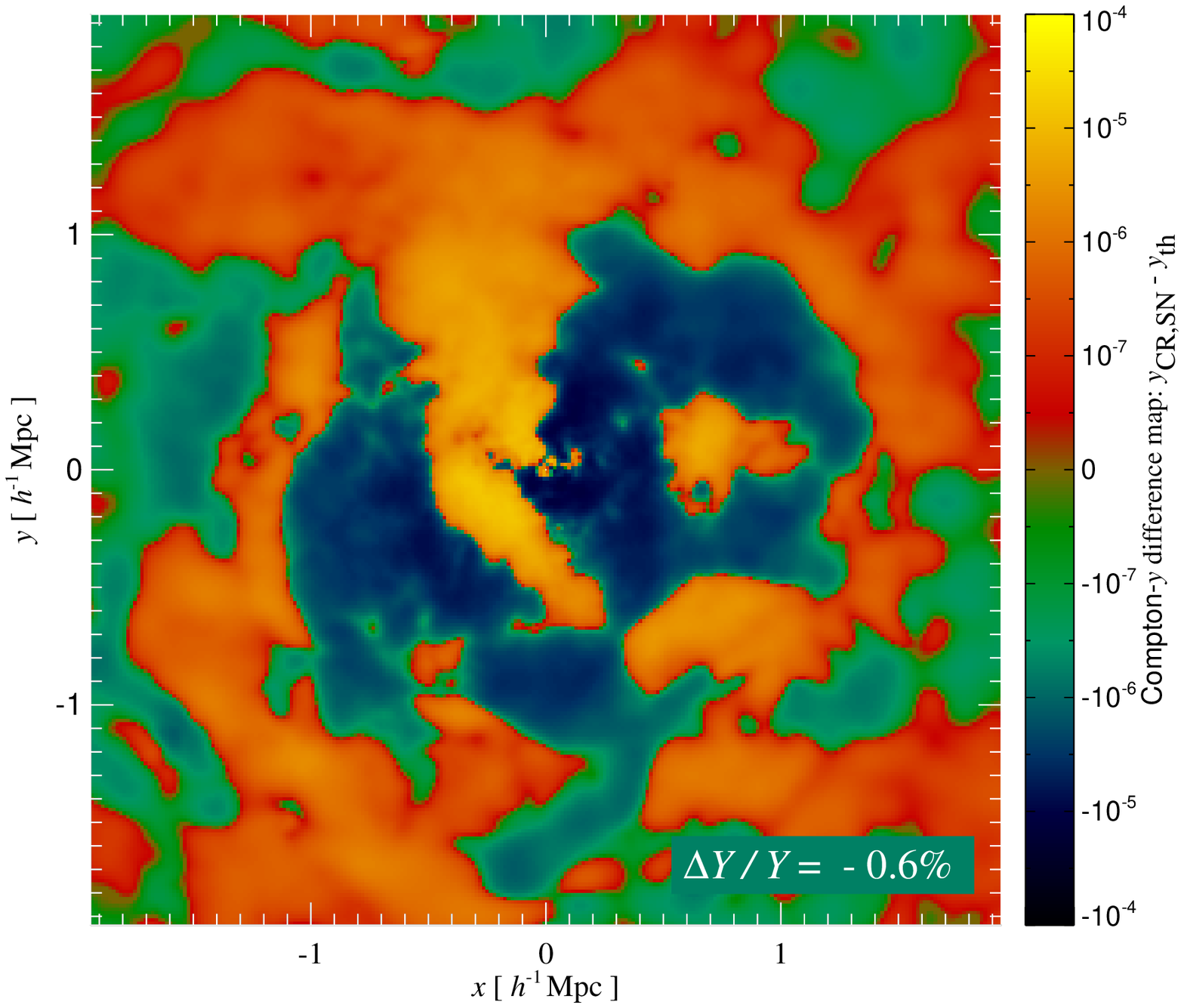}}\\
\end{center}
  \caption{Difference of the Compton-$y$ parameter in a radiative simulation
  with CRs and the corresponding reference simulation without CRs.  Left-hand
  side: small cool core cluster with a virial mass of $8.8\times
  10^{13}\,h^{-1}\,$Mpc at $z=0$, right-hand side: large cool core cluster with a virial
  mass of $1.8\times 10^{15}\,h^{-1}\,$Mpc. Upper panels: simulations only
  account for CR acceleration at structure formation shocks, lower panels:
  additionally, CRs are also injected by supernova shocks. The relative
  difference of the integrated Compton-$y$ parameter is given in the inlay.}
  \label{fig:ydiff}
\end{figure*}

The amplitude of the thermal Sunyaev-Zel'dovich effect is given by the
Compton-$y$ parameter and is proportional to the line-of-sight integral of the
thermal pressure which is obtained by replacing $\alpha$ in
Eqn.~(\ref{eq:projection}) by
\begin{equation}
  \label{eq:y_projection}
  a_\alpha = \frac{\sigma_\rmn{T}}{m_\e\, c^2}\,
  (\gamma_\th-1)\, X_\rmn{H}\, x_{\e,\alpha}\, \mu\, u_\alpha\, \rho_\alpha,
\end{equation}
where $X_\rmn{H}=0.76$ is the primordial hydrogen mass fraction,
$x_{\e,\alpha}$ is the ratio of electron and hydrogen number densities of the
SPH particle (labelled with $\alpha$) which we dynamically track in our
radiative simulations, $\mu = 4 /(3 X_\rmn{H} + 1 +
4\,X_\rmn{H}\,x_{\e,\alpha})$ denotes the mean molecular weight of partly
ionised gas, $u_\alpha$ is the internal energy per unit mass, and $\rho_\alpha$
is the physical mass density of the SPH particle.

Figure~\ref{fig:ydiff} shows the difference between the Compton-$y$ parameter
in a radiative simulation with CRs and the corresponding reference simulation
without CRs.  The left-hand side shows a small cool core cluster with a virial
mass today of $8.8\times 10^{13}\,h^{-1}\,$Mpc (cluster 10) and the right-hand
side displays a large cool core cluster with a virial mass today of $1.8\times
10^{15}\,h^{-1}\,$Mpc (cluster 1). While the upper panels show the simulations
that only account for CR acceleration at structure formation shocks, the
simulations displayed in the lower panels take also CRs from supernovae into
account. The relative difference of the integrated Compton-$y$ parameter is
given in the inlay, appropriately colour-coded such that the CR effect on the
integrated Compton-$y$ parameter can easily be read off from the colour bar.
In the small cluster simulation, CR feedback leads to an enhancement of the
central Compton-$y$ parameter followed by a ring of a (much weaker)
decrement. The more massive system shows a more asymmetric difference map with
a larger contrast at the centre that is roughly one order of magnitude larger
compared to the smaller system, as expected for the larger virial mass. As
analysed in Sect.~\ref{sec:rad_profiles}, the high fraction of pressure support
provided by CRs in the central regions results in a higher compressibility of
the composite fluid, which in turn requires a larger overdensity in order to
balance the gravitational pressure from the dark matter and gas. This increases
the central density of the cluster and leads to a pressure enhancement,
provided the hydrostatic equation is applicable, $\dd P /(\dd r) =
\rho_\rmn{gas}\, G\, M(<r)\, r^{-2}$. Hence, the Compton-$y$ parameter in our
CR simulations shows a profile that is more peaked compared to our reference
case. The increased central Compton-$y$ parameter comprises a larger area in
our CR-shock model compared to the complete CR model. This is due to the
additional CR injection from SN in our complete CR model that reduces the
amount of thermal SN feedback and thus reduces the Compton-$y$ parameter.

Table~\ref{tab:CReffect} confirms this finding in a representative cross
section of our clusters by showing the central Compton-$y$ parameter $y_0$ as
well as the relative difference of $y_0$.  In practice, this quantity will be
smoothed by the experimental beam which we did not try to model here.  The behaviour
of $\Delta y_0 / y_0$ is much more regular in our shock-CR model since the
CR effects are integrated over a longer timescale. In contrast, there is a
larger variation of $\Delta y_0 / y_0$ in our complete CR model, even among
nearly similarly evolving clusters (cluster 10 and cluster 11), due to the
central CR injection by supernovae which reflects the current central star
formation rate.

The total Compton-$y$ parameter $Y$ of a galaxy cluster  is given by the
Compton-$y$ parameter integrated over the solid angle element subtended by the
cluster,
\begin{equation}
  \label{eq:Y}
  Y = D_\rmn{ang}^2\,\int_{\Omega_\rmn{cluster}}\!\!\!\dd \Omega\, y(\btheta) = 
  \int_{A_\rmn{cluster}}\!\!\!\dd^2\! x\,\, y(\vecbf{x}_\bot) = 
  \frac{\sigma_\rmn{T}}{\gamma_\th - 1}\,\frac{E_\th}{m_\e c^2},
\end{equation}
where $D_\rmn{ang}$ is the angular diameter distance to the cluster.
Interestingly, the total Compton-$y$ parameter seems to be robust with respect
to CR feedback because it reflects the cluster's total thermal energy $E_\th$.
We confirm this finding by looking at a representative cross section of
clusters in our sample in Table~\ref{tab:CReffect}. In large CC clusters, CR
feedback has a negligible effect on the total Compton-$y$ while in merging
clusters, the total Compton-$y$ parameter can either be slightly enhanced or
decreased, depending on the merger virialisation progress of the induced shock
waves leading to temporary inhomogeneities in the ICM. In small CC clusters,
the value of $Y$ is only slightly decreased in our CR-shock model while this
effect is more pronounced in our complete CR model where $Y$ is systematically
decreased by 4 per cent due to the additional CR injection from SN that reduces the
Compton-$y$ parameter.


\section{Summary}

We performed high-resolution simulations of a sample of 14 galaxy clusters that
span a mass range from $5\times 10^{13} h^{-1}\, \rmn{M}_\odot$ to $2\times 10^{15}
h^{-1}\, \rmn{M}_\odot$ to study the effects of cosmic rays (CRs) on thermal cluster
observables such as X-ray emission and the Sunyaev-Zel'dovich effect.  In a
systematic study, we analyse the CR effects on the intra-cluster medium while
simultaneously taking into account the cluster's dynamical state as well as the
mass of the cluster.  The modelling of the cosmic ray physics includes adiabatic
CR transport processes such as compression and rarefaction, and a number of
physical source and sink terms which modify the cosmic ray pressure of each
particle. The most important sources considered are injection by supernovae (in
our radiative simulations) and diffusive shock acceleration at cosmological
structure formation shocks, while the primary sinks are thermalization by
Coulomb interactions, and catastrophic losses by hadronic interactions.  While
the relative pressure contained in CRs within the virial radius is of the order
of 2 per cent averaged over our cluster sample in non-radiative simulations, their
average contribution rises to 32 per cent in our simulations with dissipative gas
physics including radiative cooling, star formation, and supernova feedback.

Our main findings for the CR pressure relative to the
thermal pressure $X_\CR = P_\CR / P_\th$ can be summarised as follows. The CR
distribution in the dilute outskirts of cluster is dominated by the effective
CR acceleration at strong structure formation shocks such as accretion and
merger shocks.  The relative CR pressure $X_\CR$ decreases as we move inwards
to smaller cluster radii that are characterised by an increase of the sound
speed and the density of the intracluster medium (ICM): (1) weak central flow
shocks are inefficient in accelerating CRs, (2) adiabatic compression of a
composite of CRs and thermal gas disfavours the CR pressure relative to the
thermal pressure due to the softer equation of state of CRs, and (3) CR loss
processes are more important at the dense centres.

Interestingly, if we include radiative losses of the thermal gas, the
relative CR pressure increases strongly towards the cluster centre and
inside dense galactic substructures within the ICM, because of the
long CR cooling times compared to those of the thermal gas in these
cooling environments. The thermal gas cools much faster radiatively
and diminishes its pressure support while it condenses out and forms
stars that do not participate in the pressure balance.  In contrast,
at a fixed density, the CR cooling time of an aged CR population
remains almost constant as the thermal gas is cooling.  High fraction
of pressure support provided by CRs yields a higher compressibility of
the composite fluid that requires a larger overdensity in order to
balance the gravitational pressure from the dark matter and gas. This
increases the density of each galactic substructure as well as in the
cluster centre where it leads to a pressure enhancement, provided the
hydrostatic equation is applicable as it is the case in cool core
clusters. This higher density leads to a higher rate of star formation
and thus increases the central baryon fraction strongly because gas
from larger scales has to replenish the condensed gas in order to
maintain hydrostatic equilibrium. We note that in real galaxy clusters
this effect is expected to be somewhat weaker, because here strong
cooling flows are suppressed by some non-gravitational heating
source. Nevertheless, this shows that CR populations injected from
supernovae shocks on galactic scales and by structure formation shocks
reinforce the cooling flow problem in contemporary simulations rather
than solving it.  CRs from supernovae shocks that are injected near
the cluster centre could have in principle triggered convective
instabilities in the lighter relativistic CR component, but this
effect is not really seen in our simulations.

The dynamical state of the cluster crucially determines the relative CR
pressure $X_\CR$: (1) In the case of ongoing merger activity, the relative CR
pressure is boosted due to strong merger shock waves that effectively inject
CRs and mix the highly CR-enriched intergalactic medium outside clusters with
the ICM. (2) In a post-merger state, weak virialisation shocks traversing the
cluster after the merger thermalize random gas motions, thereby increasing the
thermal energy and yielding a decrease of $X_\CR$. CR loss processes and adiabatic
compression furthermore decrease $X_\CR$ in intermediate cluster regions with
a comparatively long thermal cooling time scale.  (3) Small cool core clusters
are characterised by a higher value of $X_\CR$ than more massive clusters,
albeit the integrated $X_\CR$ is significantly reduced compared to our merging
systems. This is mainly due to the nature of the virialisation process where
weaker shocks are responsible for the energy dissipation in larger
systems. This leads to a more efficient CR energy injection in small systems
and hence to a smaller value of $X_\CR$. Altogether, we observe a strong
correlation between merging activity and the level of $X_\CR$.

How does a CR component affect thermal cluster observables? We found out that
central cluster regions of low-mass clusters are most strongly affected by the
CR component which leads to an enhanced central density and thermal
pressure. Since the X-ray emission scales as the square of the gas density, the
X-ray luminosity in low mass clusters of our CR simulations is boosted by up to
$\Delta L_X / L_X \simeq 40$ per cent. The amplitude of the thermal
Sunyaev-Zel'dovich effect is also modified by CR feedback. While the central
Compton-$y$ parameter in these systems is increased by $\Delta y_0 / y_0 \simeq
25$ per cent, the integrated Sunyaev-Zel'dovich effect appears to be remarkably
robust and the total flux decrement only slightly reduced by typically 2 per
cent.  In merging clusters, the sign and the magnitude of the CR-induced
effects are quite irregular due to the merger induced inhomogeneities in the
ICM.

\section*{Acknowledgments}
It is a pleasure to thank Dick Bond, Subhabrata Majumdar, Jon Sievers, Yoram
Lithwick, and Stefano Borgani for stimulating discussions and an anonymous
referee for carefully reading the manuscript.  All computations were performed
on CITA's McKenzie cluster \citep{2003...McKenzie} which was funded by the
Canada Foundation for Innovation and the Ontario Innovation Trust.

\bibliography{bibtex/gadget} 
\bibliographystyle{mn2e}

\appendix
\section{Schwarzschild criterion for convective stability}
\label{app:Schwarzschild}

This section follows the derivation by \citet{2004ApJ...616..169C} and derives a
dimensionless criterion for convective stability in the presence of a composite
fluid of thermal gas and CRs. Consider a parcel of thermal gas and CRs
initially at a distance $r = r_0$ from the cluster centre. The parcel is
displaced by a radial distance $\Delta r$ to $r = r_1 = r_0 + \Delta r$. For
simplicity thermal conduction and CR diffusion into and out of the parcel are
neglected so that the parcel expands adiabatically from the gas density $\rho$
to $\rho'$. At the final stage, the parcel is mixed with its surroundings. The
volume occupied by the thermal gas and that occupied by the CRs expands by the
same amount, implying
\begin{equation}
\label{eq:expansion}
\frac{\rho'}{\rho_0} = \frac{\rho_\CR'}{\rho_{\CR,0}} = 1 + \delta.
\end{equation}
The values of $\rho$, $P_\th$, and $P_\CR$ in the parcel initially at the same
average values at $r = r_0$ which are denoted by $\rho_0$, $P_{\th,0}$, and
$P_{\CR,0}$. After the parcel is displaced radially outward, the new fluid
quantities within the parcel are denoted $\rho'$, $P_\th'$, and $P_\CR'$. The
average fluid quantities (outside the parcel) at $r = r_1$ are denoted
$\rho_1$, $P_{\th,1}$, and $P_{\CR,1}$.  The total pressure is given by the sum
of the partial pressures and denoted without any subscript, $P = P_\th +
P_\CR$.  The difference between the thermal gas density in the displaced parcel
and its surroundings at $r = r_1$ is denoted by $\Delta\rho = \rho' -
\rho_1$. Other fluid quantities such as the thermal pressure are denoted
similarly, $\Delta P_\th = P_\th' - P_{\th,1}$, etc. Since the parcel expands
adiabatically, we have
\begin{equation}
\label{eq:adiabatic_exp}
P_\th' = P_{\th,0}\, (1+\delta)^{\gamma_\th}
\qquad\mbox{and}\qquad
P_\CR' = P_{\CR,0}\, (1+\delta)^{\gamma_\CR}.
\end{equation}
Assuming that the turbulent velocities are subsonic, the total pressure inside
the parcel  approximately matches the average value outside the parcel, $P_1 =
P'$. We assume a small displacement $|\Delta r| \ll r$ and expand the average
values of the density and the pressure outside the parcel at $r = r_1$ into
their Taylor series,
\begin{equation}
\label{eq:Taylor}
\rho_1 \simeq \rho + \frac{\dd \rho}{\dd r}\,\Delta r \qquad\mbox{and}
\qquad P_1 \simeq P + \frac{\dd P}{\dd r}\,\Delta r.
\end{equation}
Employing adiabatic expansion of the total pressure within the parcel
(Eqn.~\ref{eq:adiabatic_exp}), to lowest order in $\Delta r / r$ we arrive at
the following set of equations,
\begin{eqnarray}
\label{eq:Schwarzschild1}
\delta &=& 
\frac{1}{\gamma_\eff\, P}\,\frac{\dd P}{\dd r}\,\Delta r, \\
\label{eq:Schwarzschild2}
\frac{\Delta \rho}{\rho} &=& 
  \left(\frac{1}{\gamma_\eff\, P}\,\frac{\dd P}{\dd r}
  - \frac{1}{\rho}\frac{\dd \rho}{\dd r}\right) \,\Delta r, \\
\label{eq:Schwarzschild3}
\frac{\Delta A}{A} &=& \left(\frac{1}{P}\,\frac{\dd P}{\dd r}
  - \frac{\gamma_\eff}{\rho}\frac{\dd \rho}{\dd r}\right) \,\Delta r.
\end{eqnarray}
Here, we introduced for convenience the effective adiabatic index of the
composite fluid,
\begin{equation}
\label{eq:gammaeff2}
\gamma_\rmn{eff} \equiv \frac{\dd \log (P_\rmn{th} + P_\CR)}
{\dd \log \rho} = 
\frac{\gamma_\rmn{th}P_\th + \gamma_\CR\, P_\CR}
{P_\th + P_\CR},
\end{equation}
and the effective entropic function $A = (P_\rmn{th} + P_\CR)\,
\rho^{-\gamma_\eff}$. Since we kept only the lowest orders, it is not necessary
to specify in Eqns.~(\ref{eq:Schwarzschild1}) to (\ref{eq:Schwarzschild3})
whether the hydrodynamic quantities are evaluated at $r_0$ or $r_1$, and hence
we dropped the subscripts on $\rho$, $P$, etc. Since the displacement $\Delta
r$ is a signed quantity, the criterion for convective stability is 
\begin{equation}
\label{eq:convective_stability}
\frac{\Delta \ln A}{\Delta r} > 0.
\end{equation}
For negligible CR pressure contribution, this criterion is equivalent to the
classical Schwarzschild criterion
\begin{equation}
\label{eq:_classical_convective_stability}
\frac{\Delta \ln A_\th}{\Delta r} > 0.
\end{equation}
Interestingly, the ICM can be convectively unstable even when the specific
entropy of the thermal gas increases outward, as long as $\dd P_\CR/ \dd r$ is
negative and its absolute value is sufficiently large!

\bsp

\label{lastpage}

\end{document}